\documentclass[12pt,a4paper]{article}

\usepackage{amsmath}
\usepackage{amssymb}
\usepackage{mathrsfs}
\usepackage{graphicx}
\usepackage[pdftex,colorlinks=true,linkcolor=blue,citecolor=blue,urlcolor=blue]{hyperref}
\usepackage{enumerate}

\numberwithin{equation}{section}

\def\half{\frac{1}{2}}
\def\eq#1 { \begin{equation} #1 \end{equation} }
\def\eqn#1{ \begin{eqnarray} #1 \end{eqnarray} }
\def\nn { \nonumber }
\def\Reals{\mathbb{R}}
\def\Re{\mathop{\rm Re}}
\def\Im{\mathop{\rm Im}}

\def\s{\sigma}
\def\d{\partial}
\def\cL{\mathcal{L}}
\def\scI{\mathscr{I}}
\def\cT{\mathcal{T}}
\def\cH{\mathcal{H}}
\def\cO{\mathcal{O}}
\def\cD{\mathcal{D}}
\def\cM{\mathcal{M}}
\def\Op{\mathscr{O}}
\def\C#1{\left\langle #1 \right\rangle}
\def\ket#1{\left| #1 \right\rangle}
\def\bra#1{\left\langle #1 \right|}

\def\braf#1{{\hspace{-4pt}{\phantom{\rangle}}_f\hspace{-2pt}\left\langle
      #1 \right|}}
\def\brak#1#2{\left\langle #1 | #2 \right\rangle}
\def\braki#1#2{\hspace{-4pt}{\phantom{\rangle}}_i\hspace{-2pt}\left\langle
    #1 | #2 \right\rangle}
\def\brakf#1#2{\hspace{-4pt}{\phantom{\rangle}}_f\hspace{-2pt}\left\langle
    #1 | #2 \right\rangle}
\def\brakii#1#2{\hspace{-4pt}{\phantom{\rangle}}_i\hspace{-2pt}\left\langle
    #1 | #2 \right\rangle_i}
\def\brakfi#1#2{\hspace{-4pt}{\phantom{\rangle}}_f\hspace{-2pt}\left\langle
    #1 | #2 \right\rangle_i}
\def\ONbrakfi#1#2{\hspace{-4pt}{\phantom{\rangle}}_{\phantom{O}f}^{ON}\hspace{-4pt}\left\langle
    #1 | #2 \right\rangle_i}

\def\ONbrakf#1#2{\hspace{-4pt}{\phantom{\rangle}}_{\phantom{O}f}^{ON}\hspace{-4pt}\left\langle
    #1 | #2 \right\rangle}

\def\brakif#1#2{\hspace{-4pt}{\phantom{\rangle}}_i\hspace{-2pt}\left\langle
    #1 | #2 \right\rangle_f}
\def\brakff#1#2{\hspace{-4pt}{\phantom{\rangle}}_f\hspace{-2pt}\left\langle
    #1 | #2 \right\rangle_f}
\def\dg{{\dagger}}
\def\KG{\overleftrightarrow{\nabla}}
\def\vy{{\vec{y}}}
\def\vx{{\vec{x}}}
\def\vL{{\vec{L}}}
\def\vK{{\vec{K}}}
\def\bx{{\overline{x}}}
\def\by{{\overline{y}}}
\def\tPhi{\widetilde{\Phi}}
\def\tR{\widetilde{R}}
\def\Op{\mathscr{O}}

\begin{document}

\title{\vspace{-0.65in}Perturbative S-matrix for massive scalar fields in global de Sitter space}

\author{
  Donald Marolf${{}^{1,2}}$\thanks{\href{mailto:marolf@physics.ucsb.edu}
    {marolf@physics.ucsb.edu}},~
  Ian A. Morrison${{}^3}$\thanks{\href{mailto:i.morrison@damtp.cam.ac.uk}
    {i.morrison@damtp.cam.ac.uk}},~
  and Mark Srednicki${{}^1}$\thanks{\href{mailto:mark@physics.ucsb.edu}
    {mark@physics.ucsb.edu}} \\ \\
  {\it ${}^1$Department of Physics, University of California at Santa Barbara,}\\
    {\it Santa Barbara, CA 93106, U.S.A.} \\
  {\it ${}^2$Department of Physics, University of Colorado,}\\
    {\it Boulder, CO 80309, U.S.A.} \\
      {\it ${}^3$DAMTP, Centre for Mathematical Sciences,
    University of Cambridge,}\\ {\it Wilberforce Road,
    Cambridge CB3 0WA, U.K.}
}

\date{\today}

\maketitle

\begin{abstract}
 We construct a perturbative  S-matrix for interacting massive scalar fields
in global de Sitter space.  Our S-matrix is formulated in terms of asymptotic particle states
in the far past and future, taking appropriate care for light fields whose wavefunctions decay only very slowly near the de Sitter conformal boundaries.  An alternative formulation expresses this S-matrix in terms of residues of poles in analytically-continued
Euclidean correlators (computed in perturbation theory),
making it clear that the standard Minkowski-space result is obtained in the flat-space limit.
Our S-matrix transforms properly
under CPT, is invariant under the de Sitter isometries and perturbative field redefinitions, and
is unitary.  This unitarity implies a de Sitter version of the optical theorem.
We explicitly verify these properties to second
order in the coupling for a general cubic interaction, including both tree- and loop-level contributions. Contrary to other statements in the literature, we find that
a particle of any positive mass may decay at tree level to
any number of particles, each of arbitrary positive masses.
In particular, even very light fields (in the complementary series of de Sitter representations)
are not protected from tree-level decays.
\end{abstract}

\newpage

\tableofcontents


\newpage

\section{Introduction}
\label{sec:intro}

The S-matrix is an invaluable tool for studying quantum field theory (QFT) on Minkowski space.  Even leaving aside the all-important connection to experiments, its gauge invariance and invariance under field redefinitions make the S-matrix a powerful way to organize our understanding of quantum fields.  The Coleman-Mandula \cite{Coleman:1967ad}, Haag-Lopuszanski-Sohnius \cite{Haag:1974qh}, and Weinberg-Witten \cite{Weinberg:1980kq} theorems are prime examples of the utility of an S-matrix approach to field theory, and the above properties allow the S-matrix to be well-defined even in perturbative (and perhaps also nonperturbative) string theory.

At a more mundane level, the fact that all correct calculations of an S-matrix element must agree allows a clean comparison of different approaches, choices of gauge, etc., that helps to resolve potential controversies and hastens the advance of knowledge. In contrast, the lack of an S-matrix-like object has been sorely felt in years of controversy regarding quantum fields in de Sitter space.  Arguments continue over the interpretation of calculations in both simple self-interacting scalar theories on a fixed de Sitter background (compare e.g. \cite{Polyakov:2007mm,Polyakov:2009nq,Krotov:2010ma,Polyakov:2012uc,Marolf:2010zp,Marolf:2010nz,Marolf:2011aa}) and also in the more complicated case of gravitational theories (compare e.g. \cite{Allen:1986ta,Miao:2010vs,Giddings:2010ui,Miao:2011fc,Miao:2011ng,Mora:2012zi,Higuchi:2011vw}).

In this work, we introduce an S-matrix for
weakly-coupled quantum field theories in de Sitter space that can be computed order-by-order in perturbation theory.  The associated spaces of asymptotic states are defined using the interacting Hartle-Hawking state $\ket{\Omega}$ as the vacuum.
For theories of massive ($M^2 >0$) scalar fields, we show our S-matrix to be unitary, de Sitter invariant, and invariant under perturbative field redefinitions.  It also transforms properly under CPT and reduces to the usual S-matrix in the flat-space limit.  Our analysis is strictly perturbative and we consider in detail only theories of interacting scalars.  Our final discussion (section \ref{sec:discussion}) will comment briefly on extensions to gauge fields and why gauge-invariance is to be expected.  Perhaps our construction will also be of use in perturbative string theory on (likely meta-stable) de Sitter backgrounds.

\begin{figure}[t!]
  \centering
  \includegraphics[height= 4.8cm]{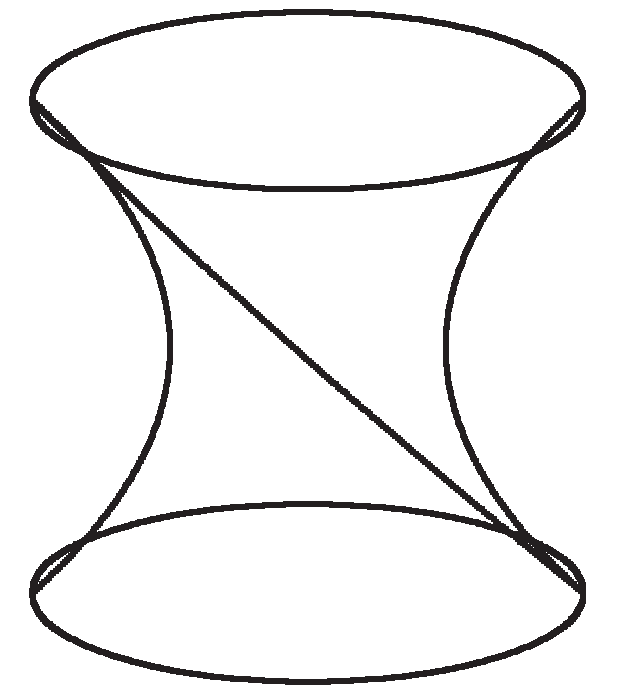}
  \includegraphics[height= 5cm]{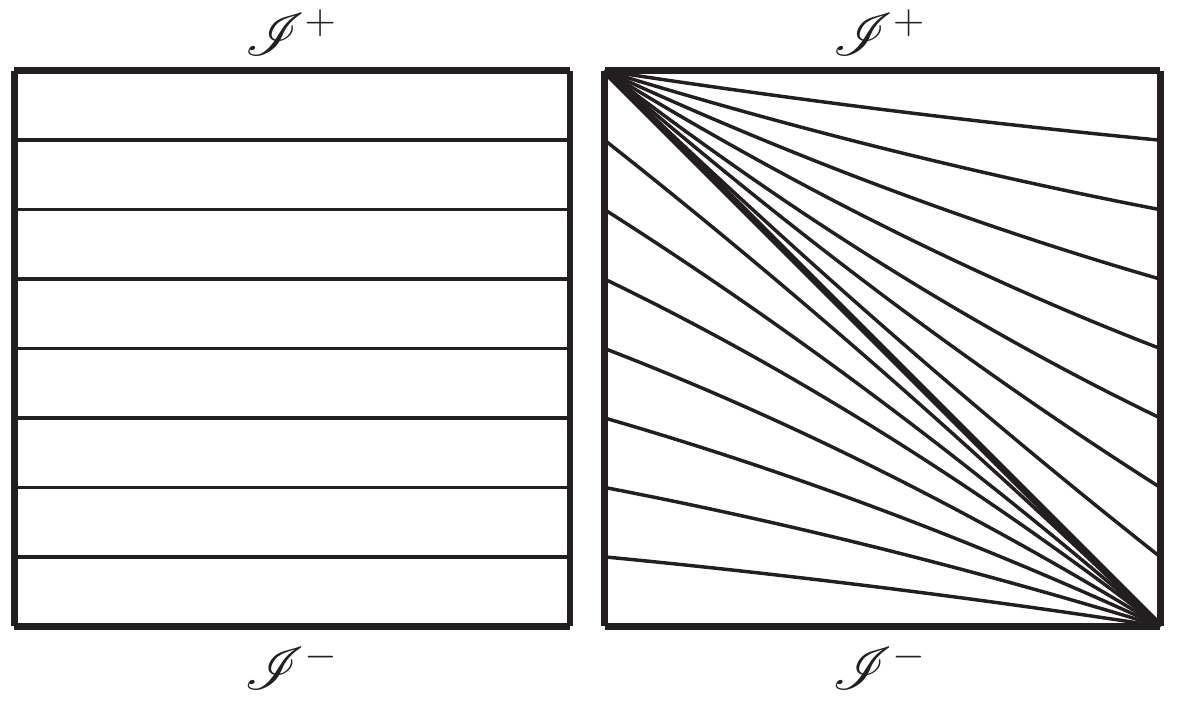}
  \caption{{\bf Left:} A finite piece of global $D$-dimensional
    de Sitter space represented as a timelike hyperboloid in $D+1$
    Minkowski space. The diagonal line is a cosmological horizon.
    {\bf Center:} A conformal (Carter-Penrose) diagram for global de
    Sitter marked with horizontal cross-sections, each representing
    an $S^{D-1}$. The left and right edges are the poles.
    {\bf Right:} The dS conformal diagram showing a cosmological
    horizon ${\cal H}$. Each point on the diagram represents an
    $S^{D-2}$ which contracts to zero size at the left and right edges.  }
    \label{fig:gdS}
    \label{fig:charts}
\end{figure}

To be specific, in this work we use the term ``de Sitter (dS)'' to refer to {\it global} de Sitter space, including both the contracting and expanding cosmological regions; see figure \ref{fig:gdS}.   Thus we formulate a global de Sitter S-matrix below, though we believe that an analogous S-matrix can be defined for both the Poincar\'e patch (also known as the $k=0$ cosmological patch), and the hyperbolic ($k=-1$) cosmological patch of dS.  Furthermore, these S-matrices should all be closely related through appropriate analytic continuations\footnote{While there should also be an S-matrix for the static patch, it would naturally be defined using asymptotic particle states built on the static vacuum in contrast to the Hartle-Hawking state used in our construction.}.

One sometimes
hears the claim that there can be no S-matrix on de Sitter space. We are aware of the following concerns regarding potential S-matrices:

\begin{enumerate}[i)]

\item \label{inout} The Minkowski S-matrix is defined using in/out perturbation theory, but it is well-known that in/out perturbation theory in dS suffers from infrared (IR) divergences.  So this definition
does not work in de Sitter space.

\item \label{unbounded} There is no positive-definite energy-like conserved quantity in de Sitter space.  As a result,
one-particle states can decay and all particles are unstable; thus there are no viable asymptotic states.  This is directly related to the concerns of e.g. \cite{Nachtmann:1968aa,Myhrvold:1983hx,Boyanovsky:1996ab,Boyanovsky:2011xn}.

\item \label{blue} The contracting phase of global de Sitter space tends to blueshift particles to high energies.  In a theory with dynamical gravity,  many states which are weakly-coupled near past infinity induce large gravitational back-reaction near the minimal-radius sphere (the de Sitter ``neck,'' $\eta =0$ in the coordinates of \eqref{eq:globalg} below).  Semi-classically, this should result in gravitational collapse to a cosmological singularity.  There is then no reason to expect late-time behavior described by weakly-coupled asymptotic states near the future de Sitter boundary.

\item \label{string} At least in string theory, all known de Sitter vacua are at best meta-stable.  So one expects that mere particle excitations of a de Sitter background cannot provide a complete set of outgoing states.

\item  \label{causal} The causal structure of global de Sitter space, and in particular the fact that its past/future boundary is spacelike, prevents any one observer from interacting with a complete set of ingoing/outgoing states.  This means that the S-matrix is not experimentally accessible to a single observer and hence need not necessarily be a well-defined object in a fundamental theory.

\end{enumerate}

The reader will note that comments (\ref{unbounded})-(\ref{causal}) are not directly relevant at the level we wish to work.  This is particularly manifest for issues (\ref{blue}), (\ref{string}), and (\ref{causal}), which concern dynamical gravity, string theory, or other supposed fundamental theories.  Issues (\ref{blue}) and (\ref{string}) (and arguably also (\ref{causal})) are also intrinsically non-perturbative, and so do not obstruct the more modest goal of formulating a perturbative S-matrix even for gravity or string theory.  While issue (\ref{causal}) implies that a direct connection to experiment is unlikely, a de Sitter S-matrix should nevertheless provide a useful theoretical tool along the lines noted above.

Let us briefly elaborate on the irrelevance of issue (\ref{unbounded}), the expected decay of all particles in de Sitter space.  First recall the usual situation in flat space:  at the perturbative level, each free field is associated with an appropriate
set of asymptotic particle states, whether or not these particles turn out to be stable.
In the unstable case there is a nonzero 1-to-2 (or more) amplitude
that can be computed order by order, and this amplitude is related
by the optical theorem to the self-energy correction for that particle. While this implies certain IR divergences in the
computation of the strict order-by-order S-matrix,\footnote{That is, without resummation of the 1PI self-energy corrections to the propagator.}
the physics of these divergences is well-understood.
We will show that all of this works analogously in de Sitter space with our definition of the perturbative
S-matrix.

It thus remains only to address the more technical concern (\ref{inout}).  Dealing with the associated potential IR divergences (in this case, those not associated with self-energy corrections) is in fact the main focus of our work below.  But the basic idea is simply that while most textbooks do use in/out perturbation theory to construct the Minkowski S-matrix, this choice is far from unique.  The LSZ formalism allows one to extract the S-matrix from vacuum correlators, no matter how they have been computed \cite{Haag:1992aa}.  In particular, one may define the Minkowski S-matrix by applying LSZ to time-ordered correlators computed using in/in perturbation theory via the closed-time-path formalism \cite{Schwinger:1960qe,Keldysh:1964ud}; see e.g. \cite{Chou,Landsman:1986uw} for reviews. To do so, one considers a path integral that begins at past infinity, runs up to future infinity, and then back to past infinity.  One then inserts operators near past infinity (say, at the beginning of the closed-time path) and also at future infinity (in the middle of the closed-time path).  Applying the LSZ formalism yields an S-matrix.  The result is manifestly equal to the usual S-matrix, except that the out-vacuum has been replaced by the in-vacuum.  But the positivity of the Minkowski conserved energy means that the in- and out-vacua agree, so that this closed-time-path definition does indeed reproduce the usual in/out S-matrix.  Since in/in perturbation theory is well-defined in dS and leads to good asymptotic behavior (e.g., an analogue of cluster decomposition) \cite{Marolf:2010nz,Marolf:2010zp,Hollands:2010pr,Marolf:2011aa}, the de Sitter analogue can lead to a good S-matrix.

\begin{figure}[t!]
  \centering
  \includegraphics{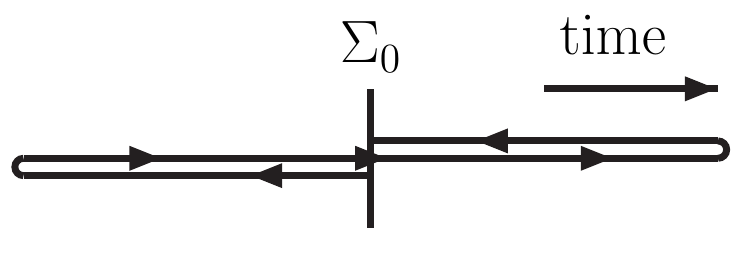}
  \caption{The time contour used to construct our global de Sitter S-matrix.
    The surface $\Sigma_0$ represents a cosmological horizon. See section
    \ref{sec:Lorentz} for details.\label{fig:SKcontour}
    }
\end{figure}

One complication is that in/in perturbation theory in de Sitter space is IR finite only when the closed time path begins and ends at a finite time.  In particular, it is best behaved when the path begins and ends on a cosmological horizon so that it constructs correlators in the interacting Hartle-Hawking state (called $|\Omega\rangle$ below); see \cite{Higuchi:2010aa} for a proof valid for scalar theories with $M^2 > 0$.  Thus the natural de Sitter analogue of the flat-space construction just outlined would be an S-matrix for the part of dS to the future of a cosmological horizon.  This region is called the Poincar\'e patch of dS by some authors and the expanding $k=0$ cosmological patch by others.  While we believe that this object does in fact exist, we save its construction and analysis for future work.  Instead, we construct an S-matrix for global de Sitter space below -- roughly speaking by gluing together two copies of the Poincar\'e patch construction just described, one for the expanding patch and one for the corresponding contracting patch to the past of the horizon -- along the de Sitter horizon.  More precisely, we use an out/out-in/in closed time path which begins on a de Sitter horizon, travels backward to past infinity, returns to the de Sitter horizon and keeps on going to future infinity, and then finally returns to the de Sitter horizon once again; see figure \ref{fig:SKcontour}. Operators are then inserted near past and future infinity and a de Sitter version of LSZ is used to extract an S-matrix.

This final LSZ-like step requires some care,  as one must sort out further IR issues to find an appropriate dS analogue of the flat-space LSZ formalism.  Some of these issues are already documented in the literature (see e.g. \cite{Floratos:1987aa,Akhmedov:2008pu,Higuchi:2008tn,Alvarez:2010te}).
The point is again that several procedures which are equivalent in flat space do not agree in dS, and that some seemingly natural dS generalizations lead to IR divergences for light (so-called complementary series) fields with masses $0 < M^2\ell^2 < (D-1)^2/4$; here $\ell$ is the de Sitter length scale and $D$ is the spacetime dimension.   Nevertheless, we show in section \ref{sec:goodOps} that a particular definition (based on extracting an S-matrix from poles in correlation functions) is free of unphysical IR divergences for generic\footnote{Our formulation fails for a measure zero subset of masses where self-energy corrections to the locations of poles are always large, even at small coupling $g$; see section \ref{sec:IntRev}.} $M^2 > 0$.  An alternate more explicit (but ultimately equivalent) procedure is described in appendix \ref{app:goodOps2}. While we find it enlightening to understand this explicit structure, it quickly becomes cumbersome at higher orders in perturbation theory.

Before proceeding, we should warn the reader of two further technical issues.  The first is that our choice of time contour and the use of the associated Schwinger-Keldysh-like formalism means that perturbation theory involves several distinct types of vertices and propagators.  This makes explicit computations more complicated than in flat space.  It may well turn out that other computational techniques, such as analytic
continuation from the Euclidean sphere or Euclidean AdS, will prove more
efficient for computing scattering amplitudes.

The second technical issue is that the lack of a positive-definite energy-like conserved quantity makes representation theory of the de Sitter group somewhat less powerful than its flat-space analogue.  In particular, de Sitter representation theory alone does not guarantee the orthogonality of what we call distinct multi-particle asymptotic states.  While such states are orthogonal for free theories, they cease to be orthogonal for general interacting theories.  Even states with different particle numbers develop non-zero inner products.  This, however,
is not an obstacle to defining a useful S-matrix.  We view it as essentially an accounting issue, though admittedly one that provides a sense in which our asymptotic states are not `free.'
A formal construction of
orthonormal asymptotic states (e.g., by the Gram-Schmidt procedure) is always possible. However,
such a construction obscures the physics, and so we will not emphasize it below.

The remainder of the paper is organized as follows.
We begin by reviewing basic properties of de Sitter space and de Sitter
QFTs in \S\ref{sec:prelims}.
We discuss the construction of asymptotic states in \S\ref{sec:states} and show that the resulting S-matrix has the desired properties in \S\ref{sec:Smatrix}.
The tools necessary to actually calculate S-matrix elements in Lorentz signature
are provided in \S\ref{sec:Lorentz}. Using these tools we compute
the S-matrix of a model cubic theory of heavy fields to second order in \S\ref{sec:Sheavy}.
We then revisit our model theory in
\S\ref{sec:light} allowing for fields of arbitrary
positive mass.
One interesting result is that, consistent with
\cite{Boyanovsky:2004gq,Boyanovsky:2012qs} but
in contrast to the claims of \cite{Bros:2006gs,Bros:2008sq,Bros:2009bz}, we find that complementary series particles generically decay.
In all cases, we explicitly verify unitarity by showing that our S-matrix satisfies the de Sitter version of the optical theorem. We conclude with a summary and discussion of open issues in \S\ref{sec:discussion}.

\section{Preliminaries}
\label{sec:prelims}

We begin by briefly reviewing de Sitter space and some relevant aspects of de Sitter quantum field theory.

\subsection{de Sitter geometry}
\label{sec:dS}

The $D$-dimensional de Sitter manifold $dS_D$ may be defined as the
single-sheet hyperboloid in an ambient $(D+1)$-dimensional
Minkowski space:
\eq{
  dS_D =
  \left\{ X \in \Reals^{D,1}\; | \; X \cdot X = \ell^2 \right\} .
}
The line element of de Sitter space may written in the
following convenient coordinates:
\eq{ \label{eq:globalg}
  \frac{ds^2}{\ell^2} = \left[ - \frac{1}{1+\eta^2} d\eta^2
    + (1+\eta^2) d\Omega_{D-1}^2 \right] ,
  \quad \eta \in \Reals.
}
Here $\ell$ is the de Sitter radius and $d\Omega_{D-1}^2$ denotes the
line element on unit $S^{D-1}$.
The time coordinate $\eta$ is related to the more familiar global
de Sitter time coordinate $t$ with $g_{tt} = -1$ via
$\eta = \sinh(t/\ell)$. In these coordinates the volume element is
$\sqrt{-g(x)} d^Dx = \ell^D (1+\eta^2)^{(D-2)/2} d\eta \,d\Omega_{D-1}(\vx)$.
The metric \eqref{eq:globalg} describes spatial sections which are spheres $S^{D-1}$ of
radius $\ell^2 (1+\eta^2)$. The conformal boundary of the chart consists of two disjoint spheres at $\eta \to \pm \infty$ denoted $\scI^\pm$.

At times it will be useful to use a second coordinate chart with line element
\eq{ \label{eq:Poincareg}
  \frac{ds^2}{\ell^2} = \tau^2 \left[ - \frac{1}{\tau^4} d\tau^2
    + d\vx^2 \right], \quad \tau \in \Reals ,
}
which also covers the entire manifold.
Here $d\vx^2$ denotes the line element on $\mathbb{R}^{D-1}$.
This chart has a coordinate singularity at $\tau=0$ corresponding
to a preferred cosmological horizon $\cH$.
In the regions $\tau < 0$ ($\tau > 0$) one may use instead the time
coordinate $\lambda = \pm \tau^{-1}$ to recover the more familiar
Poincar\'e coordinate chart (a.k.a. the expanding cosmological
chart). Penrose diagrams which depict the above charts are shown at center ($\tau$-chart) and right ($\eta$-chart) of
Fig.~\ref{fig:charts}.

While dS is maximally symmetric, it is important to remember that
 it has no globally timelike Killing vector field.
In particular, neither $\d_\eta$ nor $\d_\tau$ are Killing vectors.
As a result, global de Sitter does not possess a conserved quantity
associated with flow in only timelike directions; in this sense there is no conserved ``energy.''
For further details of de Sitter spacetime we refer the reader to
\cite{Hawking:1973uf,Birrell:1982ix,Spradlin:2001pw}.

\subsection{Scalar fields on dS}
\label{sec:dSQFT}

Free scalar QFT on a fixed de Sitter
background is a well-understood subject with many
good references (e.g., \cite{Birrell:1982ix,Allen:1985ux,Hollands:2010pr}).  Below we focus only on establishing notation and on certain group theoretical
aspects which will be useful later.  At the end of this section, we comment briefly on interacting fields.

\begin{figure}[t!]
  \centering
  \includegraphics[width = 2in]{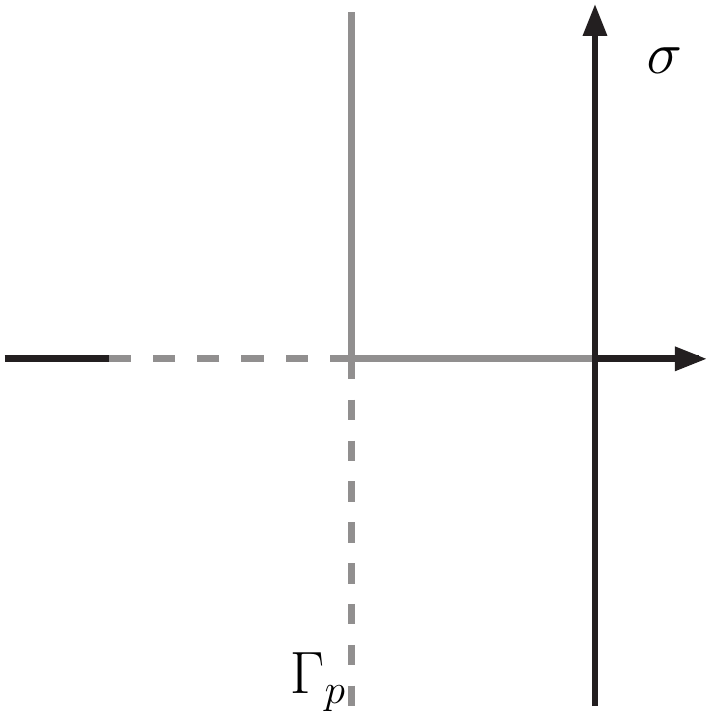}
  \caption{Scalar de Sitter UIRs are depicted by the solid gray line in the complex $\s$ plane.  For complementary series representations the weight $\s$ takes
    values along the negative real axis $\s \in (-(D-1)/2, 0)$
    while for the principal series $\s$ takes complex values
    $\s = -(D-1)/2 + i \rho$, $\rho \ge 0$.
    For each series the `conjugate weights' $-(\s+D-1)$ are depicted with a
    dashed gray line.
    We denote by $\Gamma_p$ the $\Re\s = -(D-1)/2$ contour.
    Representations with $\s$ values to the left of $\Gamma_p$
    are reducible; they may be represented as an integral
    over the principal series UIRs. Representations with $\Re \s > -(D-1)/2$
    and $\Im \s \neq 0$ are not unitary.
    \label{fig:dSreps}
}
\end{figure}

Scalar fields on de Sitter are associated with representations of the de Sitter
isometry group $SO(D,1)$. For instance, the one-particle states of a free scalar field $\phi_\s(x)$
described by the canonical Lagrangian
\eq{ \label{eq:Lfree}
  \cL_0[\phi_\s] = -\half \nabla_\mu\phi_\s \nabla^\mu\phi_\s(x)
  - \frac{M^2}{2}\phi_\s^2(x),
}
form a unitary irreducible representation (UIR) of $SO(D,1)$.
We define the \emph{weight} $\s$ via
\eq{
  M^2(\s) \ell^2 = -\s(\s+D-1) .
}
The right-hand side is invariant under $\s \to - (\s+D-1)$;
we choose $\s$ to be given by
\eq{
  \s = -\frac{(D-1)}{2} + \left[ \frac{(D-1)^2}{4} - M^2\ell^2 \right]^{1/2} .
}
The UIRs of the de Sitter group may be classified
as follows \cite{Vilenken:1991aa}:
\begin{enumerate}
  \item
    principal series:
    \eqn{
      \frac{(D-1)^2}{4} \le M^2\ell^2, \quad\Rightarrow\quad
      \s = -\frac{(D-1)}{2} + i \rho, \;\;
      \rho \in \Reals,\;\; \rho \ge 0 ,
    }
  \item
    complementary series:
    \eq{
      0 < M^2\ell^2 < \frac{(D-1)^2}{4}, \quad\Rightarrow\quad
      \s \in \left( -\frac{(D-1)}{2}, 0\right) ,
    }
  \item for some $D$, there is also a
    discrete series:
    \eq{
      M^2\ell^2 = - n(n+D-1) \;{\rm for}\; n\in\mathbb{N}_0,
      \quad\Rightarrow\quad
      \s = n .}

\end{enumerate}
``Heavy'' fields belong to the principal series while
extremely light fields with masses of order $\ell^{-2}$ belong to the
complementary series.  These will suffice for our purposes as we limit our analysis below to fields with mass $M^2 > 0$.
In contrast, UIRs of the discrete series are associated with $M^2 =0$ and certain tachyonic masses.
It is useful to visualize the de Sitter representations in the complex
$\s$ plane -- see Fig.~\ref{fig:dSreps}.

For a given mass $M^2 \ge 0$ solutions to the Klein-Gordon (KG) equation
$(\Box - M^2)\phi_\s(x) = 0$ are endowed with a positive-definite
``Klein-Gordon norm.'' Working in the global chart, there is
a natural basis of Klein-Gordon modes $u_{\s \vL}(x)$
which are distinguished by their angular momenta $\vL$ on the spatial
$S^{D-1}$ and are orthonormal with respect to the Klein-Gordon norm
\eqn{ \label{eq:KGnorm}
  ( u_{\s \vL_1},\; u_{\s \vL_2})_{\rm KG}
  &:=&
  - i \int d\Sigma^\nu(x) \left[ u_{\s \vL_1}(x) \KG_\nu u_{\s \vL_2}^*(x)
  \right] \bigg|_{\eta={\rm const.}}
  \nn \\
  &=&
  - i \ell^{D-2} (1+\eta^2)^{D/2} \int d\Omega_{D-1}(\vx) \,
  \left[ u_{\s \vL_1}(x) \KG_\eta u_{\s \vL_2}^*(x)
  \right] \bigg|_{\eta={\rm const.}}
  \nn \\ &=& \delta_{\vL_1\vL_2} .
}
The Klein-Gordon modes may be written explicitly as
\eq{ \label{eq:u}
  u_{\s \vL}(x) = \ell^{(2-D)/2} f_{\s L}(\eta) Y_\vL(\vx)  ,
}
where $Y_\vL(\vx)$ are spherical harmonics on $S^{D-1}$
parametrized by angular momenta $\vL$ with total angular momentum $L$
and the functions $f_{\s L}(\eta)$ may be written, e.g., as
\eqn{\label{eq:f}
  f_{\s L}(\eta) &=& N_{\s L} (1+\eta^2)^{L/2}
  {}_2F_1\left[L-\s, L+\s+D-1; L+ \frac{D}{2} ; \frac{1-i\eta}{2}\right]
  ,
  \\
  \label{eq:N}
  N_{\s L} &=& \frac{2^{-L-(D-1)/2} }{\Gamma\left(L+\frac{D}{2}\right)}
  \left[\Gamma(L-\s)\Gamma(L+\s+D-1)\right]^{1/2} ,
}
where ${}_2F_1(a,b;c;z)$ is the Gauss hypergeometric function
\cite{Slater:1966}. The behavior of the mode functions in
the asymptotic regions $|\eta|\to\infty$ is given by \cite{Marolf:2011aa}
\eqn{ \label{eq:fLargeEta}
  f_{\s L}(\eta)
  &=& (1+\eta^2)^{L/2}
  \left[K_{\s L}(2i\eta)^{\s-L} + K_{-(\s+D-1) L}(2i\eta)^{-(\s+D-1)-L}\right]
  \left[1 + \cO(\eta^{-2})\right] ,
  \nn \\ & &
  |\eta| \gg (L-\s) ,
}
with $K_{\s L}$ a coefficient whose value is
\eq{
  K_{\s L} = \frac{2^{\s+(D-3)/2}}{\sqrt{\pi}}
  \Gamma\left(\s+\frac{D-1}{2}\right)
  \left[ \frac{\Gamma(L-\s)}{\Gamma(L+\s+D-1)} \right]^{1/2} .
}

It is important to note, however, that the Klein-Gordon current does
not generally lead to a useful inner product between functions with support on
multiple de Sitter representations.
Integrals of the form
\eq{ \label{eq:bad}
  - i \int d\Sigma^\nu(x) \,
  \left[ u_{\s \vL_1}(x) \KG_\nu F(x)
  \right] \bigg|_{\eta={\eta_0}}
}
for general $F(x)$ are complex, depend upon $\eta_0$, and may diverge as
$|\eta_0| \to \infty$.
This includes the case where $F(x) = u^*_{\s_2\vL_2}(x)$
is a Klein-Gordon mode corresponding to a mass $\s_2 \neq \s$.
Indeed, the asymptotic behavior
(\ref{eq:fLargeEta}) shows that the expression
\eq{
  \lim_{\eta_0\to\pm\infty}\left[
  - i \ell^{D-2} (1+\eta^2)^{D/2} \int d\Omega_{D-1}(\vx) \,
  \left[ u_{\s_1 \vL_1}(x) \KG_\eta u_{\s_2 \vL_2}^*(x)
  \right] \bigg|_{\eta={\eta_0}}\right]
}
generally diverges for $\sigma_1 \neq \sigma_2$, though it is finite for $\s_1 = \s_2$ and is oscillatory for
$\s_1\neq\s_2$ both in the principal series (in which case it converges to zero as a distribution).

Free massive scalar fields in de Sitter have a unique normalizable
maximally symmetric Hadamard state $\ket{0}$ known variously as the Hartle-Hawking (HH)
state, the Euclidean vacuum, and Bunch-Davies state
\cite{Allen:1985ux,Mottola:1984ar}. The multiplicity of names can
be attributed to the multiplicity of ways to construct the state.
The first two names come from the fact that the state, defined by its
set of correlation functions, may be constructed by analytic
continuation from the Euclidean section $S^{D}$.
Alternatively, one may construct this state in either global de Sitter or the
Poincar\'e chart by imposing adiabatic or ``Bunch-Davies'' vacuum
conditions on the cosmological horizon.
Within the static chart this state has yet another interpretation
as the thermal state at the de Sitter temperature (it is
the unique thermal state regular on the cosmological horizons).

For a scalar field $\phi_\s(x)$
we denote the associated Wightman and time-ordered 2-point functions by
\eqn{
  W_\s(x_1,x_2) &:=&
  \bra{0} \phi_\s(x_1)\phi_\s(x_2) \ket{0} , \\
  G_\s(x_1,x_2) &:=& 
  \bra{0} T \phi_\s(x_1)\phi_\s(x_2) \ket{0} .
}
They may of course be written in terms of the Klein-Gordon
modes (\ref{eq:u}) as
\eqn{
  W_\s(x_1,x_2) &=& \sum_{\vL} u_{\s \vL}(x_1) u^*_{\s \vL}(x_2) ,
  \\
  G_\s(x_1,x_2) &=& W_\s(x_1,x_2) \theta(\eta_1 - \eta_2)
  + W_\s(x_2,x_1) \theta(\eta_2 - \eta_1) .
  \label{eq:GinKG}
}
In the latter expression $\theta(\eta)$ is the Heaviside step function taking values in $\{0,1\}$.

For interacting theories, perturbative corrections to such correlators often involve products of several free 2-point functions $W_\sigma$ or $G_\sigma$.  Such computations can be simplified by making use of so-called linearization formulae, which express these products as weighted integrals of $W_\mu$ or $G_\mu$ over some contour in the complex $\mu$-plane; see e.g. \cite{Marolf:2010zp}.  We will make use below of the slightly more complicated linearization formula
\eqn{
  u_{\s_1 \vL_1}(x) u_{\s_2 \vL_2}(x)
  &=& \sum_\vK \textrm{CGC}(\vL_1,\vL_2;\vK)
  \int_\mu (2\mu + D-1) \rho_{\s_1\s_2 L_1 L_2}(\mu,K) u_{\mu \vK}(x)
  \nn \\ &=:&
  \sum_\vK \int_\mu (2\mu + D-1)
  \rho_{12}(\mu,\vK) u_{\mu \vK}(x) ,
  \label{eq:uLinearization}
}
 for the Klein-Gordon mode functions themselves (which are just de Sitter harmonics). To explain this formula,
first recall that spherical harmonics obey their own linearization formula
\eq{ \label{eq:YLinearization}
  Y_{\vL_1}(\vx) Y_{\vL_2}(\vx) = \sum_{\vK} \textrm{CGC}(\vL_1,\vL_2;\vK)
  Y_{\vK}(\vx) ,
}
where $\textrm{CGC(\dots;\dots)}$ denote the generalized Clebsch-Gordon
coefficients of $SO(D)$ \cite{Junker:1993aa}.  Equation \eqref{eq:uLinearization} then follows by using \eqref{eq:u} and the analogous formula
\eq{ \label{eq:fLinearization}
  f_{\s_1 L_1}(\eta) f_{\s_2 L_2}(\eta)
  = \int_\mu (2\mu + D-1) \rho_{\s_1\s_2 L_1 L_2}(\mu,K)
  \,f_{\mu K}(\eta) .
}
for the time-dependent parts of the modes $u_{\sigma \vec L}$.  The kernel $\rho_{\s_1\s_2 L_1 L_2}(\mu,K)$ may be computed the methods of appendix A of \cite{Marolf:2010zp}.  Indeed, up to normalization the $\rho_{\sigma_1 \sigma_2}(\mu)$ found there is our $\rho_{\s_1\s_2 L_1 L_2}(\mu,K)$ evaluated at $L_1 = L_2 = K =0$.  A useful Mellin-Barnes representation \eqref{eq:rho_app} of $\rho_{\s_1\s_2 L_1 L_2}(\mu,K)$ is given in our own appendix~\ref{app:linearization}, though we will not need the details here.

In \eqref{eq:fLinearization} the symbol $\int_\mu\dots$ denotes a contour integral in the
complex $\mu$ plane with measure $d\mu/(2\pi i)$.
The contour is traversed from $\mu = -i\infty$ to $\mu = +i\infty$ within
the strip $\Re(\s_1+\s_2) < \Re \mu < 0$. The representation \eqref{eq:rho_app} can be used to show that
$\rho_{\s_1\s_2 L_1 L_2}(\mu, K)$ is a meromorphic function of $\mu$ within this strip
and that it decays sufficiently rapidly as $|\Im \mu| \to \infty$ for
the above integral to converge absolutely.  At generic arguments $\s_1,\s_2$ the function $\rho_{\s_1\s_2 L_1 L_2}(\mu, K)$ in this strip
has only simple poles at
\eqn{\label{eq:rhoPoles}
  \mu &=& \s_1 +\s_2 - 2n, \quad
  \mu = \s_1 -\s_2 - D+ 1 - 2n,
  \nn \\
  \mu &=& -\s_1 + \s_2 - D+ 1 - 2n,
  \quad
  \mu = -\s_1 - \s_2 - 2D+2 - 2n,
  \quad {\rm for}\; n \in \mathbb{N}_0 ;
}
higher order poles arise at arguments $\sigma_1,\sigma_2$ where two or more of the above poles coalesce.
These poles are required in order for the left-hand side of
(\ref{eq:fLinearization}) to have the same asymptotic behavior in the regime
$|\eta| \gg L_1 - \s_1,\; |\eta| \gg L_2 - \s_2 $ as the right-hand side.\footnote{
  This can be verified by inserting the asymptotic expansion
  (\ref{eq:fLargeEta}) for $f_{\mu K}(\eta)$ into (\ref{eq:uLinearization}),
  closing the $\mu$ integration contour appropriately and using the
  Cauchy integral formula to equate the contour integrals with the sum
  of residues due to the poles (\ref{eq:rhoPoles}).
}
It also useful to know that for $\s_1$, $\s_2$ corresponding to positive
mass-squared $\rho_{\s_1\s_2\vL_1\vL_2}(\mu,K)$ satisfies
\eq{
\label{eq:realrho}
  \left[\rho_{\s_1 \s_2 L_1 L_2}(\mu, K)\right]^*
  = \rho_{\s_1 \s_2 L_1 L_2}(\mu^*, K) .
}
so in particular $\rho_{\s_1 \s_2 L_1 L_2}(\mu, K) \in \Reals$ for $\mu$ in the
complementary series and generally complex for $\mu$ in the principal
series.  See appendix \ref{app:linearization} for further comments.

\subsection{Interacting fields}
\label{sec:IntRev}

Finally, we review some relevant features of interacting massive scalar de Sitter QFTs.  When the potential is bounded below, such theories admit a maximally symmetric state $\ket{\Omega}$ which is calculable at the level of perturbation theory, and for which perturbative correlators decay near past and future infinity.  As a result, general states in the Hilbert space have the property that local correlators approach those of $|\Omega \rangle$ in the distant past and future.  We say that $|\Omega \rangle$ is an attractor state for local operators \cite{Marolf:2010nz,Hollands:2010pr}. Since correlation functions in $\ket{\Omega}$ may be defined by Wick rotation from $S^D$, we
refer to $\ket{\Omega}$ as the Hartle-Hawking state.  However, these correlators may also be computed using the Lorentz-signature Schwinger-Keldysh
formalism (see e.g. \cite{Higuchi:2010aa}).  We delay a detailed discussion of theses techniques until
\S\ref{sec:CFs} when such technical details become necessary.

\begin{figure}[t!]
 \centering
  \includegraphics[width = 2in]{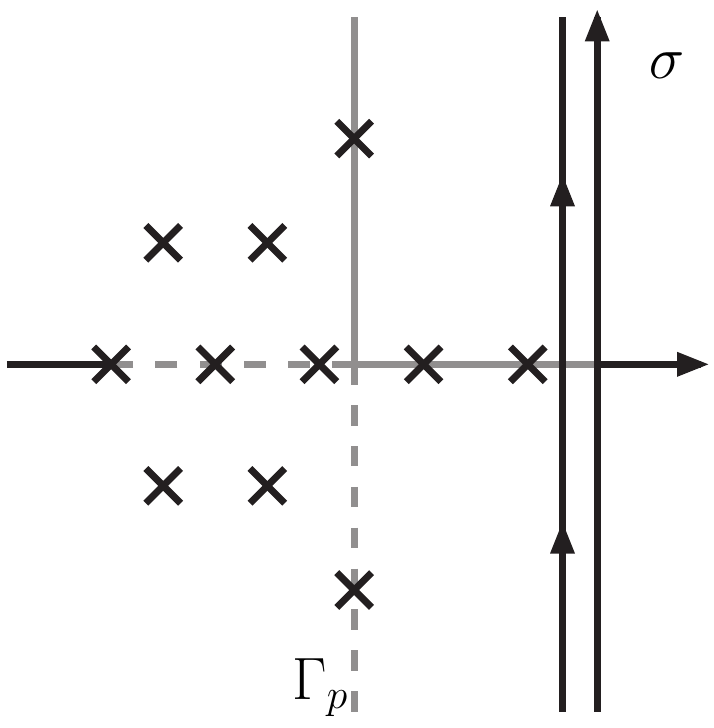}
  \caption{An example contour of integration for the Lehmann-Kall\"en
    2-point function. The contour is traversed from $-i\infty$ to $+i\infty$
    in the strip to the left of the
    imaginary axis and to the right of any singularities (x's for our example) in the K\"all\'en-Lehmann weight with $\Re \sigma < 0$.  The particular example involves the theory discussed in section \ref{sec:Sheavy}, which involves three species of scalar field.  We have shown the poles in $\langle \phi_1(x)\phi_1(y)\rangle$ at $\cO(g^2)$ for a case where $\phi_1,\phi_2$ lie in the principle series and $\phi_3$ lies in the complementary series.
    \label{fig:LK}
  }
\end{figure}

The structure of our S-matrix will be intimately tied to the de Sitter K\"all\'en-Lehmann representation of the
2-point function (and to corresponding generalizations for higher correlators).  Using the results of \cite{Marolf:2010zp,Marolf:2010nz,Hollands:2010pr}, it was argued in \cite{Hollands:2011we} that
the 2-point function of any (perhaps composite) scalar
operator $\Phi(x)$ in a general interacting theory may be written
\eqn{
  \C{\Phi(x_1)\Phi(x_2)} &:=& \langle\Omega|\Phi(x_1)\Phi(x_2)|\Omega\rangle \\
  &=& \int_\mu \rho(\mu) W_\mu(x_1,x_2) ,
\label{eq:LK}
}
where $\rho(\mu)$ is the K\"all\'en-Lehmann weight of $\Phi$ and $\int_\mu\dots$ denotes
a contour integral in the complex $\mu$ plane (corresponding to complex
operator weight $\mu$) with measure $d\mu/(2\pi i)$.
The integration contour is traversed from $-i\infty$ to $+i\infty$ to the
left of the imaginary axis and to the right of the singularities in
$\rho(\mu)$ -- see Fig.~\ref{fig:LK}. See also
\cite{Bros:1990cu,Bros:1994dn,Bros:1995js,Bros:2008sq} for earlier related results.

For example, in a free theory (\ref{eq:LK}) is simply
\eq{ \label{eq:LKfree}
  \bra{0} \phi_\s(x_1)\phi_\s(x_2) \ket{0}
  = \int_\mu \frac{(2\mu+D-1)}{(\mu-\s)(\mu+\s+D-1)} W_\mu(x_1,x_2)
  = W_\s(x_1,x_2) .
}
By deforming the contour of integration to lie along the UIRs
one may convert the contour integral in (\ref{eq:LK}) into an integral over
positive mass-squared $M^2>0$ (see Appendix A.2 of \cite{Hollands:2011we})
 but the form (\ref{eq:LK}) is more useful for our purposes.
Other 2-point functions (time-ordered, (anti-)symmetric, retarded/advanced)
may be obtained by exchanging $W_\mu(x_1,x_2)$ in the integrand of
(\ref{eq:LKfree}) with the appropriate free 2-point function.

The singularities in the K\"all\'en-Lehmann representation are particularly important.  As emphasized in \cite{Marolf:2010zp,Marolf:2010nz,Hollands:2010pr} they determine the large-$\eta$ behavior of the two-point function just as the behavior of flat-space correlators at large spacelike separations is governed by their singularities in the complex $m^2$.  Simple poles at UIRs contribute power law terms that appear in the corresponding free fields while higher poles contribute additional logarithms.  More complicated singularities
do not arise; in particular, there are no branch cuts\footnote{The familiar branch cuts of flat-space field theory arise in the limit $\ell \rightarrow \infty$ through the coalescence of an infinite family of poles.}. This fact was shown in \cite{Marolf:2010nz,Hollands:2010pr}, which analyzed the Mellin transforms of correlation functions in such theories in the perturbative expansion\footnote{The
      Mellin transform of the 2-point function is closely related to
      the K\"all\'en-Lehmann weight -- see Appendix A.2 of
      \cite{Hollands:2011we}.}.
The point may be argued without going into technical details as follows.
The Mellin transform for any correlation function of
a free field $\phi_\s(x)$ contains only simple poles.
The Mellin transform of correlation functions of composite
operators built from $\phi_\s(x)$ in the free theory are obtained
by repeated use of the linearization formula (\ref{eq:uLinearization}),
and from this one easily sees that the these transforms also contain only
poles.
Now consider an interacting theory constructed to first order in the
coupling $g$. The correlation functions in this theory obey
Schwinger-Dyson equations of the form
\eq{
  \label{eq:intSD}
  (\Box_x - M^2(\s)) \C{\phi_\s(x)\dots}^{(1)}
  = - g \C{\frac{\delta\Gamma_{\rm int}}{\delta\phi_\s(x)}\dots}^{(0)} ,
}
where $g \Gamma_{\rm int}$ is the interacting part of the effective
quantum effective action for $\phi_\s(x)$. The right-hand side is just
a correlation function of the linear theory, so it's Mellin transform
contains only poles. We obtain the Mellin transform of
the left-hand side by inverting the Klein-Gordon operator on the right-hand
side; this adds pole singularities to the MB transform.
By repeating this procedure for all correlators we deduce that
the Mellin transforms of all $\cO(g)$ correlators of $\phi_\s(x)$ contain
only poles. The transforms of correlators of composite operators at
$\cO(g)$ likewise contain only poles.  Thus the argument can be repeated at higher orders and extends to all orders in perturbation theory.

In perturbation theory about a free theory with weights $\sigma_i$, the following additional properties of $\rho(\mu)$ also hold.
Property (\ref{causality}) follows directly from causality.  The remaining properties follow from the argument above, though they can alternatively be read off from the Mellin-Barnes analysis of \cite{Marolf:2010nz,Hollands:2010pr}.
These results will play key roles in constructing our S-matrix, showing that it is finite, and demonstrating the desired properties.
\begin{enumerate}[i)]
  \item {\bf Mass gap:} All operators are ``massive,'' by which we mean that their K\"all\'en-Lehmann
    weight $\rho(\mu)$ is analytic on a strip in the complex $\mu$ plane
    $\epsilon < \Re \mu < 0$ for some $\epsilon < 0$.  As a result,
    there always exists an allowed path for the contour of integration.

  \item {\bf Causality:} \label{causality} The 2-point function of real scalar operators
    is real at spacelike separations. This requires that
    \eq{
      \left[ \rho(\mu) \right]^* = \rho(\mu^*) .
    }
    Any singularity in the lower half-plane has an image in the upper
    half-plane.

  \item {\bf Representation theory:}
    The singularities of $\rho(\mu)$ are contained in
    the region
    \eq{
      R_{U} = \left\{ \Re \mu \le -\frac{(D-1)}{2} \quad \cup \quad
        \mu \in (-(D-1)/2,0) \right\} .
    }
    This is the largest region consistent with scalar fields forming
    unitary representations of the de Sitter group. This is a direct corollary of property (\ref{generic}) below, though we state it separately to highlight its importance.

  \item {\bf Lattice of poles:}
  \label{generic}
    For any scalar operator $\Phi(x)$,
    all poles in $\rho(\mu)$ lie on the lattice built by taking linear combinations (with non-negative coefficients) of $\sigma_i -2n$, $-[(D-1)+\sigma_i] -2n$ for non-negative integers $n$ ($n \in {\mathbb N}_0$). Here $\sigma_i$ are the weights $\sigma_i$ of the free fields about which we perturb.

    We will call a theory {\em exceptional} if some weight $\sigma_i$, or some `conjugate weight'
    $\tilde \sigma_i =  -[(D-1)+\sigma_i]$ can be expressed as a non-trivial combination of the above generators with non-negative integer coefficients.  By `non-trivial,' we mean that the sum of the coefficients should be greater than $1$; i.e., that they should not just be $(1,0,0,\dots)$ or some permutation thereof.  Our definition of the de Sitter S-matrix will fail precisely for this (measure zero) set of exceptional theories.  We will call a theory {\em generic} if it is not exceptional in the above sense.

    Note that for generic theories the bare weights $\sigma_i$ may be recovered given only the lattice of poles; i.e., given only the correlators computed to some perturbative order.

 \item{\bf Smooth self-energy:} \label{comp} For generic theories, the one-particle irreducible (1PI) diagrams for two-point functions of composite operators have no poles at the bare weights $\sigma_i$ or the bare conjugate weights $\tilde \sigma_i$.  This may be shown by the same methods used in \cite{Marolf:2010nz} to show that all diagrams (even if one-particle reducible) decay at large arguments at least as the two-point function of the lightest field involved.  As a result, the corresponding self-energy $\Pi(\mu)$ (see below) is smooth at $\mu = \sigma_i, \tilde \sigma_i$.

\end{enumerate}

Finally, when $\Phi(x) = \phi_\s(x)$ is a massive operator with bare mass
$M^2(\s)$ it can be convenient to write the K\"all\'en-Lehmann form in
terms of the self-energy $\Pi(\mu)$ defined as usual by the
sum of 1PI $1 \rightarrow 1$ graphs:
\eq{ \label{eq:LK1PI}
  \C{\Phi(x_1)\Phi(x_2)}
  = \int_\mu \frac{(2\mu+D-1)}{(\mu-\s)(\mu+\s+D-1) - \Pi(\mu)}
  W_\mu(x_1,x_2) .
}
In a strict perturbative expansion of the 2-point function
about small couplings the denominator of (\ref{eq:LK1PI}) must
be expanded. Suppose the coupling constant is $g$; then the perturbative
expansion of $\Pi(\mu)$ is
\eq{ \label{eq:Piexpansion}
  \Pi(\mu) = \sum_{n=1}^\infty \Pi^{(n)}(\mu) ,
}
where $\Pi^{(n)}(\mu)$ is $\cO(g^n)$, and the expansion of the
2-point function is
\eqn{ \label{eq:LK1PIexpansion}
  \C{\Phi(x_1)\Phi(x_2)}
  &=& \int_\mu \frac{(2\mu+D-1)}{(\mu-\s)(\mu+\s+D-1)} W_\mu(x_1,x_2)
  \Bigg[
    1 + \frac{\Pi^{(1)}(\mu)}{(\mu-\s)(\mu+\s+D-1)}
    \nn \\ & &
    + \frac{\Pi^{(2)}(\mu)}{(\mu-\s)(\mu+\s+D-1)}
    - \left(\frac{\Pi^{(1)}(\mu)}{(\mu-\s)(\mu+\s+D-1)}\right)^2
    + \cO(g^3)
  \Bigg] . \ \ \ \  \ \ \ \
}
Clearly then, order by order, each term in (\ref{eq:Piexpansion})
inherits the five characteristics of $\rho(\mu)$ mentioned above.

One may also consider the 2-point function obtained by truncating the sum \eqref{eq:Piexpansion} to some finite order $N$ and inserting the result into \eqref{eq:LK1PI}.
While this object includes contributions from an infinite number
of graphs at infinite orders in the strict perturbative expansion, its study is well-motivated when, say, the physics of interest is governed by the nature of the singularities (poles) in
the K\"all\'en-Lehmann weight.   As in flat space, the self-energy can be used to compute perturbative shifts of these poles at each order.  In generic theories, the fact that $\Pi(\mu)$ is smooth at both the bare weights and the bare conjugate weights (see point (\ref{comp}) above) implies that these shifts are finite at each order.

Correlation functions of three or more fields are then computed order-by-order
via a diagrammatic expansion with the same general structure as various
flat space constructions;
see section~\ref{sec:CFs} below.
In particular, all higher-point correlation functions have a perturbative
expansion analogous to \eqref{eq:LK1PIexpansion}
\cite{Hollands:2010pr,Marolf:2010nz}.
Thus, for generic theories, the order of any pole at $\sigma_i, \tilde \sigma_i$ in any $n$-point
correlation function is controlled by the perturbative expansion in precisely the same manner
as in flat space.

\section{Asymptotic states}
\label{sec:states}

The quantum states involved in traditional scattering experiments
may be described in terms of multiple widely-separated,
non-interacting particles in the far past/future.
While the construction of such particle states is trivial
in free theories, we do not expect to be able to construct such
states in interesting interacting field theories at finite separations.
The best we can do is construct \emph{asymptotic particle states}
which enjoy a particle-state interpretation in the limits
$\eta\to \pm \infty$. In the Heisenberg picture, inner products between
past- and future-asymptotic states define the S-matrix (a more precise
definition will be provided in \S\ref{sec:Smatrix}).

The construction of asymptotic particle states will
present the main technical challenge below. We therefore devote this section to addressing
this issue in detail. Our definition of asymptotic particle states
is a straightforward generalization of definitions that appear
in standard QFT textbooks \cite{Weinberg:1995mt,Haag:1992aa,Srednicki:2007}.  However, we warn the reader that one must use caution in applying the term `non-interacting' to our asymptotic particle states.  While our states transform in a simple way under the de Sitter group (and thus may be said to evolve freely in time), for interacting theories the inner product on our space of asymptotic multi-particle states will differ from that associated with any free theory.

\subsection{General formalism}
\label{sec:formalism}

The starting point for our construction is a (bosonic) local quantum field $\Phi(x)$ and the corresponding de Sitter K\"all\'en-Lehmann
spectral representation reviewed in section \ref{sec:IntRev}.  As in flat space,
we take the masses of the asymptotic particle states to be associated
with values of $\sigma$ at which the integrand is singular.
In particular,
for interacting fields we take them to be given by the bare weights $\sigma_i$ defined by the lattice of poles; see point \ref{generic} in section \ref{sec:IntRev}.
Precisely free fields have a K\"all\'en-Lehmann representation supported only on a single dS UIR which defines the associated particles.

Our goal is to use $\Phi(x)$ to construct asymptotic particle
states labeled by the quantum numbers of a de Sitter representation
$n = (\s,\vL)$. We will give particular prescriptions for doing so
below. For now, let us simply list the desired properties of such
initial and final asymptotic states $\{ \ket{\psi}_{i/f}\}$:
\begin{enumerate}
  \item The states are normalizable:
    ${}_{i/f}\langle \psi | \psi \rangle_{i/f} < \infty$.
  \item The Hilbert-space is spanned by multi-particle states , i.e. any initial (final)
    state is a sum of states labeled by strings of quantum\footnote{
      Depending on the details of the theory more quantum numbers may
      be necessary to fully specify the state, but $n= (\s,\vL)$ will
      be sufficient for our purposes.} numbers.  So the states
    $\ket{\psi}_{i/f} = \ket{n_1, n_2,\dots,n_k}_{i/f}$ form a basis.
  \item Such basis states are symmetric with respect to their particle labels, i.e.,
    $\ket{n_1,n_2,\dots}_{i/f} = \ket{n_2,n_1,\dots}_{i/f}$.
  \item
    In the asymptotic past (future) the initial (final) asymptotic states
    transform as symmetric tensor products under the action of the (diagonal)
    de Sitter group. This requirement is most easily understood in the
    Heisenberg picture.
    If $U(g)$ is a de Sitter transformation with parameter
    $g$ then a $k$-particle asymptotic state transforms as
    \eq{ \label{eq:AStransformation}
      U(g) \ket{n_1,n_2,\dots,n_k}_{i/f}
      = \ket{g n_1, g n_2,\dots, g n_k}_{i/f} , \quad gn_i = (\s, g \vL_i).
    }
The notation $g \vL$ denotes an appropriate linear combination of angular momenta.
The weight $\s$ is unchanged by de Sitter transformations
    because it denotes the eigenvalue of the quadratic Casimir of $SO(D,1)$
    which commutes with $U(g)$.
    We say that states satisfying \eqref{eq:AStransformation} transform covariantly under the
    the de Sitter group.  We also require that \eqref{eq:AStransformation} be unitary.
\end{enumerate}

Each set of multi-particle states should include a vacuum state $\ket{v}_{i/f}$.
From criteria 1 and 4 we conclude that this state must be both normalizable
and invariant under the action of the de Sitter group:
$U(g) \ket{v}_{i/f} = \ket{v}_{i/f}$. For free fields these requirements
limit the possible choices of $\ket{v}_{i/f}$ to the so-called Mottola-Allen
vacua \cite{ Mottola:1984ar,Allen:1985ux}.
If we require
that the short-distance structure be Hadamard,
the free Hartle-Hawking state becomes the unique allowed choice .  It is thus natural to take both $\ket{v}_{i/f}$ to be the Hartle-Hawking state $\ket{\Omega}$ in the interacting case as well.
The fact that $\ket{\Omega}$ is an attractor state for local operators \cite{Marolf:2010zp,Marolf:2010nz,Hollands:2010pr,Marolf:2011aa} means that it is also a natural choice from the perspective
of local physics in the asymptotic regions\footnote{
  The authors
  \cite{Bousso:2001mw,Spradlin:2001nb,Conamhna:2003aa,Lagogiannis:2011st}
  have considered asymptotic
  states constructed out of the non-Hadamard Mottola-Allen vacua.
  Though they typically have divergent
  stress-tensors and do not reduce to the Minkowski asymptotic particle states in the
  flat-space limit \cite{Brunetti:2005pr},
  it has been argued that these asymptotic states may nevertheless play an important role
  in a dS/CFT correspondence.
}.

In the free case, the full space of states is readily constructed by acting with ladder operators
\eq{ \label{eq:ASfree}
  \ket{n_1,n_1,\dots,n_k}_{i/f} := a^\dg_{n_1} a^\dg_{n_2}\dots a^\dg_{n_k} \ket{0}_{i/f},
}
where
\eqn{ \label{eq:Afree}
  a^\dg_{\s\vL} &:=&
  -i \int d\Sigma^\nu(\vec{x}) \left[ u_{\s\vL}(x) \KG_\nu \phi_\s(x) \right]
  \bigg|_{\eta={\rm const}} ,
  \nn \\
  a_{\s\vL} &:=&
  -i \int d\Sigma^\nu(\vec{x}) \left[ \phi_\s(x) \KG_\nu u^*_{\s\vL}(x) \right]
  \bigg|_{\eta={\rm const}} ,
}
and as usual we also have
\eq{ \label{eq:CCR}
 \phi_\s(x) = \sum_\vL \left[ u_{\s\vL}(x) a_{\s\vL} +
     u^*_{\s \vL}(x) a^\dg_{\s\vL} \right] , \quad \left[ a_{\s \vL}, a_{\s \vK} \right] = 0
  = \left[ a^\dg_{\s \vL}, a^\dg_{\s \vK} \right] ,
  \quad
  \left[ a_{\s \vL}, a^\dg_{\s \vK} \right] = \delta_{\vL \vK} .
}
From the commutations relations (\ref{eq:CCR}) we readily verify that
these particle states are normalizable and symmetric, and from (\ref{eq:Afree})
we see that the ladder operators transform covariantly under the
de Sitter group: $U(g) a^\dg_{\s\vL} = a^\dg_{\s g\vL}$ where $g \vL$ denotes
the covariantly transformed angular momenta. Unitarity of $U(g)$ follows from the manifest de Sitter symmetry.  Thus the states
(\ref{eq:ASfree}) satisfy our criteria for asymptotic
particle states in both past and future.
As an added bonus, this set of states also forms an orthonormal basis.  Since $\ket{0}_i = \ket{0}_f = \ket{0}$, the S-matrix is trivial.

We will describe extensions of this construction to interacting fields in sections \ref{sec:IR} and \ref{sec:goodOps} below.  For now, we mention that the initial (final) states define two multi-particle Hilbert spaces $\cH_{i/f}$ and that for perturbation theory in small couplings we expect both
$\cH_{i/f}$ to be isomorphic to the entire Hilbert space $\cH$. So
each set of multi-particle asymptotic states should provide a complete basis for $\cH$.
We simply take this as an assumption, though our explicit verification of the optical theorem at low orders of perturbation theory in sections
\ref{sec:Sheavy} and \ref{sec:light} provides supporting evidence.

We noted above that the states \eqref{eq:ASfree} are orthonormal for free fields.
However, we have imposed no requirement on the inner product  $\brakff{a}{b}$ between two final states (or $\brakii{a}{b}$ between two initial states) for more general theories.  In particular, because requirement (4) involves only the diagonal action of the de Sitter group (with all $g$'s in \eqref{eq:AStransformation} being the same), it is much weaker than requiring our asymptotic multi-particle states to be tensor products of asymptotic single-particle states, and this in principle allows the inner products $\brakff{a}{b}$ to depend on the interactions. In Minkowski space, the continuous spectrum of the energy-momentum operator means that the associated conservation laws enforce orthogonality of distinct asymptotic particle states even in the interacting theory.
But de Sitter space does not enjoy such simple conservation laws:
there is no positive-definite conserved energy, and conservation of the spatial
momenta is governed by the addition of angular momenta (i.e., the
Clebsch-Gordon coefficients of $SO(D)$) which is less stringent than
the addition of linear momenta.
So long as $|L_1 - L_2| \le L_3 \le L_1 + L_2$, dS group theory alone does not require the vanishing of overlaps like $\brakii{n_1 n_2}{n_3}$.  Instead, we will be led to compute such inner products order by order in perturbation theory using appropriate Feynman-like diagrams.

\subsection{Interacting theories and IR divergences}
\label{sec:IR}

We now turn to the construction of asymptotic states for (massive) interacting fields; i.e., we seek a generalization of the LSZ formalism to de Sitter space.  Unfortunately, the generalization of LSZ to de Sitter space is not unique.  We describe one generalization below that is both conceptually and computationally straightforward.  It turns out to suffice for explicit computations of heavy (principal series) fields but becomes IR divergent in the presence of light (complementary series) fields.  Closely related IR divergences were noted previously in \cite{Floratos:1987aa,Akhmedov:2008pu,Higuchi:2008tn,Alvarez:2010te}.   After identifying the relevant issues below, we present a modified procedure in section \ref{sec:goodOps} that is free of unphysical divergences.  The procedure of \S\ref{sec:goodOps} is also required to give a field-redefinition-invariant definition of the S-matrix even for heavy fields.

Our first attempt at an LSZ-like prescription involves defining (now time-dependent) ladder operators
in analogy with those of the free theory (\ref{eq:Afree}):
\eq{ \label{eq:Aint}
  a^\dg_{\s\vL}(\eta) :=
  -i \int d\Sigma^\nu(\vec{x}) \left[ u_{\s\vL}(x) \KG_\nu \phi_\s(x) \right]
  \bigg|_{\eta} ,
  \quad
  a_{\s\vL}(\eta) :=
  -i \int d\Sigma^\nu(\vec{x}) \left[ \phi_\s(x) \KG_\nu u^*_{\s\vL}(x) \right]
  \bigg|_{\eta} .
}
We hope to construct initial (final) states via
\eq{ \label{eq:ASint}
  \ket{n_1,n_2,\dots,n_k}_{i/f} = \lim_{\eta\to\mp \infty}
  \left[ \prod_{j=1}^k a^\dg_{n_j}(\eta) \right] \ket{\Omega} .
}

As mentioned above,  asymptotic particle states will not generically
remain orthonormal in the presence of interactions.  But so long as the inner products are finite this need not be seen as an obstruction.  Indeed, they may be computed order by order in perturbation theory as we discuss in sections \ref{sec:Sheavy} and \ref{sec:light} below.  If desired, one may then to choose to construct orthonormal bases via e.g. the Gram-Schmidt procedure.  The resulting
orthonormal asymptotic states then enjoy all of the properties discussed in \S\ref{sec:formalism}
except that they do not have well-defined particle numbers with
respect to the field operator $\phi_\s(x)$.

However, it turns out the inner products of these states can diverge even at lowest order in a coupling
(i.e., at tree level) when complementary series fields are involved.  In such cases the Klein-Gordon
inner products used to define the ladder operators (\ref{eq:Aint}) may fail to have well-defined limits as $\eta \rightarrow \pm \infty$.
Nothing is wrong with the theory per se: the $\phi_\s(x)$ correlation
functions are well-defined and for non-coincident configurations
enjoy power-law decay as the operators are taken to large times $|\eta| \gg 1$.
The problem is simply that the Klein-Gordon inner products contain a competing
power-law growth $d\Sigma u_{\s \vL} \sim (1+\eta^2)^{D/2} \eta^\s$.
The single potentially-dangerous term cancels precisely for free fields, but not with interactions.

\subsection{The $R_\s$ projector}
\label{sec:goodOps}

The IR divergences found above appear to result from technical rather than physical issues.  In Minkowski space, treating the KG inner product in a natural distributional sense usefully extracts the residue of the operator at a simple pole and allows us to define asymptotic states. When correlators of $\phi_\sigma$ can be written as integrals over only principle series UIRs, the same is true in dS.
But this procedure fails in dS when correlators receive contributions from complementary series UIRs.
As already mentioned in \S\ref{sec:dSQFT}, in de Sitter space the KG norm
fails to converge on generic functions in
the asymptotic regions, even when these functions decay exponentially.
So it seems that the KG norm is simply not the right tool
for extracting the complementary-series residues and thus the desired particle content from the
correlation functions.

We will resolve this technical issue by
modifying (\ref{eq:Aint}) so as to remove contributions from any support away from the designated value of $\s$.  This may be accomplished by inserting the operator
\eq{ \label{eq:R}
  R_\s := \lim_{b\to\infty}
  \exp\left[ - \frac{b^2}{(\Box_x - M^2(\s) )^2}\right]
}
into the
definition of the ladder operators:
\eqn{ \label{eq:Aint2}
  a^\dg_{\s\vL}(\eta) &:=&
  -i \int d\Sigma^\nu(\vec{x})
  \left[ u_{\s\vL}(x) \KG_\nu R_\s \Phi(x) \right]
  \bigg|_{\eta} , \nn \\
  a_{\s\vL}(\eta) &:=&
  -i \int d\Sigma^\nu(\vec{x})
  \left[ R_\s \Phi(x) \KG_\nu u^*_{\s\vL}(x) \right]
  \bigg|_{\eta} .
}
Note that $R_\s$ annihilates smooth functions of $\sigma$ as well as poles elsewhere in the complex plane, though  for any finite value of $b$ it leaves the residues at $\sigma$ unchanged. Since all derivatives of $R_\sigma$ with respect to $\sigma$ vanish at the pole, the higher order coefficients in the Laurent expansion about $\sigma$ are unchanged as well. Thus the multi-particle states defined by acting on $\ket{\Omega}$ with \eqref{eq:Aint2} are normalizable when the pole at $\sigma$ is simple, and these states agree with those built from \eqref{eq:Aint} when all fields lie in the principal series.

When the pole at $\sigma$ is not simple, it is important to point out that acting on $\ket{\Omega}$ with \eqref{eq:Aint2} still generally yields non-normalizable states.  However, the remaining divergences are only logarithmic in $\eta$, as opposed to the power-law divergences that arose for light fields in section \ref{sec:IR}.
Similar logarithmic divergences are encountered in Minkowski space, where they are associated with perturbative self-energy corrections that shift the original (bare) simple poles to new locations in the complex plane.
The general structure of the 1PI expansion \eqref{eq:LK1PIexpansion} shows that for generic theories (in which the self-energy $\Pi(\mu)$ is smooth at the poles) the same is true in dS; i.e., the series \eqref{eq:LK1PIexpansion} can be summed to yield \eqref{eq:LK1PI} which has only simple poles near the original bare poles $\sigma, -[\sigma +D-1]$.  While we will not work with the resummed series below, it is clear that logarithmic divergences of this type are physically meaningful.
We therefore take them to be acceptable in an order-by-order computation of the perturbative S-matrix
that does not use resummed propagators. We then say that, despite these divergences, our S-matrix ``exists'' for generic theories.  We will comment further on the exceptional theories
(in which $\Pi(\mu)$ is not smooth at the poles)
in section \ref{sec:discussion}.

An operator $\Phi$ for which the naive LSZ expressions \eqref{eq:Aint} converge up to logarithms
will be called a ``good operator''.
As we have discussed, examples include free fields of all masses, as well as general local operators in theories
containing only principal series fields. The operator $\Phi(x) := R_\s\phi_\s(x)$ is
a good operator in any theory.
Some might object that this $\Phi(x)$
is a rather formal object, as the application of $R_\sigma$ involves taking an infinite number
of derivatives. But it is useful to
note that there is an entire class of operators that are in some sense equivalent to $\Phi(x)$.
Let us write $\tPhi(x) \cong \Phi(x)$ when the operators differ only by terms which decay faster than $\cO(\eta^{-(\s+D-1)})$, i.e.
only up to terms which do not contribute to amplitudes of asymptotic states.  Clearly any such $\tPhi(x)$ is also a good operator.  At a fixed order in perturbation theory it is straightforward (if
tedious) to construct operators $\tPhi(x)$ equivalent, in this sense,
to $R_\s\phi_\s(x)$ whose explicit expressions involve only a finite
number of terms and contain only a finite number of derivatives of $\phi_\s$.
This procedure is illustrated in appendix \ref{app:goodOps2}.

\section{The S-matrix}
\label{sec:Smatrix}

We have assumed that both the initial and final spaces of asymptotic states form complete bases for our Hilbert space.  For generic theories we then define the S-matrix to be the operator which induces the corresponding change of basis.  To be explicit, given bases $\ket{A}_f$ and $\ket{B}_i$ for the initial and final spaces of asymptotic states, we define the corresponding S-matrix $S_{BA}$ such that
\eq{ \label{eq:S}
\ket{B}_i = \sum_A S_{BA} \ket{A}_f.
}
For orthonormal bases, this gives
\eq{ \label{eq:Sorth}
S_{BA} = \brakfi{A}{B}.
}
The essential steps in computing the S-matrix are the construction of correlators in the Hartle-Hawking state $|\Omega \rangle$ and the application to those correlators of the LSZ-like operations described in sections \ref{sec:IR} and \ref{sec:goodOps}.  Perturbatively, $n$-point functions in $|\Omega \rangle$ can be computed either by analytic continuation from the Euclidean D-sphere $S^D$ or equivalently (see e.g. \cite{Higuchi:2010aa}) by using Schwinger-Keldysh perturbation theory along a time-contour that begins and ends on some cosmological horizon.  In order to insert operators near both $I^+$ and $I^-$, we take the Schwinger-Keldysh contour to be as shown in figure \ref{fig:SKcontour}.  In the $\tau$-coordinates of section \ref{sec:prelims}, this may be described as the contour which for fixed $x^i$ starts at $\tau =0$, runs backward to $\tau = - \infty$, then runs forward to $\tau  = +\infty$, and finally returns to $\tau =0$.  We will use this scheme for our explicit computations in section \ref{sec:Sheavy} and \ref{sec:light}.  The associated diagrammatic expansion will be reviewed in section \ref{sec:Lorentz} below.

However, regardless of the calculational scheme employed,  the S-matrix enjoys the following properties:
\begin{enumerate}
\item \label{vvu} {\bf The vacuum-to-vacuum amplitude is unity:}
  By construction the initial and final vacuum states are
  explicitly the same state $|\Omega\rangle$.  By definition this state is normalized to
  unity.


\item {\bf Covariance under the de Sitter group:}
  This follows from the covariance of the asymptotic states.
  E.g.,
  \eq{ \label{eq:SdS}
    \brakfi{A}{B} = \braf{A} 1 \ket{B}_f = \braf{A} U^{-1}(g) U(g) \ket{B}_f =
    \brakfi{gA}{gB} .
  }
\item {\bf Behavior under CPT:}
  Under the action of the CPT operator $\Theta$ the S-matrix transforms as
  \eq{ \label{eq:CPT}
    \Theta S = S^{-1} \Theta.
  }
  We note that for real fields like the scalar operators we consider section \ref{sec:Sheavy} and \ref{sec:light} the CPT operation is equivalent to the antipodal map $A$, so
  property (\ref{eq:CPT}) is simply a consequence of (\ref{eq:SdS}).

\item {\bf Invariance under perturbative field-redefinitions:}
  For simplicity, consider a theory defined by a single fundamental scalar field $\phi_\s$ and consider the field
  \eq{ \label{eq:redef}
    \Phi(x) = \phi_\s(x) +  \Op(x)
  }
 for some local operator $\Op(x)$.
 We assume that $\Op(x)$ may expanded in powers of $\phi_\s$.  Since taking $\Op(x)$ linear in the fundamental fields would induce only a trivial field renormalization (which might be matrix-valued if the theory involves multiple fields), we take this expansion to have no linear terms.  As noted in section \ref{sec:IntRev}, the poles in the K\"all\'en-Lehmann representation of $\Op(x)$ lie on the same lattice as those of $\phi_\sigma$. They and are thus associated with the same bare weight $\sigma$, and the same bare conjugate weight $\tilde \sigma$, and define asymptotic particles of the same mass as $\phi_\sigma$.  Since section \ref{sec:IntRev} also noted there that for generic theories the operator $\Op(x)$ has no poles at $\sigma$, $\tilde \sigma$ themselves.  It follows that $R_\sigma \Phi = R_\sigma \phi_\sigma$ and thus that $\Phi(x)$ and $\phi_\s$ define the same S-matrix.

\item {\bf Unitarity:} The physical content of S-matrix unitarity is implicit in our assumption that the initial and final spaces of asymptotic states each form complete bases for the same Heisenberg-picture Hilbert space, with the same norm.  When expressed in terms of orthonormal bases, this assumption implies that the matrix $S_{BA}$ is unitary in the standard sense that $S^\dg S = 1 = S S^\dg$.  The unitarity of such an $S$ is often
  phrased in terms of the transition matrix $\cT$ where $S = 1 + i\cT$.
  Then the unitarity of $S$ is equivalent to the optical theorem
  \eq{ \label{eq:Optical}
    2 \Im \cT = \cT^\dg \cT .
  }
While this remains an assumption at present, sections \ref{sec:Sheavy} and \ref{sec:light} will provide evidence in its favor by explicitly verifying the optical theorem at low orders of perturbation theory.

\item \label{flatlim} {\bf Flat-space limit:} Consider the flat-space limit $\ell \rightarrow \infty$ (with all masses and couplings held fixed in physical units). As mentioned above, the correlators used to define our S-matrix may obtained via analytic continuation from Euclidean signature; i.e., from the path integral on $S^D$.  The latter clearly approaches the usual flat-space result for large $S^D$.  In particular, the generating poles defined by the dS correlators approach the 1-particle poles of the Minkowski correlators.  We noted above that the recipe \eqref{eq:Aint2} relates our S-matrix to the residues of such poles. Since this is analogous to the construction of the Minkowski S-matrix from residues of 1-particle poles, in the flat-space limit they must agree up to possible normalizations.  The agreement of such normalizations will be verified by explicitly calculating the S-matrix for a $\phi^3$ theory in section \ref{sec:Sheavy}.  We will find there that the de Sitter corrections are exponentially small in $M^2\ell^2$, and are thus nonperturbative in $1/\ell$.
\end{enumerate}

%
%

\section{Lorentz-signature perturbation theory}
\label{sec:Lorentz}

The above sections give a general definition of our de Sitter S-matrix.  In order to demonstrate the properties (\ref{vvu}-\ref{flatlim}) from section \ref{sec:Smatrix},  sections \ref{sec:Sheavy} and \ref{sec:light} below will explicitly calculate the S-matrix for cubic theories through 2nd order.  We choose to use Lorentz-signature Schwinger-Keldysh perturbation theory \cite{Schwinger:1960qe,Keldysh:1964ud} (see also
\cite{Jordan:1986ug,Paz:1990jg,Giddings:2010ui,Weinberg:2005vy}).
Since we are unaware of a presentation in the literature  of this technique
in the global de Sitter chart we give a
brief introduction in section \ref{sec:CFs} below\footnote{
The Lorentz signature discussion of \cite{Goldstein:2003qf} involves a similar 3-legged contour but uses the
de Sitter ``neck'' $\eta = 0$ as the initial and final surface.  In order to successfully produce the Hartle-Hawking correlators, this approach requires an independent computation of interacting correlators at $\eta =0$.  This latter step is not required in our approach.  See \cite{Higuchi:2010aa} for further comments on the use of spacelike initial surfaces.}.
Section \ref{sec:amplitudes} then  states the associated diagrammatic rules for computing
our scattering amplitudes. For now, these rules are valid only
for sufficiently heavy fields such that there are no IR divergences.

\subsection{Schwinger-Keldysh correlation functions}
\label{sec:CFs}

The generating functional for time-ordered Hartle-Hawking correlators in global de Sitter may be constructed as follows.
Consider for simplicity a theory of a single scalar field $\phi(x)$.
We note that
\eq{\label{eq:unity}
  1 = \langle \Omega | \Omega \rangle
  = \sum_f \sum_p  \langle \Omega | f \rangle \langle f | p \rangle
  \langle p | \Omega \rangle ,
}
where in the second equality we have inserted two complete sets of
states $\{\ket{f}\}$ and $\{\ket{p}\}$ defined at a time in the far
future and past, respectively, of an initial surface $\Sigma_0$.
We take $\Sigma_0$ to be a cosmological horizon so
that regions to ``future''/``past'' correspond to expanding/contracting Poincar\'e
charts.
The state $\ket{\Omega}$ may be defined by imposing adiabatic
boundary conditions on $\Sigma_0$ (see e.g. \cite{Higuchi:2010aa}).
The expression (\ref{eq:unity}) may then be understood as a path
integral with a three-legged time contour of integration:
reading (\ref{eq:unity}) from right to left, the time contour begins
at the horizon, is traversed into the far past, then to the far future,
then back to the horizon.
In the coordinates of (\ref{eq:Poincareg}),  $\Sigma_0$ is
simply the $\tau=0$ hypersurface, the ``future''/``past'' are
the regions $\tau > 0$/$ \tau < 0$, and the time contour is
$\tau = 0 \to - L \to + L \to 0$ for any sufficiently large $L$;
see Fig.~\ref{fig:SKcontour}.

By inserting classical sources $J_f, J_0, J_p$ into the three legs of the path integral
we obtain the generating functional for $\phi(x)$ correlation
functions with respect to $\ket{\Omega}$:
\eqn{
  Z[J_f,J_o,J_p] &:=&
  \sum_f \sum_p  \langle \Omega | f \rangle_{J_f} \langle f | p \rangle_{J_0}
  \langle p | \Omega \rangle_{J_p}
  \\
  &=&
  \int [\cD \phi_f][\cD \phi_o][\cD \phi_p]
  \exp \bigg[
  - i \left(S[\phi_f] + \int_x \phi_f(x)J_f(x)\theta_f(x)\right)
  \nn \\ & & \phantom{  N \int [\cD \phi_f][\cD \phi_o][\cD \phi_p]
  \exp \bigg[}
    + i \left(S[\phi_o] + \int_x \phi_o(x)J_o(x)\right)
  \nn \\ & & \phantom{  N \int [\cD \phi_f][\cD \phi_o][\cD \phi_p]
  \exp \bigg[}
    - i \left(S[\phi_p] + \int_x \phi_p(x)J_p(x)\theta_p(x)\right)
  \bigg] .
}
Here $\theta_f(x)$, $\theta_p(x)$ are step functions with unit support
in the future/past of the horizon, e.g. $\theta(\tau)$, $\theta(-\tau)$.
Time-ordered correlation functions of $\phi(x)$ are generated via
\eq{
\label{eq:fopcorr}
  \bra{\Omega}T \phi(x_1)\dots\phi(x_n)\ket{\Omega}
  =
  \frac{\delta}{i \delta J_o(x_1)} \dots
  \frac{\delta}{i \delta J_o(x_n)}
  Z[J_p,J_o,J_f] \bigg|_{J_f=J_o=J_p=0,\;\phi_f=\phi_o=\phi_p=\phi} .
}
To generate diagrammatic rules one may, in the standard fashion,
re-write the interacting part of the action $S_{\rm int}[\phi]$
in terms of functional derivatives, then perform the Gaussian path
integral over the fields.\footnote{We assume the Lagrangian
has the decomposition $\cL_0 + \cL_{\rm int}$ with $\cL_0$ the
canonical free field Lagrangian (\ref{eq:Lfree}).}
The result is
\eq{ \label{eq:Zint}
  Z[J_f, J_o, J_p] =
  \exp\left(
    - i S_{\rm int}\left[- \frac{\delta}{i\delta J_f}\right]
    + i S_{\rm int}\left[  \frac{\delta}{i\delta J_o}\right]
    - i S_{\rm int}\left[- \frac{\delta}{i\delta J_p}\right]
  \right)
  Z_0[J_f, J_o, J_p] ,
}
with
\eqn{
  Z_0[J_f, J_p, J_o] &:=&
  \exp \int_x \int_{\bx} \bigg[
    -\half J_o(x) G(x,\bx) J_o(\bx)
    -\half J_f(x) G^*(x,\bx) J_f(\bx) \theta_f(x) \theta_f(\bx)
    \nn \\ & & \phantom{\exp \int_x \int_{\bx} \bigg[}
    -\half J_p(x) G^*(x,\bx) J_p(\bx) \theta_p(x) \theta_p(\bx)
    + J_o(x) W(x,\bx) J_p(\bx)\theta_p(\bx)
    \nn \\ & & \phantom{\exp \int_x \int_{\bx} \bigg[}
    - J_f(x) \theta_f(x) W(x,\bx) J_p(\bx)\theta_p(\bx)
    + J_f(x) \theta_f(x) W(x,\bx) J_o(\bx)
    \bigg] . \nn \\
}
For correlation functions with arguments contained in
a single Poincar\'e chart the generating function (\ref{eq:Zint}) reduces
to the more familiar
generating functions of ``in-in'' or ``out-out'' closed-time-path
perturbation theory
\cite{Jordan:1986ug,Paz:1990jg,Calzetta:1986ey,Calzetta:1986cq}.
Likewise, for arguments contained within a static chart
(\ref{eq:Zint}) reduces to the generating function of thermal
field theory  \cite{Landsman:1986uw}.

%
%

The generating function (\ref{eq:Zint}) differs from
generating functional of ``in-out'' perturbation theory in that it has
two additional path integrals (or ``time contours''). These may
be interpreted as providing the necessary corrections to $\bra{\Omega}$
and $\ket{\Omega}$ respectively. Their origin lies in the fact
that we have defined $\bra{\Omega}$ and $\ket{\Omega}$ at an
intermediate time rather than at the conformal boundaries
$\eta = \pm\infty$. In terms of Feynman graphs, these additional contours
lead to extra vertices
(which we label by $f$ and $p$) in addition to those (which label $o$, for ordinary) arising ``in-out'' perturbation theory; i.e., we label a vertex $f,o,p$ to indicate the term in \eqref{eq:Zint} from which it arises.

The resulting graphs can contain
legs which are Wightman functions rather than time-ordered Green's
functions. These have the important role of restricting
the region of spacetime integration involved in the construction
of any correlator to the union of i) the past light
cones of operators in the upper Poincar\'e chart,  and ii) the future light cones of operators in the lower Poincar\'e chart.
As a result, the perturbative construction of correlation functions
via (\ref{eq:Zint}) is manifestly causal.

\subsection{Scattering amplitudes}
\label{sec:amplitudes}

The above prescription for computing
vacuum correlators leads directly to diagrammatic rules for scattering
amplitudes.
We work in the asymptotic particle bases and use the ladder operators
(\ref{eq:Aint}), if necessary evaluating the states \eqref{eq:ASint} at a finite cutoff $\eta = \pm \eta_0$. This allows us to deal with potential logarithmic divergences.
The $\eta_0 \rightarrow \pm \infty$ limits give rise to our S-matrix so long as the fields used in \eqref{eq:fopcorr} are good operators in the sense of section \ref{sec:goodOps}.  In particular, the rules below may be used if all fields involved are sufficiently heavy or if, in the case of complementary series fields, we have performed a field redefinition equivalent in the sense of section \ref{sec:goodOps} to $\phi \rightarrow R_\sigma \phi$; see appendix \ref{app:goodOps2} for an explicit example.

With the above understanding, we may interchange the order of the KG inner
products and these spacetime integrals so as to first evaluate the KG
inner products between mode functions and the Green's functions connected
to external points, then take the $\eta \rightarrow \pm \infty$ limit,
and finally evaluate the rest of the Feynman diagram. This is
precisely what one does in the traditional Minkowski setting.

Let us focus on the amplitudes of types
$\brakii{a}{b}$, $\brakff{a}{b}$, or $\brakfi{a}{b}$ for which we
may use the time-ordered field correlators. Such amplitudes involve the $p$, $o$, and $f$-type vertices defined above according to the following prescription. First draw every Feynman diagram that contributes to
an amplitude, paying no regard to different varieties of vertices.
Then for each diagram:
\begin{enumerate}
\item \label{r1} For every vertex connected directly to a final bra, include only
  $p$ and $o$ varieties, and replace the Green's function connecting
  to a final bra with $u_n^*(y)$, where $y$ is the integration variable.
\item  \label{r2} For every vertex connected directly to a final ket, include only the
  $f$ variety, and replace the Green's function connecting
  to a final bra with $u_n(y)$.  See also rule \ref{r5} below.
\item  \label{r3} For every vertex connected directly to an initial bra, include only the
  $p$ variety, and replace the Green's function connecting
  to a final bra with $u_n^*(y)$.
\item  \label{r4} For every vertex connected directly to an initial ket, include only the
  $f$ and $o$ varieties, and replace the Green's function connecting
  to a final bra with $u_n(y)$. See also rule \ref{r5} below.

\item \label{r5} The diagram vanishes if it contains a vertex connected directly to both a final bra and final ket (in the computation of some $\brakff{a}{b}$) or to both an initial bra and an initial ket (in the computation of some $\brakii{a}{b}$).  This removes any conflicts between rules \ref{r1} and \ref{r2} and between rules \ref{r3} and \ref{r4}.

\item For every vertex that is not connected to an external state,
  include all varieties $o$, $f$, $p$.
\item Include multiplicative factors of $(ig)$ for every $o$ vertex
  and $-(ig)$ for every $f$ and $p$ vertex in the diagram.
\item Each $f/p$ vertex includes a theta function $\theta_{f/p}(y)$
  where $y$ is the integration variable.
\item The Green's functions for lines between vertices are
  as follows:
  \begin{center}
  \begin{tabular}{ccc}
    $x$ vertex & $y$ vertex & Green's function\\
    \hline
    $o$ & $o$ & $G_\s(x,y)$ \\
    $o$ & $f$ & $W_\s(y,x)$ \\
    $o$ & $p$ & $W_\s(x,y)$ \\
    $f$ & $f$ & $G^*_\s(x,y)$ \\
    $f$ & $p$ & $W_\s(x,y)$ \\
    $p$ & $p$ & $G^*_\s(x,y)$ \\
  \end{tabular}
  \end{center}
\end{enumerate}
These rules allow us to write scattering amplitudes as multiple spacetime
integrals whose integrands involve products of KG modes,
Green's functions, and the theta functions $\theta_{f/p}(y)$.
These integrals are not easily computed in closed form but they
they do reveal key properties of the amplitudes.

A useful fool for analyzing amplitudes is the antipodal map $A$
which maps a point $x$ in global de Sitter to its antipodal point $A x$.
For a fixed coordinate chart (\ref{eq:Poincareg}), if $x$ is in the
past Poincar\'e chart then $A x$ is in the future Poincar\'e chart.
$A$ acts upon our basic ingredients as
\eq{
  A: Y_\vL(\vx) \mapsto (-1)^L Y_\vL(\vx), \quad
  A: f_{\s L}(\eta) \mapsto f_{\s \vL}(-\eta) . \quad
}
Complex conjugation acts upon our ingredients as
\eq{
  Y^*_{\vL}(\vx) = (-1)^m Y_\vL(\vx), \quad
  f^*_{\s L}(\eta) = f_{\s L}(-\eta) . \label{eq:fconj}
}
Here $m$ is the quantum number associated with the azimuthal direction
in the standard spherical coordinates.\footnote{Our conventions for
  spherical harmonics are those of \cite{Higuchi:1986wu}. All we really
  need for our analysis is that $Y_\vL(\vx) \propto e^{im\varphi}$
  where $\varphi$ is the azimuthal coordinate.
}
Combining these results we obtain the relations
\eqn{ \label{eq:AonU}
  u_{\s \vL}(A x) &=& (-1)^{L+m} u^*_{\s \vL}(x) ,
  \\
  \label{eq:AonW}
  W_\s(A x, A y) &=& W_\s(y,x) = \left[ W_\s(x,y) \right]^* ,
  \\
  \label{eq:AonG}
  G_\s(A x, A y) &=& G_\s(x,y) ,
  \\
  \theta_{p/f}(A x) &=& \theta_{f/p}(x) ,
  \\
  \sqrt{-g(Ax)} &=& \sqrt{-g(x)} .
}
The final equality follows from the fact that $A$ is a de Sitter
isometry.

\section{Example with heavy fields}
\label{sec:Sheavy}

We now provide an explicit example of the construction of section \ref{sec:Smatrix}.
Consider a model theory of three
massive scalars with a cubic interaction described by the Lagrangian
\eqn{ \label{eq:model}
  \cL[\vec{\phi}] &=& \sum_{i=1}^3 \cL_0[\phi_i] + \cL_{\rm int}[\vec{\phi}]
  + \cL_{\rm c.t.}[\vec{\phi}] ,
  \\
  \cL_{\rm int}[\vec{\phi}] &=& g \phi_3\phi_2\phi_1(x) ,
  \\
  \cL_{\rm c.t.}[\vec{\phi}] &=& \sum_{i=1}^3
  \left[ -\frac{(Z_{\phi_i} - 1)}{2} \nabla_\mu \phi_i \nabla^\mu \phi_i(x)
    - \frac{(Z_{M_i}-1)M^2_i}{2} \phi_i^2(x) \right] + \cO(g^3) .
  \label{eq:Lct}
}
We work in perturbation theory with $g \ll 1$, and we will examine
S-matrix elements through $\cO(g^2)$.
The dS Hartle-Hawking correlators of this theory were studied previously \cite{Marolf:2010zp}.
The field and mass renormalization counterterms may be chosen to
be $Z_{\phi_i} = 1 + \cO(g^2)$ and $Z_{M_i} = 1 + \cO(g^2)$; no other
counterterms are necessary until $\cO(g^3)$.

In this section we restrict the bare weights to the regime
\eq{ \label{eq:weights}
  \Re(\s_1+\s_2+\s_3) < -(D-1) .
}
This condition is automatically satisfied by principal series fields (for which $\Re \s = -(D-1)/2$), but it can also be satisfied when some or all of the fields are sufficiently heavy members of the complementary series.
To $\cO(g^2)$ in the above theory, the condition \eqref{eq:weights} will suffice to qualify the fields as ``heavy."  In particular, the naive LSZ prescription \eqref{eq:Aint} gives finite results and the diagrammatic rules of section \ref{sec:amplitudes} may be applied directly. We will revisit this theory in \S\ref{sec:light}
in order to lift the restriction \eqref{eq:weights}.

Below, we use subscripts $1,2,3$ to indicate which of the 3 flavors of scalar field in \eqref{eq:model} are described by a given object.  E.g. $u_1, u_1'$ denote wavefunctions of $\phi_1$-particles, $n_2, n_2'$ denote occupation numbers for $\phi_2$-particles, etc.  More primes will be added as needed.

\subsection{$\cO(g)$ amplitudes}
\label{sec:OgAmplitudes}

\begin{figure}[t!]
  \centering
  \includegraphics{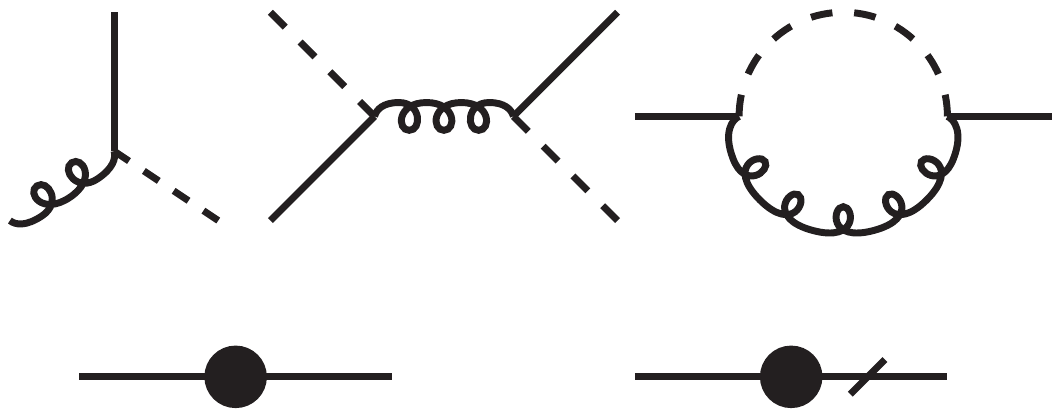}
  \caption{Feynman graphs contributing to connected corrections
    at $\cO(g)$ and $\cO(g^2)$ in $g\phi_3\phi_2\phi_1(x)$ theory.
    We use solid lines for $\phi_1(x)$, dashed lines for
    $\phi_2(x)$, and curly lines for $\phi_3(x)$.
    In the top row, from left to right: the corrections
    $\C{\phi_3(x_3)\phi_2(x_2)\phi_1(x_1)}^{(1)}$,
    $\C{\phi_2(\bx_2)\phi_1(\bx_1)\phi_2(x_2)\phi_1(x_1)}^{(1)}$, and
    $\C{\phi_1(\bx)\phi_1(x)}^{(2)}$.
    On the bottom row are the mass and field renormalization counterterms
    present in $\C{\phi_1(\bx)\phi_1(x)}^{(2)}$.
    There are also graphs of the same topology but with different
    permutations of the fields.
    \label{fig:phi3graphs}
    }
\end{figure}

Contributions to the S-matrix at $\cO(g)$
arise from the 3-point function
$\C{\phi_3(x_3)\phi_2(x_2)\phi_1(x_1)}^{(1)}$.  Here and below we use
parenthesized superscripts (e.g., ${}^{(1)}$) to denote the order in $g$ to which a quantity is calculated.
We begin with the amplitude
\eq{
  \brakfi{n_3 n_2}{n_1}^{(1)}
  = o {\rm\; term}
  = i g \int_y u_3^* u_2^* u_1(y),
  \label{eq:1to2}
}
where we abbreviate
$u_1 u_2 \dots u^*_k(y) = u_{n_i}(y) u_{n_2}(y) \dots u_{n_k}^*(y)$.
In the first equality we note, following the rules of \S\ref{sec:amplitudes},
that only the `o' vertex contributes.
The final integral expression is simply the de Sitter analogue of the familiar
Minkowski result.

 The integral may be decomposed as
\eq{
  \brakfi{n_3 n_2}{n_1}^{(1)} = ig
  \int_{-\infty}^{+\infty} d\eta (1+\eta^2)^{(D-2)/2}
  f^*_{\s_3 L_3}f^*_{\s_2 L_2}f_{\s_1 L_1}(\eta)
  \int_\vx Y_{\vL_3}^*Y_{\vL_2}^*Y_{\vL_1}(\vx) .
}
The integral over $\vx$ provides an $SO(D)$ Clebsch-Gordon coefficient (CGC).
These CGCs are always real, and this particular CGS is non-vanishing when
the angular momenta satisfy the triangle inequalities \cite{Junker:1993aa}:
\eq{
  |L_i - L_j| \le L_k \le L_i + L_j, \quad \forall \{i,j,k\} = \{1,2,3\} .
}
The remaining integral over $\eta$ is also real, as can be seen by
changing integration variables $\eta \to -\eta$ and using
(\ref{eq:fconj}).\footnote{More generally one can use time reversal
  and parity to show that integrals of the form
  \eq{
    I := \int_y \prod_{i=1}^k u_i(y) \prod_{j=1}^l u^*_j(y) ,
    \quad \Re\left( \sum_{i=1}^k \s_i + \sum_{j=1}^l \s_j \right) < -(D-1) ,
  }
  are real and vanish when either
  $L_{\rm tot} = \sum_{i=1}^{k+l} L_i$ or $m_{\rm tot} = \sum_{i=1}^{k+l} m_i$
  is odd.
}
Thus, the entire amplitude (\ref{eq:1to2}) is imaginary.
Note that this integral converges only for masses satisfying
(\ref{eq:weights}).

To obtain an explicit expression for (\ref{eq:1to2}) we will compute the  integral
\eq{ \label{eq:I1}
  I_1 := \int_y u_3 u_2 u_1^*(y) ,
}
where we have taken a complex conjugate relative to \eqref{eq:1to2} so that we may more directly employ the linearization formula \eqref{eq:fLinearization} to replace the product $u_3 u_2$ with a weighted integral over a single mode function.

We will also make use of two further tricks.  The first involves
the cutoff integral
\eq{
  \int_{y,\eta_0} u_1 u^*_2(y) :=
  \int_{-\eta_0}^{+\eta_0}d\eta (1+\eta^2)^{(D-2)/2} \int_\vx u_1 u^*_2(y)  .
}
Because $u_{1,2}$ satisfy the Klein-Gordon equation with mass
$M_{1,2}^2$ respectively we have
\eqn{ \label{eq:cutoffIntegral}
  \int_{y,\eta_0} u_1 u^*_2(y)
  &=& \frac{1}{M_1^2 - M_2^2}
  \int d\Sigma^\nu(y) \left[ [u_1 \KG_\nu u^*_2(y)]_{\eta = \eta_0}
    - [u_1 \KG_\nu u^*_2(y)]_{\eta= -\eta_0}\right] .
}
This formula is valid for all complex values of $\s_1$ and $\s_2$
apart from $\s_i = L_i + n$ or $\s_i = -L_i - (D-1) -n$ where $n$
is a non-negative integer.

The other trick is to note that when $|\eta| > 1$ the functions
$f_{\s L}(\eta)$ may be split into two functions each of which
solves the Klein-Gordon equation individually but contains only one asymptotic behavior.\footnote{The function
$f_{\s L}(\eta)$ may be decomposed this way for $|\eta| > 1$,
but the asymptotic behavior does not dominate the behavior until
$|\eta| \gg L-\s$ \cite{Marolf:2011aa}.}
We label the two behaviors ``fast'' and ``slow'':
\eq{
  \label{eq:ffastslow}
  f_{\s L}(|\eta| > 1) = f^{\rm s}_{\s L}(\eta) + f^{\rm f}_{\s L}(\eta) ,
  \quad
  f^{\rm s}_{\s L}(\eta) = \cO(\eta^\s),
  \quad
  f^{\rm f}_{\s L}(\eta) = \cO(\eta^{-(\s+D-1)}) ,
}
\eq{ \label{eq:ffsReflection}
  f^{\rm s}_{-(\s+D-1) L}(\eta) = f^{\rm f}_{\s L}(\eta),
  \quad
  f^{\rm f}_{\s L}(\eta) = f^{\rm s}_{-(\s+D-1) L}(\eta) .
}
The explicit form of $f^{s/L}_{\s L}(\eta)$ may be found using
identities of the Gauss hypergeometric function, but we will only
need the leading behaviors given in (\ref{eq:fLargeEta}).
Complex conjugation affects $f^{\rm s/f}_{\s L}(\eta)$
differently depending on the series to which $\s$ belongs:
\eqn{ \label{eq:ffsConjugation}
  \left[f^{\rm s}_{\s L}(\eta)\right]^*
  &=& f^{\rm s}_{\s L}(-\eta) , \quad
  \left[f^{\rm f}_{\s L}(\eta)\right]^*
  = f^{\rm f}_{\s L}(-\eta) , \quad \s \in \textrm{c.~series} ,
  \nn \\
  \left[f^{\rm s}_{\s L}(\eta)\right]^*
  &=& f^{\rm f}_{\s L}(-\eta) , \quad
  \left[f^{\rm f}_{\s L}(\eta)\right]^*
  = f^{\rm s}_{\s L}(-\eta) , \quad \s \in \textrm{p.~series} .
}

Returning to (\ref{eq:I1}), introducing the time cutoff $|\eta| < \eta_0$, and
using the linearization formula for $u_2 u_3(y)$ yields:
\eqn{
  I_1(H) &=& \sum_\vK \int_\mu (2\mu+D-1) \rho_{2 3}(\mu,\vK)
  \int_{y,\eta_0} u_{\mu\vK} u_1^*(y)
  \nn \\
  &=& \int_\mu (2\mu+D-1) \rho_{2 3}(\mu,\vL_1)
  \int_{y,\eta_0} u_{\mu \vL_1} u_1^* (y) ,
  \label{eq:I1a}
}
 where the final step is just orthogonality of distinct spherical harmonics.
The integration contour may be traversed anywhere in the strip
$\Re(\s_2 + \s_3) \le \mu < 0$; for concreteness let the contour
traverse $\Re \mu = -\epsilon$ for infinitesimal positive $\epsilon$.
We then integrate over $y$ using (\ref{eq:cutoffIntegral})
to obtain
\eqn{
  I_1(H) &=& - \int_\mu \bigg\{ \frac{(2\mu+D-1)}{(\mu-\s_1)(\mu+\s_1+D-1)}
  \rho_{2 3}(\mu,\vL_1)
  \nn \\ & & \phantom{- \int_\mu \bigg\{\;}
  \times \int d\Sigma^\nu(y)
  \left[ [u_{\mu\vL_1} \KG_\nu u^*_{1}(y)]_{\eta = \eta_0}
    - [u_{\mu \vL_1} \KG_\nu u^*_{1}(y)]_{\eta= -\eta_0}\right]\bigg\} .
  \label{eq:I1b}
}
Next we split $f_{\mu K}(\eta)$ into fast and slow parts as in
(\ref{eq:ffastslow}). The $f^{{\rm s}*}_{\mu L_1}(\eta)$ contributes
terms $\cO(H^{\s_1+D-1-\epsilon})$ while the $f^{{\rm f}*}_{\mu L_1}(\eta)$
contributes terms $\cO(H^{\s_1-\epsilon})$. Thus
the latter, ``fast-decay'' terms vanish in the limit $\eta_0 \to \infty$
and we need only keep only the contributions of $f^{{\rm s}*}_{\s L_1}(\eta)$.
We now deform the integration contour to the left from $\Re \mu = -\epsilon$
to $\Re\mu = \Re(\s_2 +\s_3)+\epsilon$. From (\ref{eq:weights}) we know
\eq{
  \Re(\s_2 +\s_3) < \Re -(\s_1+D-1) \le \Re \s_1 < 0 ,
}
so the only poles we encounter in moving the contour to
$\Re\mu = \Re(\s_2 +\s_3)+\epsilon$ are the simple poles at
$\mu = \s_1$ and $\mu = - (\s_1+D-1)$ (examine the denominator (\ref{eq:I1b})).
The remaining contour integral  behaves like
$\cO(\eta_0^{\Re(\s_2 +\s_3+\s_1)+D-1+\epsilon})$ and so vanishes as $\eta_0 \to \infty$.
Thus
\eqn{
  & & I_1(\eta_0) =
  - \bigg\{ \rho_{2 3}(\s_1,\vL_1)
  \int d\Sigma^\nu(\vy)
  \left[ [u_{\mu \vL}^{\rm s} \KG_\nu u^*_1(y)]_{\eta = \eta_0}
    - [u_{\mu\vL}^{\rm s} \KG_\nu u^*_1(y)]_{\eta= -\eta_0}\right]
  \nn \\ & &
  + \rho_{2 3}(-(\s_1+D-1),\vL_1)
  \int d\Sigma^\nu(\vy)
  \left[ [u_{\mu \vL}^{\rm f} \KG_\nu u^*_1(y)]_{\eta = \eta_0}
    - [u_{\mu\vL}^{\rm f} \KG_\nu u^*_1(y)]_{\eta= -\eta_0}\right]
  \bigg\} .\nn \\
}
It is straight-forward to evaluate the remaining KG norms and take
the limit $\eta_0\to\infty$. We finally obtain
\eqn{
  I_1 &=& -\cot\left[\pi\left(\s_1+\frac{D-1}{2}\right)\right]
  \left[ \rho_{23}(\s_1,\vL_1) -
    \rho_{23}(-(\s_1+D-1),\vL_1) \right] .
  \label{eq:I1final}
}
For $\s_1$ in the complementary series $\cot\left[ \pi \left(\s+\frac{D-1}{2}\right)\right]$
is real and so too is the second term in brackets, as
$\rho_{23}(\mu\in\Reals,L_1) \in \Reals$. For $\s_1$ in the
principal series $\cot\left[ \pi \left(\s+\frac{D-1}{2}\right)\right]$ is imaginary
and so too is the term in brackets which is equal to
$2 i \Im \rho_{23}(\s_1,\vL_1)$.  So $I_1$ is real in all cases.
Buried within $\rho_{2 3}(\s_1,\vL)$ is the $SO(D)$ Clebsch-Gordon
coefficient.

The most salient feature of the result (\ref{eq:I1final}) is that
the amplitude $\brakfi{n_3 n_2}{n_1}^{(1)} = i g I_1$ is generically nonzero
whenever allowed by addition of angular momentum. This is true for
\emph{any} set of masses satisfying $\Re(\s_1 + \s_2 + \s_3) < (D-1)$.
This includes configurations in which some of the fields
belong to the complementary series.
The expression (\ref{eq:I1final}) contains $\rho_{2 3}(\mu,\vL)$
evaluated in the region of the complex $\mu$-plane for which
$\rho_{2 3}(\mu,\vL)$ is analytic and nontrivial; thus, it can
vanish at most at isolated points.  
We emphasize this point in order to contrast with the statements of  \cite{Bros:2006gs,Bros:2008sq} that complementary series particles can decay only in what we have called exceptional theories. Instead, our conclusions support the analysis of \cite{Boyanovsky:2004gq,Boyanovsky:2012qs}; the discrepancy with \cite{Bros:2006gs,Bros:2008sq} will be explained in section \ref{sec:discussion}.  Due to the detailed form of $\rho_{\s_1\s_2L_1L_2}(\mu,K)$, the amplitude decays exponentially at large masses;
see Appendix~\ref{app:linearization}.  We have checked these results by direct numerical integration of \eqref{eq:I1}.

There is also a non-vanishing connected contribution to the amplitude
\eq{
  \brak{\Omega}{n_3 n_2 n_1}^{(1)}_i
  = f + o {\rm\; terms}
  = i \int_y \theta_p(y) u_3 u_2 u_1(y) .
  \label{eq:0to3i}
}
This amplitude may be understood as both a transition
amplitude for a $3\to 0$ process and as a correction to the
overlap between two initial states.
As with the previous example (\ref{eq:0to3i}) converges absolutely
only for masses in the regime (\ref{eq:weights}).
The same physics may be accessed different ways: we
may compute
\eq{
  \braki{n_3 n_2 n_1}{\Omega}^{(1)}
  = p {\rm \; term}
  = - i g \int_y \theta_p(y) u^*_3 u^*_2 u^*_1(y),
}
which is clearly the complex conjugate of (\ref{eq:0to3i}), or
we could compute the amplitudes
\eqn{
  \brakf{n_3 n_2 n_1}{\Omega}^{(1)} &=& o + p {\rm \; term}
  = i g \int_y \theta_f(y) u_3^* u_2^* u_1^*(y),
  \\
  \brak{\Omega}{n_3 n_2 n_1}^{(1)}_f &=& f {\rm \; term}
  = - i g \int_y \theta_f(y) u_3 u_2 u_1(y) , \label{eq:0to3f}
}
which are also clearly complex conjugates. Moreover, by letting
$y \to Ay$ in any of these expressions we verify that
\eq{
  \braki{n_3 n_2 n_1}{\Omega}^{(1)} = \brak{\Omega}{n_3 n_2 n_1}^{(1)}_f,
  \quad {\rm and} \quad
  \brakf{n_3 n_2 n_1}{\Omega}^{(1)} = \brak{\Omega}{n_3 n_2 n_1}^{(1)}_i,
}
as is required by CPT.

Evaluating these
amplitudes explicitly is rather awkward because they are expressed
as an integral of \emph{global} KG modes over a single Poincar\'e chart.
We will be content to show that the amplitude is generically non-zero
when $L_1 + L_2 + L_3$ is even.
Consider the imaginary part of the amplitude,
\eq{
\label{eq:I2}
  2 i\, \Im \brak{\Omega}{n_3 n_2 n_1}^{(1)}_i
  = \brak{\Omega}{n_3 n_2 n_1}^{(1)}_i - \brak{\Omega}{n_3 n_2 n_1}^{(1)}_f
  = i g \int_y u_3 u_2 u_1(y) =: i g I_2 .
}
The integral $I_2$ may be evaluated by following the same procedure
as was used to compute $I_1$. The result is
\eq{
  I_2 = \frac{e^{-i\pi D} \cos\left[\pi\left(\frac{D-1}{2}\right)\right]}
  {\sin\left[\pi\left(\s+\frac{(D-1)}{2}\right)\right]}
  \left[ \rho_{23}(\s_1,\vL_1) -
    \rho_{23}(-(\s_1+D-1),\vL_1) \right] .
  \label{eq:I2final}
}
Indeed, this expression is generically non-zero. We plot an example
in Fig.~\ref{fig:vacuumPlots} below.

Note that all other corrections to either the initial or final inner products (e.g. $\brakii{n_1}{n_2n_3}$ or $\brakff{n_1}{n_2n_3}$) vanish by rule \ref{r5} of section \ref{sec:amplitudes}.

 \begin{figure}[t!]
   \centering
   \includegraphics{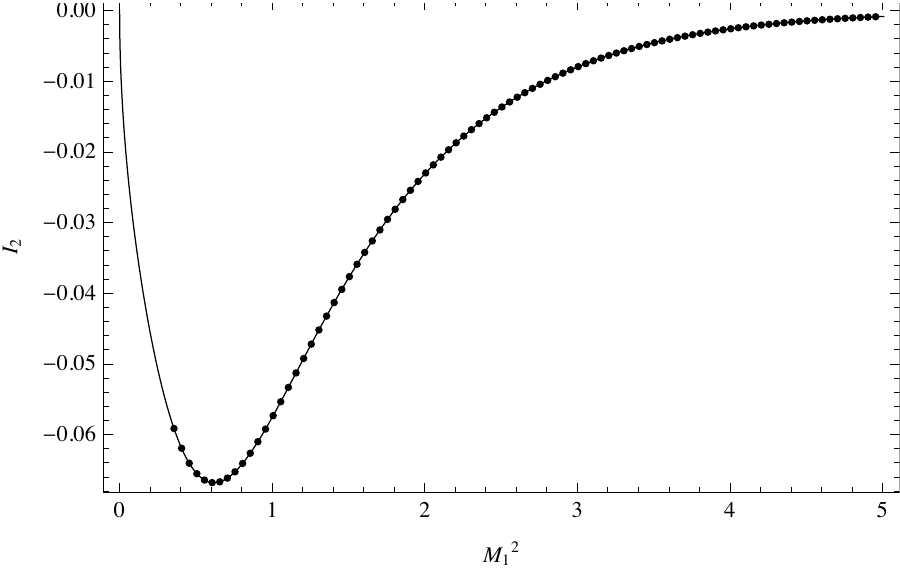}
   \caption{
    The amplitude $I_2$ in $D=3$ as a function of $M_1^2$ with
    $\vL_1=\vL_1=\vL_3=0$, $M_2^2 =2/\ell^2$, and $M^2_3 = 1.25/\ell^{2}$. The continuous curve is plotted using \eqref{eq:rho_app} while the dots represent direct numerical integration of \eqref{eq:I2}.  The amplitude is
    small, tends rapidly to zero as $M_1^2$ increased, and approaches zero for $M_1^2 \to 0$. Since $M_2$ and $M_3$ lie in the principal series, \eqref{eq:I2} converges for all $M_1^2 > 0$ --  though the slow convergence near  $M_1^2 \approx 0$ makes direct numerical evaluation of \eqref{eq:I2} difficult in that regime.
   \label{fig:vacuumPlots}
    }
\end{figure}

\subsection{$\cO(g^2)$ state corrections}
\label{sec:Oggheavy}

We now compute corrections to the initial state inner products at $\cO(g^2)$.  The corresponding final state corrections are then easily obtained by applying CPT.
Since explicit expressions for amplitudes at $\cO(g^2)$ are both lengthy and time-consuming
to attain, we will compute results only at the level needed to verify the optical theorem in section \ref{sec:Ogheavy}.  Due to the way that the various terms enter the optical theorem, it will be convenient to delay computation of the actual S-matrix elements until section \ref{sec:Ogheavy}.

At $\cO(g^2)$ there are two types of connected corrections to vacuum
correlators functions: tree-level corrections to 4-point functions and
1-loop corrections to the 2-point functions (recall Fig.~\ref{fig:phi3graphs}).
Tree-level corrections to 4-point functions contribute to the connected
part of the amplitudes
\eqn{ \label{eq:in40}
  & & \braki{n_1 n_2 n_1' n_2'}{\Omega}^{(2)}_C
  = pp \;\textrm{term} \nn \\
  & & =
  (ig)^2 \int_{y_1} \int_{y_2} \theta_p(y_1) \theta_p(y_2)
  \left[ u_1 u_2(y_1) u_1'u_2'(y_2)
    + u_1 u_2'(y_1) u_1'u_2(y_2) \right]
  G^*_3(y_1,_2) ,
}
as well as to amplitudes obtained by permuting the species.  As we will see in section \ref{sec:Ogheavy}, such vacuum-to-many amplitudes cancel out completely in any check of the optical theorem.
There is also a tree-level correction to the
connected part of the 2-particle amplitude
${}_i\brak{n_2' n_1'}{n_2 n_1}_{i,C}^{(2)}$. In principle both s- and u-channel
graphs may contribute, though the u-channel contributions vanish by rule \ref{r5} of section \ref{sec:amplitudes}.   Thus we compute only
\eqn{
   \braki{n_1' n_2'}{n_1 n_2}_{i \,s-{\rm channel}}^{(2)}
   &=& pf + po \;\textrm{terms}
  \nn \\
  &=& g^2 \int_{\by} \int_y \theta_p(\by) \theta_p(y)
  u_1'^* u_2'^*(\by) u_1 u_2(y) W_3(y,\by)
  \nn \\
  &=& \sum_{\vK}
  \left[ (ig) \int_{\by} \theta_p(\by) u_1'u_2'u_{\s_3\vK}(\by) \right]^*
  \left[(ig) \int_{y} \theta_p(y) u_1 u_2 u_{\s_3 \vK} (y) \right] .
  \nn \\
  \label{eq:in22}
}
These amplitudes are UV-finite because the $SO(D)$ CGCs limit the
range of the sum over $\vK$ to
\eq{
  {\rm min}\left\{ |L_2' - L_1'|,\; |L_{2} - L_{1}| \right\}
    \le K \le
  {\rm max}\left\{ L_2' + L_1',\; L_{2} + L_{1} \right\} .
}
The right-hand side can correct amplitudes
both on and off the diagonal $\delta_{\vL_1'\vL_1}\delta_{\vL_2'\vL_2}$.
On the diagonal the right-hand side is a sum of absolute values
of integrals which were shown to be generally non-vanishing above.

The 1-loop correction to the 2-point function $\C{\phi_1(\bx)\phi_1(x)}^{(2)}$
alters the normalization of the 1-particle states $\brakii{n_1}{n_1}^{(2)}$, though
conservation of angular momentum prevents it contributing to off-diagonal
1-particle amplitudes.
The 1-loop correction is UV-divergent for $D > 3$, so we must include
the standard perturbative counterterms\ that
follow from (\ref{eq:Lct}). Only the field renormalization counterterm
contributes to this amplitude:
\eqn{
  \brakii{n_1}{n_1}^{(2)}
   &=& po + pf {\rm \; terms} + (Z_{\phi_1}-1)
   \nn \\
   &=& - (ig)^2 \int_\by \int_y \theta_p(\by) \theta_p(y) u_1^*(\by) u_1(\by)
   W_3(y,\by) W_2(y,\by)
   + (Z_{\phi_1}-1)
   \nn \\
   &=& \sum_{\vL_2} \sum_{\vL_3}
  \left[ (ig) \int_\by \theta_p(\by) u_3 u_2 u_1(\by)\right]^*
  \left[ (ig) \int_y \theta_p(y) u_3 u_2 u_1(y)\right]
  + (Z_{\phi_1}-1) .\ \ \ \ \
}
This expression is IR-finite but the sum over momenta does not terminate
and the expression contains an ultraviolet divergences for $D\ge 6$. We choose $Z_{\phi_1}$ to cancel this divergence.  All of this  is very much analogous to the computation in Minkowski space.
The 1-loop correction $\C{\phi_1(\bx)\phi_1(x)}^{(2)}$ also contributes
a correction to the 2-to-vacuum amplitude
$\brak{\Omega}{\s_i \vec{0}, \s_i \vec{0}}_i^{(2)}$, which will again cancel out entirely in any check of the optical theorem.

There are of course analogous formulae for similar amplitudes obtained by permuting the flavors $(1,2,3)$ above, as well as those between
final states which may be obtained by using CPT symmetry and conjugation
to relate final state amplitudes to the results above. For the remaining inner products between initial bras and initial kets which have yet to be mentioned, the $O(g^2)$ contributions vanish either manifestly (there are no relevant diagrams) or by rule \ref{r5} of section \ref{sec:amplitudes}.  In particular, we have
\eq{
\label{eq:vanish}
  \braki{n_2' n_1' n_2}{n_1}^{(2)}_{i\,C} = 0, \quad
  \braki{n_2' n_2}{n_1' n_1}^{(2)}_{i\,C} = 0.
}

\subsection{The Optical Theorem}
\label{sec:Ogheavy}

We wish to explicitly verify the optical theorem, and thus the unitarity of our S-matrix, to order $g^2$ for our model theory. In terms of an orthonormal basis, the optical theorem takes the standard form
\eq{
  - 2 \Re \widetilde {\brakfi{B}{A} }= \sum_C
  \widetilde { \brakif{B}{C} } \widetilde  {\brakfi{C}{A} },
\label{eq:OTH}
}
where $\widetilde { \brakfi{B}{A} } :=  \brakfi{B}{A} - \delta_{AB}.$
We may restrict attention to amplitudes involving at
most 4-particles as only these have fully-connected contributions.

To verify \eqref{eq:OTH}, we first construct orthonormal bases from our multi-particle asymptotic states.  At $\cO(g)$ the 1- and 2-particle states require no modification; we need only remove the
overlap of 3-particle states with the vacuum, as well as overlaps
between higher-number particle states due to disconnected contributions.  We therefore define
\eq{ \label{eq:3ortho}
  \ket{n_3 n_2 n_1}^{ON}_i = \ket{n_3 n_2 n_1}_i
  - \ket{\Omega}\brak{\Omega}{n_3 n_2 n_1}^{(1)}_i + \cO(g^2) .
}

The only non-zero connected contributions to the S-matrix at $\cO(g)$ are the amplitudes $\brakfi{n_3 n_2}{n_1}$ which were computed in section \ref{sec:OgAmplitudes}.   Recall that these amplitudes are imaginary, so that \eqref{eq:OTH} is satisfied at $\cO(g)$.

The vacuum-to-three amplitude in the orthonormal bases vanishes by
construction:
\eq{
  \ONbrakfi{n_3 n_2 n_1}{\Omega}^{ON(1)} = 0 .
}
Quite generally, converting in this way from a particle basis to
an orthonormal basis removes contributions from any diagram which is
attached to only the final bra or initial ket (in the particle basis).
It is natural to think of such diagrams as computing corrections to
the relevant bra or ket, rather than representing genuine particle
scattering processes, so it is not surprising that these diagrams do not ultimately
contribute to the S-matrix despite being non-zero.

At $\cO(g^2)$, transforming to orthonormal bases once again
removes overlaps between particle states and the vacuum; the transition
also removes overlaps between non-identical pairs of 2-particle states and corrects normalizations.  Below we use the orthonormal states
\eqn{
\label{eq:ONstates}
  \ket{n_1}_i^{ON} &=& \ket{n_1}_i\left[1 - \half \brakii{n_1}{n_1}^{(2)} \right]
  + \cO(g^3) , \\
  \label{eq:637}
 \ket{n_1' n_1}_i^{ON} &=& \ket{n_1' n_1}_i
 \left[1 - \half \brakii{n_1' n_1}{n_1'n_1}^{(2)} \right]
  + \cO(g^3) , \\
  \label{eq:638}
 \ket{n_2 n_1}_i^{ON} &=& \ket{n_{2} n_1}_i
 -  \frac{1}{2} \sum_{n_1',n_2'} \ket{n_1'n_2'}_i \brakii{n_1'n_2'}{n_2 n_1}^{(2)}
  + \cO(g^3) .
}

Returning to \eqref{eq:OTH}, we begin with the $\phi_1\to\phi_1$ scattering process.  Again, angular momentum conservation sets the off-diagonal terms to zero so that the non-trivial part is just
\eq{
  \ONbrakfi{n_1}{n_1}^{ON(2)} =
  \brakfi{n_1}{n_1}^{(2)}
  - \half\left[ \brakii{n_1}{n_1}^{(2)} + \brakff{n_1}{n_1}^{(2)} \right]
  = ({\rm graph}) + ({\rm c.t.})
}
Here $({\rm graph})$ denotes the contribution from the 1-loop
diagram while $({\rm c.t.})$ denotes contributions from the mass-
and field-renormalization counterterms. Explicitly these are:
\eqn{
  ({\rm graph}) &=&
  of + oo + po + pf -\half\left[ po + pf + of + pf \right] {\rm \; terms}
  \nn \\
  &=& oo + \half\left[ of + po \right] {\rm \; terms}
  \nn \\
  &=& (ig)^2 \int_\by \int_y u_1^*(\by) u_1(y) \cM(\by,y) ,
}
where
\eq{
  \cM(\by),y := G_3(y,\by)G_2(y,\by)
    - \half W_3(y,\by)W_2(y,\by) \left(\theta_f(y) + \theta_p(\by) \right) ,
}
and
\eq{
  ({\rm c.t.}) = o {\rm \; term} = i(Z_{M_1} + Z_{\phi_1}-2)
  \int_y u_1^* u_1(y) .
  \label{eq:ct}
}
Note that the integrand in \eqref{eq:ct} is positive-definite and behaves at large $|\eta| \gg 1$ like
$\eta^{2\s_1}$ so that the integral is IR-divergent for all values of
$\s_1$.
This divergence turns out to cancel a similar divergence that appears
in $({\rm graph})$ under precisely the same conditions that UV counterterms are necessary.
There is also a logarithmic IR divergence proportional to $\Pi^{(1)}(\sigma_1)$ as expected from the second term in \eqref{eq:LK1PIexpansion}, which we therefore regard as physical.  Recall that a similar divergence in Minkowski space is associated with a finite decay rate per unit time (``Fermi's golden rule'') and plays a key role in satisfying the optical theorem.

To verify our dS the optical theorem we compute
\eqn{ \label{eq:blah}
  - 2 \Re \ONbrakfi{n_1}{n_1}^{ON(2)} &=&
  - ({\rm graph}) - ({\rm graph})^*
  \nn \\ &=&
  - (ig)^2 \int_\by \int_y u_1^*(\by) u_1(y)
  \left[ \cM(\by,y) + \cM^*(A\by, Ay) \right] .
}
In the second equality we have used (\ref{eq:AonU}). From
(\ref{eq:AonG}) and (\ref{eq:AonW}) it follows that
\eqn{
  \cM(\by,y) + \cM^*(A\by,A y)
  &=&
  G_3(y,\by)G_2(y,\by) + G^*_3(\by,y)G^*_2(\by,y)
  - W_3(y,\by)W_2(y,\by)
  \nn \\
  &=& W_3(\by,y)W_2(\by,y) .
\label{eq:645}}
Inserting \eqref{eq:645} into (\ref{eq:blah}) we obtain
\eqn{
\label{eq:OTH1to1}
  - 2 \Re \ONbrakfi{n_1}{n_1}^{ON(2)} &=&
  - (ig)^2 \int_\by \int_y u_1^*(\by) u_1(y) W_3(\by,y)W_2(\by,y)
  \nn \\
  &=&  \sum_{\vL_2} \sum_{\vL_3} \left| (ig) \int_y u_1 u_2^* u_3^*(y) \right|^2
  =  \sum_{\vL_2} \sum_{\vL_3} \left| \brakfi{n_2n_3}{n_1}^{(1)} \right|^2 .
}
The final equality is precisely what is required by the optical
theorem.

\begin{figure}[t!]
  \centering
  \includegraphics[width = 2cm]{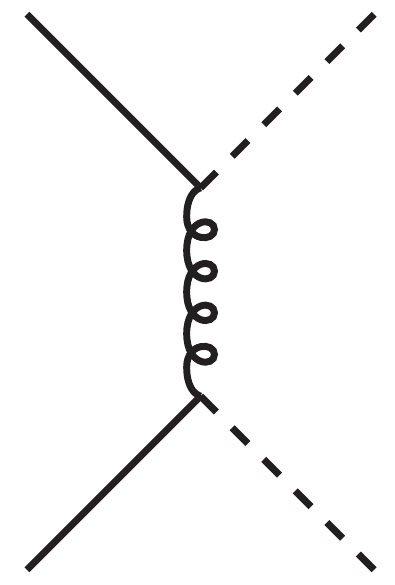}\hspace{2cm}
  \includegraphics[width = 2cm]{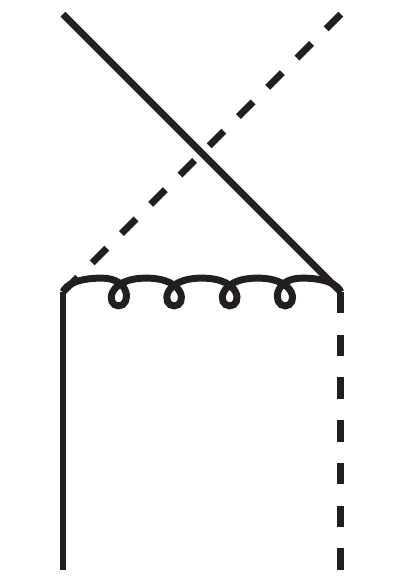}\hspace{2cm}
  \includegraphics[width = 2cm]{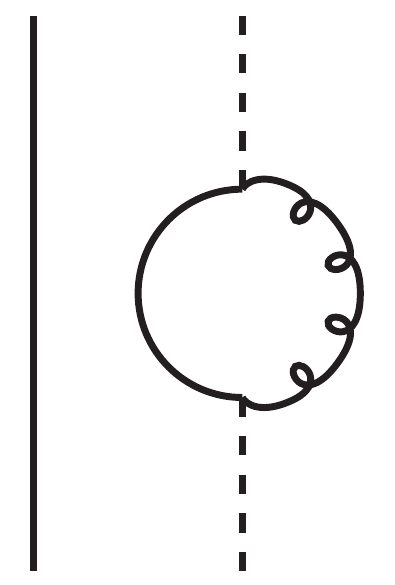}\hspace{2cm}
  \includegraphics[width = 2cm]{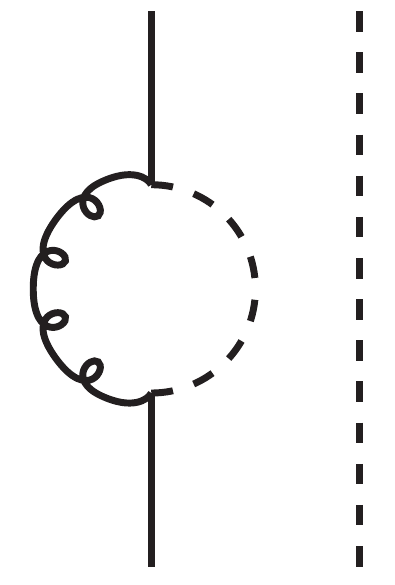}
  \caption{Corrections to $\phi_2\phi_1 \to \phi_{2}\phi_{1}$
    scattering at $\cO(g^2)$. From left to right are the s-channel,
    the u-channel, and two disconnected
    1-loop corrections.
    \label{fig:2to2}}
\end{figure}

Next we examine the $\phi_2\phi_1 \to \phi_{2}\phi_{1}$ amplitude
\eq{
  \ONbrakfi{n_1' n_2'}{n_1 n_2}^{ON(2)}
  = \brakfi{n_1' n_2'}{n_1 n_2}^{(2)}
  - \half \left[\brakff{n_1' n_2'}{n_1 n_2}^{(2)}
    + \brakii{n_1' n_2'}{n_1 n_2}^{(2)} \right] .
}
This expression naturally splits into a connected part
$\brakf{n_1' n_2'}{n_1 n_2}^{(2)}_{i\,C}$ which receives tree-level
corrections and a disconnected part that receives 1-loop corrections; see Fig.~\ref{fig:2to2}. The connected part may further be split
into s-channel and u-channel contributions with the former including
the state corrections $- \half \left[\brakff{n_1' n_2'}{n_1 n_2}^{(2)}
    + \brakii{n_1' n_2'}{n_1 n_2}^{(2)} \right]$.
Using our diagrammatic rules it is straightforward to compute
\eqn{ \label{eq:2to2a}
  \ONbrakf{n_1' n_2'}{n_1 n_2}^{ON(2)}_{i\,s-{\rm channel}}
  &=& + oo + of + po + pf  - \half\left[ pf + po + pf + of \right]
  \nn \\
  &=& + oo + \half( of + po )
  \nn \\
  &=& (ig)^2 \int_{\by} \int_{y} u_1'^*u_2'^*(\by) u_1 u_2(y)
  \cM(\by,y) ,
}
with
\eq{
  \cM(\by,y) :=
  G_3(\by,y) - \half W_3(y,\by)\left( \theta_p(\by) + \theta_f(y)\right) .
}
To verify the optical theorem we compute
\eq{
  - 2 \Re \ONbrakf{n_1' n_2'}{n_1 n_2}^{ON(2)}_{i\,s-{\rm channel}}
  = - (ig)^2 \int_{\by} \int_{y} u_1'^*u_2'^*(\by) u_1 u_2(y)
  \left[ \cM(\by,y) + \cM^*(A\by,y) \right] ,
  \label{eq:2to2b}
}
where in the second term we have once again used (\ref{eq:AonU}). From
(\ref{eq:AonG}) and (\ref{eq:AonW}) it follows that
\eq{
  \cM(\by,y) + \cM^*(A\by,y)
  = G_3(y,\by) + G^*_3(y,\by) - W_3(y,\by) = W_3(\by,y) ,
}
and inserting this into (\ref{eq:2to2b}) we obtain
\eqn{
  - 2 \Re \ONbrakf{n_1' n_2'}{n_1 n_2}^{ON(2)}_{i\,s-{\rm channel}}
  &=& - (ig)^2 \int_{\by} \int_{y} u_1'^*u_2'^*(\by) u_1 u_2(y)
  W_3(\by,y)
  \nn \\
  &=& \sum_{\vL_3} \left[ -(ig) \int_\by u_3 u_2'^* u_1'^*(\by) \right]
  \left[ (ig) \int_y u_3^* u_{2} u_{1}(y) \right]
  \nn \\
  &=& \sum_{\vL_3}
  \brakif{n_2' n_1'}{n_3}^{(1)} \brakfi{n_3}{n_2 n_1}^{(1)} ,
}
where the summand vanishes unless
\eq{
    {\rm max}\left\{ |L_2' - L_1'|,\; |L_{2} - L_{1}| \right\}
    \le L_3 \le
  {\rm min}\left\{ L_2' + L_1',\; L_{2} + L_{1} \right\} .
}
This is indeed a piece of the necessary result.
To complete the desired check of unitarity, we must show that the u-channel
amplitude combines with the 1-loop contributions to yield
\eqn{
  & & -2 \Re \ONbrakf{n_1' n_2'}{n_1 n_2}^{ON(2)}_{i\,u-{\rm channel}
    + {\rm 1-loop}}
  \nn \\ & &
  = \sum_{\vL_1'''} \sum_{\vL_3''} \sum_{\vL_1''}
  \brakif{n_2' n_1'}{n_1''' n_3'' n_1''}^{(1)}
  \brakfi{n_1''' n_3'' n_1''}{n_{2} n_{1}}^{(1)}
  \nn \\ & &
  + \sum_{\vL_2'''} \sum_{\vL_3''} \sum_{\vL_2''}
  \brakif{n_2' n_1'}{n_2''' n_3'' n_2''}^{(1)}
  \brakfi{n_2''' n_3'' n_2''}{n_{2} n_{1}}^{(1)} .
  \label{eq:fullOpt}
}
It is easy to repeat the above steps for the
u-channel, as there is only one term.  Doing so verifies that it
contributes the part of the right-hand side of (\ref{eq:fullOpt})
arising from a cut tree diagram.
Our earlier result for $-2\Re\brakfi{N_1}{N_1}^{(2)}$ shows that
it indeed provides the part of the right-hand side of corresponding
to cut loop diagrams. Thus the optical theorem is satisfied.

\section{Light fields}
\label{sec:light}
\label{sec:R}


Section \ref{sec:Sheavy} restricted study of the
model theory \eqref{eq:model}-\eqref{eq:ct} to bare weights satisfying $\s_1 + \s_2 + \s_3 < - (D-1)$. We now extend the treatment of this model to arbitrarily small positive masses using the $R_\s$ projector of section \ref{sec:goodOps}. As before, our goal is to verify the optical theorem to $\cO(g^2)$.
A more explicit but ultimately equivalent method of computing the S-matrix for light fields is described in appendix \ref{app:goodOps2}, which studies the same model theory to $\cO(g^2)$.

Let us begin by analyzing amplitudes at $\cO(g)$.
It is important to note that the actions of $R_\s(x)$ and time-ordering
do not commute, e.g.:
\eq{
  R_1(x) R_1(\bx) \C{T \phi_1(x) \phi_1(\bx)} =
  \C{T\left[ R_1\phi_1(x) R_1\phi_1(\bx)\right]} + ({\rm contact}) .
}
While these contact terms ultimately do not affect scattering amplitudes,
they do complicate the computations. We find it simpler to first
apply the $R_i(x)$ operators on the Wightman functions of the
theory, and then to use the diagrammatic rules of \S\ref{sec:amplitudes}
to compute amplitudes. It is useful to define $\tR_i(x)$ through
\eq{
  R_\s(x) = 1 + \tR_\s(x)(\Box_x - M^2(\s)) ,
}
so that
\eqn{
  & &\hspace{-2cm}
  R_3(x_3)R_2(x_2)R_1(x_1) \C{\phi_3(x_3)\phi_2(x_2)\phi_1(x_1)}^{(1)}
  \nn \\
  &=& \C{\phi_3(x_3) \phi_2(x_2) \phi_1(x_1)}^{(1)}
  - g \tR_3(x) \C{\phi_2\phi_1(x_3) \phi_2(x_2) \phi_1(x_1)}^{(0)}
  \nn \\ & &
  - g \tR_2(x) \C{\phi_3(x_3) \phi_3\phi_1(x_2) \phi_1(x_1)}^{(0)}
  - g \tR_1(x) \C{\phi_3(x_3) \phi_2(x_2) \phi_3\phi_2(x_1)}^{(0)}
  \nn \\ &=&
  \C{\phi_3(x_3) \phi_2(x_2) \phi_1(x_1)}^{(1)}
  - g \tR_3(x) W_2(x_3,x_2) W_1(x_3,x_1)
  \nn \\ & &
  - g \tR_2(x) W_3(x_3,x_2) W_1(x_2,x_1)
  - g \tR_1(x) W_3(x_3,x_1) W_2(x_2,x_1) ,
  \label{eq:R3pt}
}
where we abbreviate $R_i := R_{\sigma_i}$.

Let us compute the amplitude $\brakfi{n_3 n_2}{n_1}^{(1)}$.
We may time order the first term in (\ref{eq:R3pt})
and use our diagrammatic rules. The contributions of the remaining
terms in (\ref{eq:R3pt}) are easily evaluated because they are
free theory correlators:
\eqn{
  \brakfi{n_3 n_2}{n_1}^{(1)}
  &=& i g \int_{y,\eta_0} u^*_3 u^*_2 u_1(y)
  - g \int d\Sigma^\nu \left[ u_1 \KG_\nu (\tR_1 u^*_3 u^*_2 ) \right]_{-\eta_0}
  \nn \\ & &
  - \frac{g}{2}
  \left[ \int d\Sigma^\nu \left[(\tR_3 u_2^* u_1) \KG_\nu u^*_3 \right]_{+\eta_0}
  + (2 \leftrightarrow 3) \right] .
}
Note that we have temporarily implemented a cutoff $|\eta| < \eta_0$, and
that we have explicitly symmetrized across the final state operators for
convenience.
We may clean up this expression by writing the second term as an
integral over the regulated space:
\eqn{  \brakfi{n_3 n_2}{n_1}^{(1)}
  &=& i g \int_{y,\eta_0} u^*_3 u^*_2 u_1(y)
  + i g \int_{y,\eta_0} u_1 (\Box - M^2_1) \tR_1(u^*_3 u^*_2)
  \nn \\ & &
- \frac{g}{2}
  \left[ \int d\Sigma^\nu \left[(\tR_3 u_2^* u_1) \KG_\nu u^*_3 \right]_{+\eta_0}
  + (2 \leftrightarrow 3) \right]
 + g \int d\Sigma^\nu \left[ u_1 \KG_\nu (\tR_1 u^*_3u_2^*) \right]_{+\eta_0}
 \nn \\
&=& i g \int_{y,\eta_0} \left(R_1 u^*_3 u^*_2\right) u_1(y)
- \frac{g}{2}
  \left[ \int d\Sigma^\nu \left[(\tR_3 u_2^* u_1) \KG_\nu u^*_3 \right]_{+\eta_0}
  + (2 \leftrightarrow 3) \right]
  \nn \\ & &
 + g \int d\Sigma^\nu \left[ u_1 \KG_\nu (\tR_1 u^*_3u_2^*) \right]_{+\eta_0} .
 \label{eq:2to1R}
}

We need to take a moment to analyze integrals of the form
\eq{
  I_3 := \int_y (R_1 u_3^* u_2^*) u_1(y) .
}
This integral is finite for all positive masses. When the weights
satisfy $\Re(\s_1+\s_2+\s_3) < -(D-1)$ we may integrate by parts
to make $R_1(y)$ act on $u_1(y)$; noting that $R_1(x) u_1(x) = u_1(x)$
it follows that
\eq{
  I_3 = \int_y (R_1 u_3^* u_2^*) u_1(y)
  = \int_y u_3^* u_2^* u_1(y)
  = I_1 ,\quad {\rm for\;} \Re(\s_1+\s_2+\s_3) < -(D-1) .
}
Using the linearization formula for $u_3^* u_2^*$ we may
compute $I_3$ in precisely the same way we computed $I_1$ in
\S\ref{sec:OgAmplitudes}.  The only difference is that now the $R_1(x)$ annihilates any contributions
from poles in the complex $\mu$ plane other than the two at
$\mu=\s$ and $\mu=-(\s+D-1)$. Thus for all positive masses
$I_3$ is given by
\eq{
  I_3 = ({\rm RHS\;of\;eq.}\; \eqref{eq:I1final}) \quad \forall M_i^2 > 0 .
}

It follows that in fact
\eq{
  I_3 = \int_y (R_1 u_3^* u_2^*) u_1(y)
  = \int_y (R_2 u_3^* u_1) u_2^*(y)
  = \int_y (R_3 u_2^* u_1^*) u_3^*(y) . \label{eq:I3equalities}
}
All three expressions are analytic (up to poles at the exceptional theories).
So \eqref{eq:I3equalities} is clear from the fact that the integrals agree for
$\Re(\s_1+\s_2+\s_3) < -(D-1)$ when the actions of $R_1, R_2, R_3$ are trivial.

We now return to the amplitude (\ref{eq:2to1R}).  Noting that
\eqn{
  \int_{y,\eta_0} (R_1 u_3^* u_2^*) u_1(y)
  &=& \int_{y,\eta_0} u_3^* u_2^* u_1(y)
  \nn \\ & &
  + \int d\Sigma^\nu \left[ u_1 \KG_\nu (\tR_1 u_3^* u_2^*) \right]_{-\eta_0}
  - \int d\Sigma^\nu \left[ u_1 \KG_\nu (\tR_1 u_3^* u_2^*) \right]_{+\eta_0} .
  \label{eq:I3surface}
}
and using (\ref{eq:I3equalities}), one sees that the surface integrals in (\ref{eq:2to1R}) cancel each other completely so that the final expression for the amplitude is given by $ig$ times \eqref{eq:I1final} for all $M_i^2 > 0$. We see that the the amplitudes are analytic in the weights $\sigma_i$, though with poles at the exceptional theories\footnote{These poles precisely match those found in \cite{Bros:2006gs,Bros:2008sq}.  The coefficients $c_k$ in \eqref{eq:alternateRedefinition} also have poles at these theories.} defined in point \ref{generic} of section \ref{sec:IntRev}.  As in section \ref{sec:Sheavy}, the $\cO(g)$ S-matrix element $\brakfi{n_3 n_2}{n_1}^{(1)}$
is imaginary in agreement
with the optical theorem.

A similar analysis (with corresponding results) can be performed for the remaining amplitudes
at $\cO(g)$. In particular, for all positive masses we have
\eq{
  2 i \Im \brak{\Omega}{n_3 n_2 n_1}_i = (ig)({\rm R.H.S.\; of \; \eqref{eq:I2final}})
  \quad \forall \; M_i^2 > 0 ,
}
see Fig.~\ref{fig:vacuumPlots} for an example.  The construction of an orthonormal basis also proceeds just as for heavy fields and so again has no effect on the 1- and 2-particle states.

We now proceed to $\cO(g^2)$.  Once the $R_\sigma$ operators are in place the calculation is similar to that described in section \ref{sec:Ogheavy} for heavy fields.  We will thus proceed rather quickly.

The inner products which vanish for heavy fields continue to do so for
lights fields as well, e.g. $\braki{n_2' n_1' n_2}{n_1}^{(2)}_{i\,C} = 0$, so it is natural to again define orthonormal states via the analogues of \eqref{eq:ONstates}.  Since the quantities of
interest for the optical theorem are the real parts of the $\cO(g^2)$
S-matrix amplitudes $\brakfi{A}{B}^{(2)}$, we focus only on these objects below.

We begin with the $1\to 1$ S-matrix element $\ONbrakfi{n_1}{n_1}^{ON(2)}$.
The real part may be expressed as
\eq{
  - 2 \Re \ONbrakfi{n_1}{n_1}^{ON(2)}
  = \brakii{n_1}{n_1}^{(2)} + \brakff{n_1}{n_1}^{(2)}
  - \brakfi{n_1}{n_1}^{(2)} - \brakif{n_1}{n_1}^{(2)} .
}
Integrating by parts and using the fact that the $\phi_1(x)$ 2-point
function is de Sitter-invariant, one may recast the right-hand side
as
\eq{
  - 2 \Re \ONbrakfi{n_1}{n_1}^{ON(2)}
  = \int_y u_1(y) (\Box_y -M_1^2) \int_\by u^*(\by)  (\Box_\by - M_1^2)
  R_1(y) R_1(\by)
  \C{\phi_1(\by) \phi_1(y)}^{(2)} .
}
We may freely commute the Klein-Gordon operators through the $R_1$
projectors and let them act on the correlation function. The
KG operators reduce the correlator to an $\cO(g^0)$ correlator:
\eqn{
  - 2 \Re \ONbrakfi{n_1}{n_1}^{ON(2)}
  &=& + g^2 \int_y u_1(y) \int_\by u^*(\by)
  R_1(y) R_1(\by)
  \C{\phi_3\phi_2(\by) \phi_3\phi_2(y)}^{(0)}
  \nn \\
  &=& + g^2 \int_y u_1(y) \int_\by u^*(\by)
  R_1(y) R_1(\by) \left[W_3(\by,y) W_2(\by, y)\right]
  \nn \\
  &=& \sum_{\vL_2} \sum_{\vL_3}
  \left| ig \int_y (R_1 u_3^* u_2^*) u_1(y) \right|^2 .
  \label{eq:1to1Unitary}
}
The right-hand side is precisely what is required by the optical
theorem.

We conclude this section by verifying the optical theorem for
the $2\to 2$ S-matrix element $\ONbrakfi{n_2' n_1'}{n_2 n_1}^{ON(2)}$.
We focus on the s-channel contribution as this is the most non-trivial;
the u-channel contribution to the optical theorem involves fewer diagrams,
and the disconnected contributions follow from (\ref{eq:1to1Unitary}).
The real part of $\ONbrakfi{n_2' n_1'}{n_2 n_1}^{ON(2)}$ may be written
\eqn{
   -2 \Re \ONbrakfi{n_2' n_1'}{n_2 n_1}^{ON(2)} &=&
  \frac{1}{4}\int_{\bx_2} \int_{\bx_1} \int_{x_2} \int_{x_1} \bigg\{
  u_1'{}^*(\bx_1)u_2'{}^*(\bx_2)u_{2}(x_2)u_{1}(x_1)
  \nn \\ & &
  (\Box_{\bx_2} - M_2^2)(\Box_{\bx_1} - M_1^2)
  (\Box_{x_2} - M_2^2)(\Box_{x_1} - M_1^2)
  \nn \\ & &
  R_2(\bx_2)R_1(\bx_1)R_2(x_1)R_1(x_1)
  \nn \\ & &
  \C{\{\phi_2(\bx_2),\, \phi_1(\bx_1)\}\{\phi_2(x_2)
    ,\,\phi_1(x_1)\}}^{(2)}
  \bigg\} .
  \label{eq:2to2R}
}
We have explicitly symmetrized the states for convenience.
Because the correlator is $\cO(g^2)$, only two $R_i(x)$ operators
act non-trivially on the correlator. Let us define the
``s-channel part'' to be the part for which
$\phi_2(\bx_2)$ and $\phi_1(\bx_1)$ are connected by a vertex.
Then the s-channel contribution to (\ref{eq:2to2R}) is
\eqn{
  -2 \Re \ONbrakf{n_2' n_1'}{n_2 n_1}^{ON(2)}_{i\,s-{\rm channel}}
  &=&
  \frac{1}{4}\int_{\bx_2} \int_{\bx_1} \int_{x_2} \int_{x_1} \bigg\{
  u_1'{}^*(\bx_1)u_2'{}^*(\bx_2)u_{2}(x_2)u_{1}(x_1)
  \nn \\ & &
  (\Box_{\bx_2} - M_2^2)(\Box_{\bx_1} - M_1^2)
  (\Box_{x_2} - M_2^2)(\Box_{x_1} - M_1^2)
  \nn \\ & &
  \left[ R_2(\bx_2) R_2(x_2) + R_2(\bx_2) R_1(x_1)
    + R_1(\bx_1) R_2(x_2) + R_1(\bx_1) R_1(x_1) \right]
  \nn \\ & &
  \C{\{\phi_2(\bx_2), \,\phi_1(\bx_1)\}\{\phi_2(x_2),\,
    \phi_1(x_1)\}}^{(2)}_{s-{\rm channel}}
  \bigg\} . \label{eq:moremess}
}
Once again we may freely commute the Klein-Gordon operators
past the $R_i(x)$ operators and let them act on the correlator.
Note that
\eqn{
  & &
  (\Box_{\bx_2} - M_2^2)(\Box_{\bx_1} - M_1^2)
  (\Box_{x_2} - M_2^2)(\Box_{x_1} - M_1^2)
  \C{\{\phi_2(\bx_2),\, \phi_1(\bx_1)\},\,\{\phi_2(x_2)
    \phi_1(x_1)\}}^{(2)}_{s-{\rm channel}}
  \nn \\ & &
  = +g^2 \delta(\bx_2, \bx_1)\delta(x_2,x_1)
  \left[
    W_3(\bx_2,x_2) + W_3(\bx_2,x_1) + W_3(\bx_1,x_2) + W_3(\bx_1,x_1)
  \right] \label{eq:mess}
}
Upon inserting (\ref{eq:mess}) into (\ref{eq:moremess}) we obtain a
number of terms. In each term we may use the delta functions to integrate
over one of the $\bx_i$ and one of the $x_i$. For each term we expand the
Wightman function $W_3(\bx_i,x_j)$ in a sum over harmonics. Cleaning
up a bit and using (\ref{eq:I3equalities}) yields the desired
result:
\eqn{
  -2 \Re \ONbrakfi{n_2' n_1'}{n_2 n_1}^{ON(2)} &=&
  g^2 \int_\by (R_3 u_2'u_1'(\by))
  \int_y (R_3 u_{2}u_{1}(y)) W_3(\by,y)
  \nn \\
  &=& \sum_{\vL_3} \left| (ig) \int_y (R_3 u_{2}u_{1}) u^*_3(y) \right|^2 ,
}
which explicitly verifies our optical theorem.

\section{Discussion}
\label{sec:discussion}

We have introduced a
perturbative
 S-matrix for interacting theories of massive ($M^2 >0$) scalar fields in global de Sitter space.  Our S-matrix reduces in the flat-space limit to the usual Minkowski S-matrix, is invariant under perturbative field redefinitions, and is unitary in the sense stated in section \ref{sec:Smatrix}.  In particular, we have verified the associated optical theorem through order $g^2$ in a $g\phi^3$ theory by explicit calculation.  While there are certain qualitative differences from Minkowski space (lack of branch cuts, non-orthogonality of distinct asymptotic multi-particle states),  we expect our S-matrix to be a useful tool for better understanding interacting de Sitter quantum fields.

As expected from the lack of a positive-definite energy-like conserved quantity, we find that all particles generically decay for any $M^2 >0$; i.e., all particles in dS are unstable.
This result agrees with \cite{Boyanovsky:2004gq,Boyanovsky:2012qs} which followed a rather different approach.
However, let us pause briefly to explain the discrepancy with \cite{Bros:2006gs,Bros:2008sq,Bros:2009bz}, which claim that general complementary series particles can decay only in theories that are exceptional in the sense of section \ref{sec:IntRev}.  Although refs. \cite{Bros:2006gs,Bros:2008sq,Bros:2009bz}  do not define a full S-matrix, they study what they call the inclusive decay rate $\Gamma$ per unit time at $O(g^2)$.    They compute the result in the principle series (where the IR issues are straightforward) and analytically continue to the complementary series.
Their $\Gamma$ is defined by generalizing certain formulae from Minkowski space, and in the principle series it agrees precisely with the right-hand-side of our $O(g^2)$ optical theorem \eqref{eq:OTH1to1}. In the principle series it is thus $- 2 \Re \ONbrakfi{n_1}{n_1}^{ON(2)}.$ The technical discrepancy with our work then arises because while the full amplitude $\ONbrakfi{n_1}{n_1}^{ON(2)}$ is analytic in our formalism, the real part alone fails to be analytic outside the principle series.  The issue is analogous to attempting to compute $\Re e^{i z} = 1/e$ at $z = i$ by analytically continuing from the real axis the function $\cos(x) =  \Re e^{i x}$ and obtaining $\cos(i) = e + e^{-1}$.  The forthcoming work \cite{PS} will comment further on this issue.

In Minkowski space, unstable particles are ultimately unsuitable as asymptotic states.  In particular, this issue can be seen in perturbation theory once self-energy corrections are included in external lines representing the unstable particles.  Before resumming 1PI graphs, the correlation functions develop higher order poles that lead to IR divergences in the S-matrix.  After resumming 1PI graphs and interpreting the result as a self-energy correction, the relevant poles move off the real axis and the associated wavefunctions are no longer normalizable.

We find similar behavior for unstable particles in the principal series of de Sitter representations; i.e., with $M^2 \ell^2 > (D-1)^2/4$. Thus, a theory with only principal-series particles apparently does not have a well-defined S-matrix outside of
perturbation theory. In Minkowski space, the notion of a generalized S-matrix with unstable particles as additional asymptotic states has been sometimes been considered (see, e.g., section 4.9 of \cite{ELOP:1966}). In de Sitter space with principle-series
fields only, such a formalism could be used to define a generalized S-matrix despite the lack of stable particles.

The behavior of complementary series fields is more subtle.  As always, the optical theorem relates the 1-to-2 decay amplitude to the imaginary part of the loop-corrected 1-to-1 amplitude.  However, as will be explained in detail in \cite{PS}, the fact that the wavefunctions do not oscillate near infinity means that the self-energy remains real; i.e., the imaginary part of the 1-to-1 amplitude is not then given by the imaginary part of the self-energy, and the poles remain on the real axis.  This raises the possibility that one might be able to define a finite S-matrix using normalizable asymptotic states built out of complementary-series fields, though this seems far from straightforward; see \cite{PS} for additional comments.

With the above understanding, our S-matrix is free of unphysical IR divergences for generic masses.  However, extra IR divergences do arise when the self-energy $\Pi$ is singular at some generating pole; i.e., for
the exceptional theories described in section \ref{sec:IntRev}.  As a result,  the locations of the poles fail to admit a good perturbative expansion.  In this sense, such exceptional theories are always strongly coupled even for small $g$. Our exceptional theories coincide
precisely with the special cases previously identified in \cite{Bros:2006gs,Bros:2008sq,Bros:2009bz}.

There are of course many possible generalizations of our construction above.  For example, we worked here in global de Sitter space.  We expect that one could use the future half of our Schwinger-Keldysh contour (see figure \ref{fig:SKcontour}) to define a similar S-matrix in the expanding Poincar\'e patch, though we have not explored this case in detail.  In addition, we have thus far restricted attention to scalars with $M^2 > 0$.  Massless minimally coupled ($M^2 =0$) scalars do not have a good dS Hartle-Hawking state in perturbation theory (see e.g. \cite{Allen:1987tz}), but massless higher spin fields are better behaved.  In particular, both the free Maxwell field and graviton admit de Sitter-invariant vacuum states with good analytic continuations to the Euclidean sphere (i.e., Hartle-Hawking states).  \cite{Allen:1985wd,Allen:1986tt,Higuchi:2001uv} One may
therefore hope to extend our S-matrix to such cases.  So long as the associated correlation functions decay sufficiently rapidly\footnote{This is expected
for the Maxwell field but is a matter of significant controversy for the
graviton.  See e.g. \cite{Garriga:2007zk,Tsamis:2007is,Tsamis:2011ep,Higuchi:2012vy}.},
 our use of Klein-Gordon inner products (and the invariance of such inner products under linearized gauge transformations)  suggests that the final result will be gauge-invariant. While naively the fact that $\sigma = -1$ for Maxwell fields and $\sigma =0$ for gravitons would seem to qualify these theories as exceptional, we expect that as in flat space gauge-invariance will prevent the self-energy from acquiring singularities at these values.   A discussion of such issues will appear soon \cite{gauge}.   It would be very interesting to understand if our definition could be further extended to perturbative string theory\footnote{We thank Eva Silverstein for raising this issue.}.

As a final comment, one would also like to understand any relation of our
S-matrix to quantities of interest in dS/CFT
\cite{Maldacena:2002vr,Harlow:2011ke,Anninos:2011ui}.
Since our S-matrix focuses on asymptotic quantities near future and past
infinity, one may expect some relation to exist.  However, because the
Hartle-Hawking state is not itself dual to a particular CFT with fixed
sources, it is in some sense a derived object in dS/CFT as opposed to a
fundamental one \cite{Anninos:2011ui,Ng:2012xp}.  Our reliance on the
Hartle-Hawking state will thus make
any connection somewhat indirect.

\medskip

\noindent{\bf Acknowledgements:} It is a pleasure to thank
Atsushi Higuchi, Stefan Hollands, Gary Gibbons,  Eva Silverstein, and
Aron Wall for interesting discussions. DM was supported in
part by the National Science Foundation under Grant Nos PHY11-25915
and PHY08-55415, by FQXi grant RFP3-1008, and by funds from the University of California.  He also thanks the University of Colorado, Boulder, for its hospitality during the final stages of this work.
MS was supported in part by the National Science Foundation under Grant No PHY07-57035. IM is supported by the Simons Foundation Postdoctoral Fellowship
Program.

\appendix

\section{KG mode linearization}
\label{app:linearization}

\def\a{\alpha}
\def\cN{\mathcal{N}}
\def\GG#1{ \Gamma\left[ #1 \right] }
\def \GGG#1#2{\,\Gamma\left[ \begin{array}{l}
      #1 \\
      #2
    \end{array} \right]}

This appendix gives some details of the function $\rho_{\s_1\s_2 L_1 L_2}(\mu,K)$ appearing in the linearization formula \eqref{eq:fLinearization}.  As noted in the main text, the results below may be derived using the methods of appendix A of \cite{Marolf:2010zp}, and in particular techniques for manipulating Mellin-Barnes representations of hypergeometric functions as described in \cite{Bateman:1955,Slater:1966}.

We introduce $\a := (D-1)/2$ as well as the notation
\begin{eqnarray}
  \GGG{a_1,a_2,\dots,a_k}{b_1,b_2,\dots,b_j}
  &:=& \frac{\Gamma(a_1)\Gamma(a_2)\dots\Gamma(a_k)}
  {\Gamma(b_1)\Gamma(b_2)\dots\Gamma(b_j)}
    \\
  \psi_{\s L}(\nu) &:=&
  \GG{L-\s+\nu,L+\s+2\a+\nu, \half-L-\a-\nu, -\nu} ,
  \\
  \mathcal{N}_{\s L} &:=& \frac{\cos\pi(\s+\a)}{2^{L+\a}\pi}
  \left\{\GG{L-\s,L+\s+2\a}\right\}^{-1/2} ,
\end{eqnarray}
and finally the terminating, ``regulated'' hypergeometric series
\eq{
  \overline{{}_3F_2}\left[\begin{array}{ccccc}
    a,& b,& -p & & \\
    & & & ; & 1 \\
    c,& d & & &
    \end{array} \right]
  := \sum_{n=0}^p \frac{(a)_n (b)_n (-p)_n}{\GG{1+n, c+n, d+n}},
}
with $(a)_n$ the Pochhammer symbol.
This series is an entire function of each parameter $a,b,c,d$.

The linearization kernel may be then expressed as the double Mellin-Barnes integral
\eqn{ \label{eq:rho_app}
  & & \rho_{\s_1\s_2 L_1 L_2}(\mu,K)
  = (-1)^{(L_1+L_2-K)/2} 2^{\a+L_1+L_2} \cN_{\s_1 L_1} \cN_{\s_2 L_2}
  \left\{
    \GG{K-\mu,K+2\a+\mu}\right\}^{1/2}
  \nn \\ & & \times
  \int_{\nu_1} \int_{\nu_2} \Bigg\{
  \psi_{\s_1 L_1}(\nu_1 -L_1) \psi_{\s_2 L_2}(\nu_2 -L_2)
  \nn \\ & & \times
  \GGG{\mu-\nu_1-\nu_2,1-\mu+\nu_1+\nu_2,
    \frac{1+K-L_1-L_2}{2} + \a + \nu_1+\nu_2}
  {\frac{K+L_1+L_2}{2} - \nu_1 - \nu_2}
  \nn \\ & & \times\;
  \overline{{}_3F_2}\left[\begin{array}{ccccc}
    \frac{1+K-L_1-L_2}{2}+\a+\nu_1+\nu_2 ,
    & 1-\left(\frac{K+L_1+L_2}{2}\right)+\nu_1+\nu_2,
    & \frac{K-L_1-L_2}{2} & & \\
    & & & ; & 1 \\
    1+\frac{K-L_1-L_2}{2} + 2\a + \mu + \nu_1 +\nu_2 ,
    & 1 +\frac{K-L_1-L_2}{2} - \mu +\nu_1+\nu_2 & & &
  \end{array} \right] \Bigg\} . \nn \\
}
The well-known asymptotic behavior of the Gamma function may be used to show
that the Mellin-Barnes integrals are absolutely convergent. Equation \eqref{eq:realrho} follows from the identity
$\left[\Gamma(x)\right]^* = \Gamma(x^*)$.

We are primarily interested locating singular points of $\rho_{\s_1\s_2L_1L_2}(\mu,K)$ as a function of $\mu$.
The singularities of the double MB integral in (\ref{eq:rho_app}) may be
determined using standard techniques presented, e.g., in Ch.~4 of
\cite{Smirnov:2004ym}.
Within the integrand there are potentially poles
due to the Gamma functions $\Gamma(\mu-\nu_1-\nu_2)$ and
$\Gamma(1 -\mu+\nu_1+\nu_2)$. However, the poles in the latter Gamma
function are canceled by zeros in each term of the hypergeometric
series, so this Gamma function does not in fact yield singularities.
Thus the only source of singularities from the double integral is the
Gamma function $\Gamma(\mu-\nu_1-\nu_2)$.
The $\nu_i$ integrals have poles at $\nu_i = \s_i -n$ and
$\nu_i = -(\s+2\a) -n$, $n\in\mathbb{N}_0$,
respectively, so it follows that the double integral contributes
poles to $\mu$ at poles at precisely the locations \eqref{eq:rhoPoles}.

The observant reader will note that the first line of (\ref{eq:rho_app})
contains combinations of gamma functions which, when
combined with an identical factor in $f_{\mu K}(\eta)$, contributes additional poles
in the integrand at
\eq{
  \mu=K+n, \quad \mu=-2\a-K-n, \quad n \in \mathbb{N}_0 .
}
We did not mention these poles in section \ref{sec:dSQFT} for two reasons:
first, they lie outside the strip in the complex $\mu$ plane in
which we utilize $\rho_{\s_1\s_2L_1 L_2}(\mu,K)$; second, these poles
do not contribute to the behavior of (\ref{eq:fLinearization})
at large $|\eta| \gg 1$, as we now show.
Recall the asymptotic form (\ref{eq:fLargeEta}) of $f_{\s L}(\eta)$
which we re-write here as
\eqn{ \label{eq:fLargeEta_app}
  f_{\s L}(|\eta|\gg 1) &=& \cN_{\s L} \left[\GG{L-\s, \half-\a-\s, 2(\s+\a)}
  \left(\frac{i\eta}{2}\right)^\s + (\s \to -(\s+2\a))\right]
  \nn \\ & & \times
  \left[1 + \cO(\eta^{-2})\right], \quad
  |\eta| \gg |L-\s| .
}
When $|\eta| \gg L_1 + L_2$, inserting (\ref{eq:fLargeEta_app})
into (\ref{eq:fLinearization}) yields
\eqn{ \label{eq:ffLargeEta}
  f_{\s_1 L_1} f_{\s_2 L_2}(|\eta| \gg L_1+L_2) &=&
  \int_\mu 2(\mu+\a) \rho_{\s_1\s_2 L_1 L_2}(\mu,K) \cN_{\mu K}
  \nn \\ & & \times
  \Bigg\{
  \GG{K-\mu, \half-\a-\mu, 2(\mu+\a)}
  \left(\frac{i\eta}{2}\right)^\mu
  \nn \\ & & \phantom{\Bigg\{ }
   +
  \GG{K+\mu+2\a, \half+\a+\mu, -2(\mu+\a)}
  \left(\frac{i\eta}{2}\right)^{-(\mu+2\a)}
  \Bigg\}
  \nn \\ & & \times
  \left[1+\cO(\eta^{-2})\right] .
}
Notice that the factor $\cN_{\mu K}$ contains the precisely the
combination of gamma functions needed to cancel the similar factor
in $\rho_{\s_1\s_2L_1 L_2}(\mu,K)$.
Upon inspection one finds that $\rho_{\s_1\s_2L_1 L_2}(\mu,K)\cN_{\mu K}$
contains zeros which cancel the poles due to $\Gamma(\pm2(\mu+\a))$.
To obtain the asymptotic form of $f_{\s_1 L_1}(\eta)f_{\s_2 L_2}(\eta)$
from (\ref{eq:ffLargeEta}) we close the integration contour with an arc
at infinity in the left-half plane for the first term, and with an arc
at infinity in the right-half plane for the second term, and use
Cauchy's integral formula to relate these two contour integrals to
the sum of residues obtained. The second term in (\ref{eq:ffLargeEta})
yields zero because the integrand contains no poles in the right-half plane.
The first term in (\ref{eq:ffLargeEta}) acquires residues from the
poles (\ref{eq:rhoPoles}), and these poles precisely reproduce
the asymptotic expansion of $f_{\s_1 L_1}(\eta)f_{\s_2 L_2}(\eta)$.

\section{The explicit form of good operators}
\label{app:goodOps2}

This appendix explores a rather brute force but very explicit method of constructing
good operators in the context of the model theory \eqref{eq:model}-\eqref{eq:ct}.  The basic idea is that, at each order, one simply chooses a perturbative
field redefinition that removes the unwanted
IR-divergent contributions but which leaves the behavior at the desired pole unchanged.  We find such explicit results enlightening, though they become cumbersome at higher orders.  As a result, we work here only to $\cO(g)$.  The results agree completely with those of section \ref{sec:light}.

Recall that the difficulty with the naive prescription \eqref{eq:Aint} was that the $\eta_0 \rightarrow \infty$ limit of
\eq{\label{eq:a7reg}
  \C{a_{\s_3 \vL_3}(\eta_0)\dots}
  -i \int d\Sigma^\nu \C{\phi_3(x)\dots} \KG_\nu u_{\s_3 \vL_3}^*(x)
  \Big|_{\eta=\eta_0}
}
is well-defined only when the correlator satisfies
\eq{
  \C{\phi_3(x)\dots} = ({\rm homogeneous}) + \cO(\eta^{-(\s+3+D-1)}) ,
  \quad \eta \gg 1 . \label{eq:asympt}
}
Here $({\rm homogeneous})$ denotes solutions to the homogeneous
Klein-Gordon equation  $(\Box - M_3^2)f(x) = 0$. We refer to the rest of $\phi_3(x)$ as the inhomogeneous part.
It is straightforward to perturbatively compute the asymptotics of the inhomogeneous part from the Schwinger-Dyson equations. For example, at $\cO(g)$ we have
\eqn{
  (\Box_3 - M_3^2) \C{\phi_3(x_3)\phi_2(x_2)\phi_1(x_1)}^{(1)}
  &=& - g \C{\phi_2\phi_1(x_3)\phi_2(x_2)\phi_1(x_1)}^{(0)}
  \nn \\
  &=& - g W_2(x_3,x_2) W_1(x_3,x_1)
  \nn \\
  &=& \Big[\cO(\eta_3^{\s_2+\s_1}) + \cO(\eta_3^{\s_2 -\s_1-(D-1)})
  + \cO(\eta_3^{-\s_2+\s_1-(D-1)}) \nn \\ & & \phantom{\bigg[}
  + \cO(\eta_3^{-\s_2 -\s_1-2(D-1)})\Big]
  \left[1 + \cO(\eta_3^{-2})\right] \quad {\rm for\;} |\eta_3| \gg 1 \ \ \ \ \ \ \ \ ,
  \label{eq:3ptAsympt}
}
where the final equality summarizes the asymptotic behavior of
the Wightman functions. Because the action of $(\Box_3 - M_3^2)$ does not change the asymptotic behavior of power laws of $\eta$, the right hand side of \eqref{eq:3ptAsympt} also describes the inhomogeneous part of $\C{\phi_3(x_3)\phi_2(x_2)\phi_1(x_1)}^{(1)}$.
This violates the condition \eqref{eq:asympt} for $-(D-1) < \Re(\s_1 + \s_2 + \s_3)$, and moreover contributes unwanted information to \eqref{eq:a7reg}.

Acting with the $R_\sigma$ projector of section \ref{sec:goodOps} is thus equivalent (in the sense of section \ref{sec:goodOps}) to constructing some $\Phi_3(x)$ whose asymptotics contain the same homogeneous part as $\phi_3(x)$ but which lacks the divergent inhomogeneous
parts.    So for $-(D-1) < \Re(\s_1 + \s_2 + \s_3)$
let $P_3=\{\s_2+\s_1,\dots\}$ be the set of weights belonging
to terms in (\ref{eq:3ptAsympt}) violating \eqref{eq:asympt} and consider
\eq{
  \Phi_3(x) = \phi_3(x) + g \sum_{p\in{P_3}} \Op_p(x) + \cO(g^2) ,
}
where $\Op_p(x)$ denotes an operator with asymptotic behavior $\cO(\eta^p) + \cO(\eta^{-(\s_3+D-1)})$ at $O(g^0)$.
We choose the $\Op_p(x)$ to be normalized so that
\eq{
  \Op_p(x) = \frac{1}{M^2(p) - M_3^2} \phi_2\phi_1(x) \Big|_{p}
 + \cO(\eta^{-(\s_3+D-1)}) ,
}
where $\phi(x)|_p$ denotes the $\cO(\eta^p)$ part of $\phi(x)$ at $\cO(g^0)$ and $M^2(p)$ is the squared-mass which gives rise to this fall off at $O(g^0)$.
The choice of operators $\Op_p(x)$ is not unique, but a natural choice
is
\eqn{ \label{eq:Op}
  \Op_{p}(x) &=& \frac{1}{M^2(p) - M_3^2}
  \left[ \prod_{q \in P_3,\,q \neq p}
   \frac{(\Box - M^2(q))}{(M^2(p) - M^2(q))}
 \right] \phi_2\phi_1(x)
 \nn \\ &=& \frac{1}{M^2(p) - M_3^2} \phi_2\phi_1(x) \Big|_{p}
 + \cO(\eta^{-(\s_3+D-1)}) .
}
With this choice for the $\Op_p(x)$ the new operator $\Phi_3(x)$
may also be written
\eq{ \label{eq:alternateRedefinition}
  \Phi_3(x) = \phi_3(x)
  + g \sum_{k=0} c_{k} \Box^k \phi_2\phi_1(x) ,
}
with appropriate coefficients $c_k$.  A similar procedure should be applied to $\Phi_2(x)$ and $\Phi_1(x)$.

It is sufficient to simply change variables from $\phi_i(x)$ to $\Phi_i(x)$ in the generating functional (the
path integral). This is most easily
accomplished with $\Phi_i(x)$ defined as in (\ref{eq:alternateRedefinition}).
The lagrangian for the $\Phi_i(x)$ contains a quadratic part in which
each $\Phi_i(x)$ has a
canonically normalized kinetic term and a mass term that agrees with that of the original field $\phi_i(x)$. At order $\cO(g)$ the lagrangian
contains the cubic interaction $c g \Phi_3\Phi_2\Phi_1(x)$, now with
an additional coefficient $c$, as well as higher-derivative cubic interactions.
These higher-derivative interactions, taken on their own, produce correlators
whose asymptotics have similar fall-off properties to the simple cubic
interaction. But these higher-derivative interactions
conspire to cancel the unwanted inhomogeneous terms in (\ref{eq:3ptAsympt})
without altering the homogeneous part. One could say that the new
higher-derivative interactions are ``IR counterterms,'' though we stress
that the theory has not been altered nor does the theory
need such terms in order for correlation functions to be well-defined.

We can now compute $\cO(g)$ amplitudes. To keep the equations
tractable let us consider the case where there is only one
problematic term at $\cO(\eta^{\s_1+\s_2+\s_3})$. In this case the $\Phi_i(x)$ defined
in (\ref{eq:Op}) are
\eqn{
  \Phi_3(x) &=& \phi_3(x) + \frac{ g }{M^2(\s_2 +\s_1) - M^2_3}\phi_2\phi_1(x)
  \nn \\
  &=& \phi_3(x)
  + \frac{ g }{(\s_3 - \s_1 - \s_2)(\s_1+\s_2+\s_3 + D-1)}\phi_2\phi_1(x)
 \label{eq:Phi3easy}
}
and likewise for $\Phi_2(x)$, $\Phi_1(x)$. The $\cO(g)$ 3-pt. function
of the corrected operators is
\eqn{
  \C{\Phi_3(x_3)\Phi_2(x_2)\Phi_1(x_1)}^{(1)}
  &=& \C{\phi_3(x_3)\phi_2(x_2)\phi_1(x_1)}^{(1)}
  \nn \\ & &
  + \frac{g}{M^2(\s_2 +\s_1) - M^2_3}
  \C{\phi_2\phi_1(x_3)\phi_2(x_2)\phi_1(x_1)}^{(0)}
  \nn \\ & &
  + \frac{g}{M^2(\s_3 +\s_1) - M^2_2}
  \C{\phi_3(x_3)\phi_3\phi_1(x_2)\phi_1(x_1)}^{(0)}
  \nn \\ & &
  + \frac{g}{M^2(\s_3 +\s_2) - M^2_1}
  \C{\phi_3(x_3)\phi_2(x_2)\phi_3\phi_2(x_1)}^{(0)} .
}
The $1\to 2$ transition amplitude is then
\eqn{
  \brakfi{n_3 n_2}{n_1}^{(1)}
  &=& i g\int_{\eta(y) \le \eta_0} u_3^* u_2^* u_1(y)
  \nn \\ & &
  - i g \int d\Sigma^\nu \left[ \half \frac{u_2^*u_1}{(M^2(\s_2 +\s_1) - M^2_3)}
    \KG_\nu u_3^*(x)
  + (2 \leftrightarrow 3) \right]_{\eta=+\eta_0}
  \nn \\ & &
  - i g
  \int d\Sigma^\nu
  \left[ u_1 \KG_\nu \frac{u_3^* u_2^*(x)}{M^2(\s_3 +\s_2) - M^2_1}
\right]_{\eta = -\eta_0} . \label{eq:2to1Phi}
}
Here we have regulated right-hand-side  by imposing $|\eta| < \eta_0$ since each term is
individually divergent as $|\eta_0| \to \infty$. We have also explicitly
symmetrized the operators of the final state.

The first term is just the regulated version $i g I_1(\eta_0)$ of the integral considered
in \S\ref{sec:OgAmplitudes}.
Recall that our computation of $I_1(\eta_0)$ for heavy fields  used the
linearization formula (\ref{eq:uLinearization}) on $u_3 u_2(x)$
then deformed the $\mu$ integration contour to
$\Re \mu < - \Re(\s_1+D-1)$. In the process we acquired residues due to
simple poles at $\mu = \s_1$ and $\mu = -(\s_1+D-1)$. A similar procedure can be used in the present case but now,
since
$- \Re(\s_1+D-1) < \Re(\s_2+\s_3)$, deforming the contour to the
left picks up an additional residue from the simple pole in the linearization kernel at
$\mu = \s_2+\s_3$. This new pole now
provides the leading behavior of $u_3 u_2(x)$. Thus
\eqn{
  I_1(\eta_0) &=& ({\rm RHS\;of\;eq.}\; \ref{eq:I1final})
  \nn \\ & &
  - \frac{1}{M^2(\s_2+\s_3) - M_1^2}
  \int d\Sigma^\nu(y)
  \bigg\{ \Big[u^*_3 u^*_2 \KG_\nu u_{1}(y)\Big]_{\eta = \eta_0}
    - \Big[ u^*_3 u^*_2 \KG_\nu u_{1}(y)
    \Big]_{\eta=-\eta_0}\bigg\}
    \nn \\ & & + \dots,
}
where the ellipses denote the remaining integral over some contour with $\Re \mu < - \Re(\s_1+D-1)$.  Contributions from this remaining contour integral vanish in the limit
$\eta_0\to\infty$.

We now examine the terms on the second and third lines of (\ref{eq:2to1Phi}).
Note that only the leading term of each KG mode
$u_{i}(x) = \cO(\eta^{\s_i})$ will contribute to these surface integrals
in the limit $\eta_0\to\infty$.  Noting that
\eq{
  \half\left[ \frac{(\s_3-\s_2-\s_1)}{M^2(\s_2+\s_1) - M_3^2}
    + (2 \leftrightarrow 3) \right]
  = -\frac{(\s_3+\s_2-\s_1)}{M^2(\s_3+\s_2)- M_1^2} ,
}
it follows that the contribution of the integrals at $\eta=+H$
may also be written as
\eqn{
  & &- i g \int d\Sigma^\nu \left[ \half \frac{u_2^*u_1}{(M^2(\s_2 +\s_1) - M^2_3)}
    \KG_\nu u_3^*(x)
    + (2 \leftrightarrow 3) \right]_{\eta=+\eta_0}
  \nn \\ & &
  = \frac{i g}{M^2(\s_3+\s_2) - M_1^2}
  \int d\Sigma^\nu \left[u_1 \KG_\nu u_3^*u_2^*(y)\right]_{\eta=+\eta_0}
  + \dots,
}
where again ellipses denotes terms that vanish as $\eta_0\to\infty$.
It follows that for $\eta_0 \rightarrow \infty$ the last two lines of (\ref{eq:2to1Phi}) precisely cancel the new residue term in $I_1(\eta_0)$. Thus the amplitude
continues to be given by $ig$ times \eqref{eq:I1final}.

It will not surprise the reader that the same result turns out to be valid for all $M_i^2 > 0$.  A detailed check shows that similar cancellations occur in the presence of further problematic terms.
For every divergent term $\cO(\eta^{\s_1+\s_2+\s_3})$,
$\cO(\eta^{\s_1+\s_2+\s_3-2})$, $\cO(\eta^{-\s_1+\s_2+\s_3-(D-1)})$, etc.,
one acquires a set of surface integrals arising from the residue of a
pole in the derivation of $I_1(\eta_0)$. These surface integrals are
exactly canceled by the contribution of the appropriate $\Op_p(x)$
operators.
It is also straightforward (if tedious) to verify similar cancellations
of the divergent terms in the remaining $\cO(g)$ amplitudes.  In agreement with section \ref{sec:light}, we see that the results are again given by the corresponding final formulae in section \ref{sec:Ogheavy}.


\addcontentsline{toc}{section}{Bibliography}
\bibliographystyle{JHEPhyper}
\bibliography{./dSSBibliography8}

\providecommand{\href}[2]{#2}\begingroup\raggedright\begin{thebibliography}{10}

\bibitem{Coleman:1967ad}
S.~R. Coleman and J.~Mandula, {\it {All possible symmetries of the S matrix}},
  {\em Phys. Rev.} \href{http://dx.doi.org/10.1103/PhysRev.159.1251}{{\bf 159}}
  \href{http://dx.doi.org/10.1103/PhysRev.159.1251}{\href{http://dx.doi.org/10%
.1103/PhysRev.159.1251}{(1967)}}
  \href{http://dx.doi.org/10.1103/PhysRev.159.1251}{\href{http://dx.doi.org/10%
.1103/PhysRev.159.1251}{\href{http://dx.doi.org/10.1103/PhysRev.159.1251}{1251%
--1256}}}.

\bibitem{Haag:1974qh}
R.~Haag, J.~T. Lopuszanski, and M.~Sohnius, {\it {All Possible Generators of
  Supersymmetries of the s Matrix}},  {\em Nucl. Phys.}
  \href{http://dx.doi.org/10.1016/0550-3213(75)90279-5}{{\bf B88}}
  \href{http://dx.doi.org/10.1016/0550-3213(75)90279-5}{\href{http://dx.doi.or%
g/10.1016/0550-3213(75)90279-5}{(1975)}}
  \href{http://dx.doi.org/10.1016/0550-3213(75)90279-5}{\href{http://dx.doi.or%
g/10.1016/0550-3213(75)90279-5}{\href{http://dx.doi.org/10.1016/0550-3213(75)9%
0279-5}{257}}}.

\bibitem{Weinberg:1980kq}
S.~Weinberg and E.~Witten, {\it {Limits on Massless Particles}},  {\em Phys.
  Lett.} \href{http://dx.doi.org/10.1016/0370-2693(80)90212-9}{{\bf B96}}
  \href{http://dx.doi.org/10.1016/0370-2693(80)90212-9}{\href{http://dx.doi.or%
g/10.1016/0370-2693(80)90212-9}{(1980)}}
  \href{http://dx.doi.org/10.1016/0370-2693(80)90212-9}{\href{http://dx.doi.or%
g/10.1016/0370-2693(80)90212-9}{\href{http://dx.doi.org/10.1016/0370-2693(80)9%
0212-9}{59}}}.

\bibitem{Polyakov:2007mm}
A.~M. Polyakov, {\it {De Sitter Space and Eternity}},  {\em Nucl. Phys.}
  \href{http://dx.doi.org/10.1016/j.nuclphysb.2008.01.002}{{\bf B797}}
  \href{http://dx.doi.org/10.1016/j.nuclphysb.2008.01.002}{\href{http://dx.doi%
.org/10.1016/j.nuclphysb.2008.01.002}{(2008)}},
  \href{http://dx.doi.org/10.1016/j.nuclphysb.2008.01.002}{no.~1-2}
  \href{http://dx.doi.org/10.1016/j.nuclphysb.2008.01.002}{\href{http://dx.doi%
.org/10.1016/j.nuclphysb.2008.01.002}{199--217}},
  \href{http://dx.doi.org/10.1016/j.nuclphysb.2008.01.002}{[\href{http://xxx.l%
anl.gov/abs/0709.2899}{{\tt arXiv:0709.2899}}]}.

\bibitem{Polyakov:2009nq}
A.~Polyakov, {\it {Decay of Vacuum Energy}},  {\em Nucl. Phys.}
  \href{http://dx.doi.org/10.1016/j.nuclphysb.2010.03.021}{{\bf B834}}
  \href{http://dx.doi.org/10.1016/j.nuclphysb.2010.03.021}{\href{http://dx.doi%
.org/10.1016/j.nuclphysb.2010.03.021}{(2010)}}
  \href{http://dx.doi.org/10.1016/j.nuclphysb.2010.03.021}{\href{http://dx.doi%
.org/10.1016/j.nuclphysb.2010.03.021}{316--329}},
  \href{http://dx.doi.org/10.1016/j.nuclphysb.2010.03.021}{[\href{http://xxx.l%
anl.gov/abs/0912.5503}{{\tt arXiv:0912.5503}}]}.

\bibitem{Krotov:2010ma}
D.~Krotov and A.~M. Polyakov, {\it {Infrared Sensitivity of Unstable Vacua}},
  {\em Nucl. Phys.}
  \href{http://dx.doi.org/10.1016/j.nuclphysb.2011.03.025}{{\bf B849}}
  \href{http://dx.doi.org/10.1016/j.nuclphysb.2011.03.025}{\href{http://dx.doi%
.org/10.1016/j.nuclphysb.2011.03.025}{(2011)}}
  \href{http://dx.doi.org/10.1016/j.nuclphysb.2011.03.025}{\href{http://dx.doi%
.org/10.1016/j.nuclphysb.2011.03.025}{410--432}},
  \href{http://dx.doi.org/10.1016/j.nuclphysb.2011.03.025}{[\href{http://xxx.l%
anl.gov/abs/1012.2107}{{\tt arXiv:1012.2107}}]}.

\bibitem{Polyakov:2012uc}
A.~Polyakov, {\it {Infrared instability of the de Sitter space}},
  \href{http://xxx.lanl.gov/abs/1209.4135}{{\tt arXiv:1209.4135}}.

\bibitem{Marolf:2010zp}
D.~Marolf and I.~A. Morrison, {\it {The IR stability of de Sitter: Loop
  corrections to scalar propagators}},  {\em Phys. Rev.}
  \href{http://dx.doi.org/10.1103/PhysRevD.82.105032}{{\bf D82}}
  \href{http://dx.doi.org/10.1103/PhysRevD.82.105032}{\href{http://dx.doi.org/%
10.1103/PhysRevD.82.105032}{(Nov, 2010)}}
  \href{http://dx.doi.org/10.1103/PhysRevD.82.105032}{\href{http://dx.doi.org/%
10.1103/PhysRevD.82.105032}{105032}},
  \href{http://dx.doi.org/10.1103/PhysRevD.82.105032}{[\href{http://xxx.lanl.g%
ov/abs/1006.0035}{{\tt arXiv:1006.0035}}]}.

\bibitem{Marolf:2010nz}
D.~Marolf and I.~A. Morrison, {\it {Infrared stability of de Sitter QFT:
  Results at all orders}},  {\em Phys. Rev.}
  \href{http://dx.doi.org/10.1103/PhysRevD.84.044040}{{\bf D84}}
  \href{http://dx.doi.org/10.1103/PhysRevD.84.044040}{\href{http://dx.doi.org/%
10.1103/PhysRevD.84.044040}{(Aug, 2011)}}
  \href{http://dx.doi.org/10.1103/PhysRevD.84.044040}{\href{http://dx.doi.org/%
10.1103/PhysRevD.84.044040}{044040}},
  \href{http://dx.doi.org/10.1103/PhysRevD.84.044040}{[\href{http://xxx.lanl.g%
ov/abs/1010.5327}{{\tt arXiv:1010.5327}}]}.

\bibitem{Marolf:2011aa}
D.~Marolf and I.~A. Morrison, {\it {The IR stability of de Sitter QFT: Physical
  initial conditions}},  {\em Gen. Rel. Grav.}
  \href{http://dx.doi.org/10.1007/s10714-011-1233-3}{\href{http://dx.doi.org/1%
0.1007/s10714-011-1233-3}{(2011)}}
  \href{http://dx.doi.org/10.1007/s10714-011-1233-3}{\href{http://dx.doi.org/1%
0.1007/s10714-011-1233-3}{1--34}},
  \href{http://dx.doi.org/10.1007/s10714-011-1233-3}{[\href{http://xxx.lanl.go%
v/abs/1104.4343}{{\tt arXiv:1104.4343}}]}.

\bibitem{Allen:1986ta}
B.~Allen, {\it {The graviton propagator in de Sitter space}},  {\em Phys. Rev.}
  \href{http://dx.doi.org/10.1103/PhysRevD.34.3670}{{\bf D34}}
  \href{http://dx.doi.org/10.1103/PhysRevD.34.3670}{\href{http://dx.doi.org/10%
.1103/PhysRevD.34.3670}{(1986)}}
  \href{http://dx.doi.org/10.1103/PhysRevD.34.3670}{\href{http://dx.doi.org/10%
.1103/PhysRevD.34.3670}{\href{http://dx.doi.org/10.1103/PhysRevD.34.3670}{3670%
}}}.

\bibitem{Miao:2010vs}
S.~Miao, N.~Tsamis, and R.~Woodard, {\it {De Sitter Breaking through Infrared
  Divergences}},  {\em Jour. Math. Phys.}
  \href{http://dx.doi.org/10.1063/1.3448926}{{\bf 51}}
  \href{http://dx.doi.org/10.1063/1.3448926}{\href{http://dx.doi.org/10.1063/1%
.3448926}{(2010)}}
  \href{http://dx.doi.org/10.1063/1.3448926}{\href{http://dx.doi.org/10.1063/1%
.3448926}{072503}},
  \href{http://dx.doi.org/10.1063/1.3448926}{[\href{http://xxx.lanl.gov/abs/10%
02.4037}{{\tt arXiv:1002.4037}}]}.

\bibitem{Giddings:2010ui}
S.~B. Giddings and M.~S. Sloth, {\it {Cosmological diagrammatic rules}},  {\em
  JCAP} \href{http://dx.doi.org/10.1088/1475-7516/2010/07/015}{{\bf 1007}}
  \href{http://dx.doi.org/10.1088/1475-7516/2010/07/015}{\href{http://dx.doi.o%
rg/10.1088/1475-7516/2010/07/015}{(2010)}}
  \href{http://dx.doi.org/10.1088/1475-7516/2010/07/015}{\href{http://dx.doi.o%
rg/10.1088/1475-7516/2010/07/015}{015}},
  \href{http://dx.doi.org/10.1088/1475-7516/2010/07/015}{[\href{http://xxx.lan%
l.gov/abs/1005.3287}{{\tt arXiv:1005.3287}}]}.

\bibitem{Miao:2011fc}
S.~Miao, N.~Tsamis, and R.~Woodard, {\it {The Graviton Propagator in de Donder
  Gauge on de Sitter Background}},  {\em Jour. Math. Phys.}
  \href{http://dx.doi.org/10.1063/1.3664760}{{\bf 52}}
  \href{http://dx.doi.org/10.1063/1.3664760}{\href{http://dx.doi.org/10.1063/1%
.3664760}{(2011)}}
  \href{http://dx.doi.org/10.1063/1.3664760}{\href{http://dx.doi.org/10.1063/1%
.3664760}{122301}},
  \href{http://dx.doi.org/10.1063/1.3664760}{[\href{http://xxx.lanl.gov/abs/11%
06.0925}{{\tt arXiv:1106.0925}}]}.

\bibitem{Miao:2011ng}
S.~Miao, N.~Tsamis, and R.~Woodard, {\it {Gauging away Physics}},  {\em Class.
  Quant. Grav.} \href{http://dx.doi.org/10.1088/0264-9381/28/24/245013}{{\bf
  28}}
  \href{http://dx.doi.org/10.1088/0264-9381/28/24/245013}{\href{http://dx.doi.%
org/10.1088/0264-9381/28/24/245013}{(2011)}}
  \href{http://dx.doi.org/10.1088/0264-9381/28/24/245013}{\href{http://dx.doi.%
org/10.1088/0264-9381/28/24/245013}{245013}},
  \href{http://dx.doi.org/10.1088/0264-9381/28/24/245013}{[\href{http://xxx.la%
nl.gov/abs/1107.4733}{{\tt arXiv:1107.4733}}]}.

\bibitem{Mora:2012zi}
P.~Mora, N.~Tsamis, and R.~Woodard, {\it {Graviton Propagator in a General
  Invariant Gauge on de Sitter}},
  \href{http://xxx.lanl.gov/abs/1205.4468}{{\tt arXiv:1205.4468}}.

\bibitem{Higuchi:2011vw}
A.~Higuchi, D.~Marolf, and I.~A. Morrison, {\it {de Sitter invariance of the dS
  graviton vacuum}},  {\em Class. Quant. Grav.}
  \href{http://dx.doi.org/10.1088/0264-9381/28/24/245012}{{\bf 28}}
  \href{http://dx.doi.org/10.1088/0264-9381/28/24/245012}{\href{http://dx.doi.%
org/10.1088/0264-9381/28/24/245012}{(2011)}}
  \href{http://dx.doi.org/10.1088/0264-9381/28/24/245012}{\href{http://dx.doi.%
org/10.1088/0264-9381/28/24/245012}{245012}},
  \href{http://dx.doi.org/10.1088/0264-9381/28/24/245012}{[\href{http://xxx.la%
nl.gov/abs/1107.2712}{{\tt arXiv:1107.2712}}]}.

\bibitem{Nachtmann:1968aa}
O.~Nachtmann, {\it {Dynamishe Stabilit\"at im de-Sitter-Raum}},  {\em Sitz.
  Ber. \"Osk. Ak. d. Wiss. II} {\bf 176} (1968) 363.

\bibitem{Myhrvold:1983hx}
N.~P. Myhrvold, {\it {Runaway particle production in de Sitter space}},  {\em
  Phys. Rev.} \href{http://dx.doi.org/10.1103/PhysRevD.28.2439}{{\bf D28}}
  \href{http://dx.doi.org/10.1103/PhysRevD.28.2439}{\href{http://dx.doi.org/10%
.1103/PhysRevD.28.2439}{(1983)}}
  \href{http://dx.doi.org/10.1103/PhysRevD.28.2439}{\href{http://dx.doi.org/10%
.1103/PhysRevD.28.2439}{\href{http://dx.doi.org/10.1103/PhysRevD.28.2439}{2439%
}}}.

\bibitem{Boyanovsky:1996ab}
D.~Boyanovsky, R.~Holman, and S.~Prem~Kumar, {\it {Inflaton decay in De Sitter
  spacetime}},  {\em Phys. Rev.}
  \href{http://dx.doi.org/10.1103/PhysRevD.56.1958}{{\bf D56}}
  \href{http://dx.doi.org/10.1103/PhysRevD.56.1958}{\href{http://dx.doi.org/10%
.1103/PhysRevD.56.1958}{(1997)}}
  \href{http://dx.doi.org/10.1103/PhysRevD.56.1958}{\href{http://dx.doi.org/10%
.1103/PhysRevD.56.1958}{1958--1972}},
  \href{http://dx.doi.org/10.1103/PhysRevD.56.1958}{[\href{http://xxx.lanl.gov%
/abs/hep-ph/9606208}{{\tt hep-ph/9606208}}]}.

\bibitem{Boyanovsky:2011xn}
D.~Boyanovsky and R.~Holman, {\it {On the Perturbative Stability of Quantum
  Field Theories in de Sitter Space}},  {\em JHEP}
  \href{http://dx.doi.org/10.1007/JHEP05(2011)047}{{\bf 1105}}
  \href{http://dx.doi.org/10.1007/JHEP05(2011)047}{\href{http://dx.doi.org/10.%
1007/JHEP05(2011)047}{(2011)}}
  \href{http://dx.doi.org/10.1007/JHEP05(2011)047}{\href{http://dx.doi.org/10.%
1007/JHEP05(2011)047}{047}},
  \href{http://dx.doi.org/10.1007/JHEP05(2011)047}{[\href{http://xxx.lanl.gov/%
abs/1103.4648}{{\tt arXiv:1103.4648}}]}.

\bibitem{Haag:1992aa}
R.~Haag, {\em Local quantum physics: fields, particles, algebras}.
\newblock Texts and monographs in physics. Springer-Verlag, 1992.

\bibitem{Schwinger:1960qe}
J.~S. Schwinger, {\it {Brownian motion of a quantum oscillator}},  {\em Jour.
  Math. Phys.} \href{http://dx.doi.org/10.1063/1.1703727}{{\bf 2}}
  \href{http://dx.doi.org/10.1063/1.1703727}{\href{http://dx.doi.org/10.1063/1%
.1703727}{(1961)}}
  \href{http://dx.doi.org/10.1063/1.1703727}{\href{http://dx.doi.org/10.1063/1%
.1703727}{\href{http://dx.doi.org/10.1063/1.1703727}{407--432}}}.

\bibitem{Keldysh:1964ud}
L.~V. Keldysh, {\it {Diagram technique for nonequilibrium processes}},  {\em
  Zh. Eksp. Teor. Fiz.} {\bf 47} (1964) 1515--1527.

\bibitem{Chou}
K.-c. Chou, Z.-b. Su, B.-l. Hoa, and L.~Yu, {\it {Equilibrium and
  nonequilibrium formalisms made unified}},  {\em Phys. Rept.}
  \href{http://dx.doi.org/10.1016/0370-1573(85)90136-X}{{\bf 118}}
  \href{http://dx.doi.org/10.1016/0370-1573(85)90136-X}{\href{http://dx.doi.or%
g/10.1016/0370-1573(85)90136-X}{(1985)}}
  \href{http://dx.doi.org/10.1016/0370-1573(85)90136-X}{\href{http://dx.doi.or%
g/10.1016/0370-1573(85)90136-X}{\href{http://dx.doi.org/10.1016/0370-1573(85)9%
0136-X}{1}}}.

\bibitem{Landsman:1986uw}
N.~P. Landsman and C.~G. van Weert, {\it {Real and Imaginary Time Field Theory
  at Finite Temperature and Density}},  {\em Phys. Rept.}
  \href{http://dx.doi.org/10.1016/0370-1573(87)90121-9}{{\bf 145}}
  \href{http://dx.doi.org/10.1016/0370-1573(87)90121-9}{\href{http://dx.doi.or%
g/10.1016/0370-1573(87)90121-9}{(1987)}}
  \href{http://dx.doi.org/10.1016/0370-1573(87)90121-9}{\href{http://dx.doi.or%
g/10.1016/0370-1573(87)90121-9}{\href{http://dx.doi.org/10.1016/0370-1573(87)9%
0121-9}{141}}}.

\bibitem{Hollands:2010pr}
S.~Hollands, {\it {Correlators, Feynman diagrams, and quantum no-hair in
  deSitter spacetime}},  \href{http://xxx.lanl.gov/abs/1010.5367}{{\tt
  arXiv:1010.5367}}.

\bibitem{Higuchi:2010aa}
A.~Higuchi, D.~Marolf, and I.~A. Morrison, {\it {On the Equivalence between
  Euclidean and In-In formalisms in de Sitter QFT}},  {\em Phys. Rev.}
  \href{http://dx.doi.org/10.1103/PhysRevD.83.084029}{{\bf D83}}
  \href{http://dx.doi.org/10.1103/PhysRevD.83.084029}{\href{http://dx.doi.org/%
10.1103/PhysRevD.83.084029}{(Apr, 2011)}}
  \href{http://dx.doi.org/10.1103/PhysRevD.83.084029}{\href{http://dx.doi.org/%
10.1103/PhysRevD.83.084029}{084029}},
  \href{http://dx.doi.org/10.1103/PhysRevD.83.084029}{[\href{http://xxx.lanl.g%
ov/abs/1012.3415}{{\tt arXiv:1012.3415}}]}.

\bibitem{Floratos:1987aa}
E.~Floratos, J.~Iliopoulos, and T.~Tomaras, {\it Tree-level scattering
  amplitudes in de sitter space diverge},  {\em Phys. Lett.}
  \href{http://dx.doi.org/10.1016/0370-2693(87)90403-5}{{\bf 197}}
  \href{http://dx.doi.org/10.1016/0370-2693(87)90403-5}{\href{http://dx.doi.or%
g/10.1016/0370-2693(87)90403-5}{(1987)}},
  \href{http://dx.doi.org/10.1016/0370-2693(87)90403-5}{no.~B3}
  \href{http://dx.doi.org/10.1016/0370-2693(87)90403-5}{\href{http://dx.doi.or%
g/10.1016/0370-2693(87)90403-5}{\href{http://dx.doi.org/10.1016/0370-2693(87)9%
0403-5}{373 -- 378}}}.

\bibitem{Akhmedov:2008pu}
E.~T. Akhmedov and P.~V. Buividovich, {\it {Interacting Field Theories in de
  Sitter Space are Non- Unitary}},  {\em Phys. Rev.}
  \href{http://dx.doi.org/10.1103/PhysRevD.78.104005}{{\bf D78}}
  \href{http://dx.doi.org/10.1103/PhysRevD.78.104005}{\href{http://dx.doi.org/%
10.1103/PhysRevD.78.104005}{(2008)}}
  \href{http://dx.doi.org/10.1103/PhysRevD.78.104005}{\href{http://dx.doi.org/%
10.1103/PhysRevD.78.104005}{104005}},
  \href{http://dx.doi.org/10.1103/PhysRevD.78.104005}{[\href{http://xxx.lanl.g%
ov/abs/0808.4106}{{\tt arXiv:0808.4106}}]}.

\bibitem{Higuchi:2008tn}
A.~Higuchi, {\it {Tree-level vacuum instability in an interacting field theory
  in de Sitter spacetime}},  {\em Class. Quant. Grav.}
  \href{http://dx.doi.org/10.1088/0264-9381/26/7/072001}{{\bf 26}}
  \href{http://dx.doi.org/10.1088/0264-9381/26/7/072001}{\href{http://dx.doi.o%
rg/10.1088/0264-9381/26/7/072001}{(2009)}}
  \href{http://dx.doi.org/10.1088/0264-9381/26/7/072001}{\href{http://dx.doi.o%
rg/10.1088/0264-9381/26/7/072001}{072001}},
  \href{http://dx.doi.org/10.1088/0264-9381/26/7/072001}{[\href{http://xxx.lan%
l.gov/abs/0809.1255}{{\tt arXiv:0809.1255}}]}.

\bibitem{Alvarez:2010te}
E.~{\'A}lvarez and R.~Vidal, {\it {Comments on the vacuum energy decay}},  {\em
  JCAP} \href{http://dx.doi.org/10.1088/1475-7516/2010/11/043}{{\bf 2010}}
  \href{http://dx.doi.org/10.1088/1475-7516/2010/11/043}{\href{http://dx.doi.o%
rg/10.1088/1475-7516/2010/11/043}{(2010)}},
  \href{http://dx.doi.org/10.1088/1475-7516/2010/11/043}{no.~11}
  \href{http://dx.doi.org/10.1088/1475-7516/2010/11/043}{\href{http://dx.doi.o%
rg/10.1088/1475-7516/2010/11/043}{043}},
  \href{http://dx.doi.org/10.1088/1475-7516/2010/11/043}{[\href{http://xxx.lan%
l.gov/abs/1004.4867}{{\tt arXiv:1004.4867}}]}.

\bibitem{Boyanovsky:2004gq}
D.~Boyanovsky and H.~J. de~Vega, {\it {Particle decay in inflationary
  cosmology}},  {\em Phys. Rev.}
  \href{http://dx.doi.org/10.1103/PhysRevD.70.063508}{{\bf D70}}
  \href{http://dx.doi.org/10.1103/PhysRevD.70.063508}{\href{http://dx.doi.org/%
10.1103/PhysRevD.70.063508}{(2004)}}
  \href{http://dx.doi.org/10.1103/PhysRevD.70.063508}{\href{http://dx.doi.org/%
10.1103/PhysRevD.70.063508}{063508}},
  \href{http://dx.doi.org/10.1103/PhysRevD.70.063508}{[\href{http://xxx.lanl.g%
ov/abs/astro-ph/0406287}{{\tt astro-ph/0406287}}]}.

\bibitem{Boyanovsky:2012qs}
D.~Boyanovsky, {\it {Condensates and quasiparticles in inflationary cosmology:
  mass generation and decay widths}},  {\em Phys. Rev.}
  \href{http://dx.doi.org/10.1103/PhysRevD.85.123525}{{\bf D85}}
  \href{http://dx.doi.org/10.1103/PhysRevD.85.123525}{\href{http://dx.doi.org/%
10.1103/PhysRevD.85.123525}{(2012)}}
  \href{http://dx.doi.org/10.1103/PhysRevD.85.123525}{\href{http://dx.doi.org/%
10.1103/PhysRevD.85.123525}{123525}},
  \href{http://dx.doi.org/10.1103/PhysRevD.85.123525}{[\href{http://xxx.lanl.g%
ov/abs/1203.3903}{{\tt arXiv:1203.3903}}]}.

\bibitem{Bros:2006gs}
J.~Bros, H.~Epstein, and U.~Moschella, {\it {Lifetime of a massive particle in
  a de Sitter universe}},  {\em JCAP}
  \href{http://dx.doi.org/10.1088/1475-7516/2008/02/003}{{\bf 0802}}
  \href{http://dx.doi.org/10.1088/1475-7516/2008/02/003}{\href{http://dx.doi.o%
rg/10.1088/1475-7516/2008/02/003}{(2008)}}
  \href{http://dx.doi.org/10.1088/1475-7516/2008/02/003}{\href{http://dx.doi.o%
rg/10.1088/1475-7516/2008/02/003}{003}},
  \href{http://dx.doi.org/10.1088/1475-7516/2008/02/003}{[\href{http://xxx.lan%
l.gov/abs/hep-th/0612184}{{\tt hep-th/0612184}}]}.

\bibitem{Bros:2008sq}
J.~Bros, H.~Epstein, and U.~Moschella, {\it {Particle decays and stability on
  the de Sitter universe}},  {\em Annales Henri Poincar}
  \href{http://dx.doi.org/10.1007/s00023-010-0042-7}{{\bf 11}}
  \href{http://dx.doi.org/10.1007/s00023-010-0042-7}{\href{http://dx.doi.org/1%
0.1007/s00023-010-0042-7}{(2010)}}
  \href{http://dx.doi.org/10.1007/s00023-010-0042-7}{\href{http://dx.doi.org/1%
0.1007/s00023-010-0042-7}{611--658}},
  \href{http://dx.doi.org/10.1007/s00023-010-0042-7}{[\href{http://xxx.lanl.go%
v/abs/0812.3513}{{\tt arXiv:0812.3513}}]}.

\bibitem{Bros:2009bz}
J.~Bros, H.~Epstein, M.~Gaudin, U.~Moschella, and V.~Pasquier, {\it {Triangular
  invariants, three-point functions and particle stability on the de Sitter
  universe}},  {\em Comm. Math. Phys.}
  \href{http://dx.doi.org/10.1007/s00220-009-0875-4}{{\bf 295}}
  \href{http://dx.doi.org/10.1007/s00220-009-0875-4}{\href{http://dx.doi.org/1%
0.1007/s00220-009-0875-4}{(2010)}}
  \href{http://dx.doi.org/10.1007/s00220-009-0875-4}{\href{http://dx.doi.org/1%
0.1007/s00220-009-0875-4}{261--288}},
  \href{http://dx.doi.org/10.1007/s00220-009-0875-4}{[\href{http://xxx.lanl.go%
v/abs/0901.4223}{{\tt arXiv:0901.4223}}]}.

\bibitem{Hawking:1973uf}
S.~W. Hawking and G.~F.~R. Ellis, {\em {The Large scale structure of
  space-time}}.
\newblock Cambridge University Press, Cambridge, UK, 1973.

\bibitem{Birrell:1982ix}
N.~D. Birrell and P.~C.~W. Davies, {\em {Quantum fields in curved space}}.
\newblock Cambridge University Press, Cambridge, UK, 1982.
\newblock 340p.

\bibitem{Spradlin:2001pw}
M.~Spradlin, A.~Strominger, and A.~Volovich, {\it {Les Houches lectures on de
  Sitter space}},  \href{http://xxx.lanl.gov/abs/hep-th/0110007}{{\tt
  hep-th/0110007}}.

\bibitem{Allen:1985ux}
B.~Allen, {\it {Vacuum States in de Sitter Space}},  {\em Phys. Rev.}
  \href{http://dx.doi.org/10.1103/PhysRevD.32.3136}{{\bf D32}}
  \href{http://dx.doi.org/10.1103/PhysRevD.32.3136}{\href{http://dx.doi.org/10%
.1103/PhysRevD.32.3136}{(1985)}}
  \href{http://dx.doi.org/10.1103/PhysRevD.32.3136}{\href{http://dx.doi.org/10%
.1103/PhysRevD.32.3136}{\href{http://dx.doi.org/10.1103/PhysRevD.32.3136}{3136%
}}}.

\bibitem{Vilenken:1991aa}
N.~Y. Vilenken and A.~U. Klimyk, {\em {Representations of Lie Groups and
  Special Functions}}, vol.~1-3.
\newblock Dordrecht: Klower Acad. Publ., 1991.

\bibitem{Slater:1966}
L.~Slater, {\em Generalized hypergeometric functions}.
\newblock University Press, 1966.

\bibitem{Mottola:1984ar}
E.~Mottola, {\it {Particle Creation in de Sitter Space}},  {\em Phys. Rev.}
  \href{http://dx.doi.org/10.1103/PhysRevD.31.754}{{\bf D31}}
  \href{http://dx.doi.org/10.1103/PhysRevD.31.754}{\href{http://dx.doi.org/10.%
1103/PhysRevD.31.754}{(1985)}}
  \href{http://dx.doi.org/10.1103/PhysRevD.31.754}{\href{http://dx.doi.org/10.%
1103/PhysRevD.31.754}{\href{http://dx.doi.org/10.1103/PhysRevD.31.754}{754}}}.

\bibitem{Junker:1993aa}
G.~Junker, {\it Explicit evaluation of coupling coefficients for the most
  degenerate representations of so(n)},  {\em Jour. Phys. A}
  \href{http://dx.doi.org/10.1088/0305-4470/26/7/021}{{\bf 26}}
  \href{http://dx.doi.org/10.1088/0305-4470/26/7/021}{\href{http://dx.doi.org/%
10.1088/0305-4470/26/7/021}{(1993)}},
  \href{http://dx.doi.org/10.1088/0305-4470/26/7/021}{no.~7}
  \href{http://dx.doi.org/10.1088/0305-4470/26/7/021}{\href{http://dx.doi.org/%
10.1088/0305-4470/26/7/021}{\href{http://dx.doi.org/10.1088/0305-4470/26/7/021%
}{1649--1661}}}.

\bibitem{Hollands:2011we}
S.~Hollands, {\it {Massless interacting quantum fields in deSitter spacetime}},
   {\em Annales Henri Poincare}
  \href{http://dx.doi.org/10.1007/s00023-011-0140-1}{{\bf 13}}
  \href{http://dx.doi.org/10.1007/s00023-011-0140-1}{\href{http://dx.doi.org/1%
0.1007/s00023-011-0140-1}{(2012)}}
  \href{http://dx.doi.org/10.1007/s00023-011-0140-1}{\href{http://dx.doi.org/1%
0.1007/s00023-011-0140-1}{1039--1081}},
  \href{http://dx.doi.org/10.1007/s00023-011-0140-1}{[\href{http://xxx.lanl.go%
v/abs/1105.1996}{{\tt arXiv:1105.1996}}]}.

\bibitem{Bros:1990cu}
J.~Bros, {\it {Complexified de Sitter space: Analytic causal kernels and
  Kallen-Lehmann type representation}},  {\em Nucl. Phys. Proc. Suppl.} {\bf
  18B} (1991) 22--28.

\bibitem{Bros:1994dn}
J.~Bros, U.~Moschella, and J.~P. Gazeau, {\it {Quantum field theory in the de
  Sitter universe}},  {\em Phys. Rev. Lett.}
  \href{http://dx.doi.org/10.1103/PhysRevLett.73.1746}{{\bf 73}}
  \href{http://dx.doi.org/10.1103/PhysRevLett.73.1746}{\href{http://dx.doi.org%
/10.1103/PhysRevLett.73.1746}{(1994)}}
  \href{http://dx.doi.org/10.1103/PhysRevLett.73.1746}{\href{http://dx.doi.org%
/10.1103/PhysRevLett.73.1746}{\href{http://dx.doi.org/10.1103/PhysRevLett.73.1%
746}{1746--1749}}}.

\bibitem{Bros:1995js}
J.~Bros and U.~Moschella, {\it {Two-point Functions and Quantum Fields in de
  Sitter Universe}},  {\em Rev. Math. Phys.}
  \href{http://dx.doi.org/10.1142/S0129055X96000123}{{\bf 8}}
  \href{http://dx.doi.org/10.1142/S0129055X96000123}{\href{http://dx.doi.org/1%
0.1142/S0129055X96000123}{(1996)}}
  \href{http://dx.doi.org/10.1142/S0129055X96000123}{\href{http://dx.doi.org/1%
0.1142/S0129055X96000123}{327--392}},
  \href{http://dx.doi.org/10.1142/S0129055X96000123}{[\href{http://xxx.lanl.go%
v/abs/gr-qc/9511019}{{\tt gr-qc/9511019}}]}.

\bibitem{Weinberg:1995mt}
S.~Weinberg, {\em {The Quantum theory of fields. Vol. 1: Foundations}}.
\newblock Cambridge Univ. Press, 1995.

\bibitem{Srednicki:2007}
M.~Srednicki, {\em Quantum Field Theory}.
\newblock Cambridge Univ. Press, 2007.

\bibitem{Bousso:2001mw}
R.~Bousso, A.~Maloney, and A.~Strominger, {\it {Conformal vacua and entropy in
  de Sitter space}},  {\em Phys. Rev.}
  \href{http://dx.doi.org/10.1103/PhysRevD.65.104039}{{\bf D65}}
  \href{http://dx.doi.org/10.1103/PhysRevD.65.104039}{\href{http://dx.doi.org/%
10.1103/PhysRevD.65.104039}{(2002)}}
  \href{http://dx.doi.org/10.1103/PhysRevD.65.104039}{\href{http://dx.doi.org/%
10.1103/PhysRevD.65.104039}{104039}},
  \href{http://dx.doi.org/10.1103/PhysRevD.65.104039}{[\href{http://xxx.lanl.g%
ov/abs/hep-th/0112218}{{\tt hep-th/0112218}}]}.

\bibitem{Spradlin:2001nb}
M.~Spradlin and A.~Volovich, {\it {Vacuum states and the S-matrix in dS/CFT}},
  {\em Phys. Rev.} \href{http://dx.doi.org/10.1103/PhysRevD.65.104037}{{\bf
  D65}}
  \href{http://dx.doi.org/10.1103/PhysRevD.65.104037}{\href{http://dx.doi.org/%
10.1103/PhysRevD.65.104037}{(2002)}}
  \href{http://dx.doi.org/10.1103/PhysRevD.65.104037}{\href{http://dx.doi.org/%
10.1103/PhysRevD.65.104037}{104037}},
  \href{http://dx.doi.org/10.1103/PhysRevD.65.104037}{[\href{http://xxx.lanl.g%
ov/abs/hep-th/0112223}{{\tt hep-th/0112223}}]}.

\bibitem{Conamhna:2003aa}
O.~A.~P. Mac~Conamhna, {\it Massive bosons and the ds/cft correspondence},
  {\em Phys. Rev.} \href{http://dx.doi.org/10.1103/PhysRevD.67.084015}{{\bf
  D67}}
  \href{http://dx.doi.org/10.1103/PhysRevD.67.084015}{\href{http://dx.doi.org/%
10.1103/PhysRevD.67.084015}{(Apr, 2003)}}
  \href{http://dx.doi.org/10.1103/PhysRevD.67.084015}{\href{http://dx.doi.org/%
10.1103/PhysRevD.67.084015}{\href{http://dx.doi.org/10.1103/PhysRevD.67.084015%
}{084015}}}.

\bibitem{Lagogiannis:2011st}
P.~Lagogiannis, A.~Maloney, and Y.~Wang, {\it {Odd-dimensional de Sitter Space
  is Transparent}},  \href{http://xxx.lanl.gov/abs/1106.2846}{{\tt
  arXiv:1106.2846}}.

\bibitem{Brunetti:2005pr}
R.~Brunetti, K.~Fredenhagen, and S.~Hollands, {\it {A remark on alpha vacua for
  quantum field theories on de Sitter space}},  {\em JHEP}
  \href{http://dx.doi.org/10.1088/1126-6708/2005/05/063}{{\bf 05}}
  \href{http://dx.doi.org/10.1088/1126-6708/2005/05/063}{\href{http://dx.doi.o%
rg/10.1088/1126-6708/2005/05/063}{(2005)}}
  \href{http://dx.doi.org/10.1088/1126-6708/2005/05/063}{\href{http://dx.doi.o%
rg/10.1088/1126-6708/2005/05/063}{063}},
  \href{http://dx.doi.org/10.1088/1126-6708/2005/05/063}{[\href{http://xxx.lan%
l.gov/abs/hep-th/0503022}{{\tt hep-th/0503022}}]}.

\bibitem{Jordan:1986ug}
R.~D. Jordan, {\it {Effective Field Equations for Expectation Values}},  {\em
  Phys. Rev.} \href{http://dx.doi.org/10.1103/PhysRevD.33.444}{{\bf D33}}
  \href{http://dx.doi.org/10.1103/PhysRevD.33.444}{\href{http://dx.doi.org/10.%
1103/PhysRevD.33.444}{(1986)}}
  \href{http://dx.doi.org/10.1103/PhysRevD.33.444}{\href{http://dx.doi.org/10.%
1103/PhysRevD.33.444}{\href{http://dx.doi.org/10.1103/PhysRevD.33.444}{444--45%
4}}}.

\bibitem{Paz:1990jg}
J.~P. Paz, {\it {Anisotropy dissipation in the early universe: Finite
  temperature effects reexamined}},  {\em Phys. Rev.}
  \href{http://dx.doi.org/10.1103/PhysRevD.41.1054}{{\bf D41}}
  \href{http://dx.doi.org/10.1103/PhysRevD.41.1054}{\href{http://dx.doi.org/10%
.1103/PhysRevD.41.1054}{(1990)}}
  \href{http://dx.doi.org/10.1103/PhysRevD.41.1054}{\href{http://dx.doi.org/10%
.1103/PhysRevD.41.1054}{\href{http://dx.doi.org/10.1103/PhysRevD.41.1054}{1054%
--1066}}}.

\bibitem{Weinberg:2005vy}
S.~Weinberg, {\it {Quantum contributions to cosmological correlations}},  {\em
  Phys. Rev.} \href{http://dx.doi.org/10.1103/PhysRevD.72.043514}{{\bf D72}}
  \href{http://dx.doi.org/10.1103/PhysRevD.72.043514}{\href{http://dx.doi.org/%
10.1103/PhysRevD.72.043514}{(2005)}}
  \href{http://dx.doi.org/10.1103/PhysRevD.72.043514}{\href{http://dx.doi.org/%
10.1103/PhysRevD.72.043514}{043514}},
  \href{http://dx.doi.org/10.1103/PhysRevD.72.043514}{[\href{http://xxx.lanl.g%
ov/abs/hep-th/0506236}{{\tt hep-th/0506236}}]}.

\bibitem{Goldstein:2003qf}
K.~Goldstein and D.~A. Lowe, {\it {Real-time perturbation theory in de Sitter
  space}},  {\em Phys. Rev.}
  \href{http://dx.doi.org/10.1103/PhysRevD.69.023507}{{\bf D69}}
  \href{http://dx.doi.org/10.1103/PhysRevD.69.023507}{\href{http://dx.doi.org/%
10.1103/PhysRevD.69.023507}{(2004)}}
  \href{http://dx.doi.org/10.1103/PhysRevD.69.023507}{\href{http://dx.doi.org/%
10.1103/PhysRevD.69.023507}{023507}},
  \href{http://dx.doi.org/10.1103/PhysRevD.69.023507}{[\href{http://xxx.lanl.g%
ov/abs/hep-th/0308135}{{\tt hep-th/0308135}}]}.

\bibitem{Calzetta:1986ey}
E.~Calzetta and B.~L. Hu, {\it {Closed Time Path Functional Formalism in Curved
  Space- Time: Application to Cosmological Back Reaction Problems}},  {\em
  Phys. Rev.} \href{http://dx.doi.org/10.1103/PhysRevD.35.495}{{\bf D35}}
  \href{http://dx.doi.org/10.1103/PhysRevD.35.495}{\href{http://dx.doi.org/10.%
1103/PhysRevD.35.495}{(1987)}}
  \href{http://dx.doi.org/10.1103/PhysRevD.35.495}{\href{http://dx.doi.org/10.%
1103/PhysRevD.35.495}{\href{http://dx.doi.org/10.1103/PhysRevD.35.495}{495}}}.

\bibitem{Calzetta:1986cq}
E.~Calzetta and B.~L. Hu, {\it {Nonequilibrium Quantum Fields: Closed Time Path
  Effective Action, Wigner Function and Boltzmann Equation}},  {\em Phys. Rev.}
  \href{http://dx.doi.org/10.1103/PhysRevD.37.2878}{{\bf D37}}
  \href{http://dx.doi.org/10.1103/PhysRevD.37.2878}{\href{http://dx.doi.org/10%
.1103/PhysRevD.37.2878}{(1988)}}
  \href{http://dx.doi.org/10.1103/PhysRevD.37.2878}{\href{http://dx.doi.org/10%
.1103/PhysRevD.37.2878}{\href{http://dx.doi.org/10.1103/PhysRevD.37.2878}{2878%
}}}.

\bibitem{Higuchi:1986wu}
A.~Higuchi, {\it {Symmetric Tensor Spherical Harmonics On The N Sphere And
  Their Application To The De Sitter Group SO(N,1)}},  {\em Jour. Math. Phys.}
  \href{http://dx.doi.org/10.1063/1.527513}{{\bf 28}}
  \href{http://dx.doi.org/10.1063/1.527513}{\href{http://dx.doi.org/10.1063/1.%
527513}{(1987)}}
  \href{http://dx.doi.org/10.1063/1.527513}{\href{http://dx.doi.org/10.1063/1.%
527513}{\href{http://dx.doi.org/10.1063/1.527513}{1553}}}.

\bibitem{PS}
D.~Marolf and I.~A. Morrison, ``{The de Sitter Optical Theorem and shifts of
  conformal weights in the Hartle-Hawking state}.'' In preparation.

\bibitem{ELOP:1966}
R.~Eden, P.~Landshoff, D.~Olive, and J.~Polkinghorne, {\em The Analytic
  S-Matrix}.
\newblock Cambridge Univ. Press, 1966.

\bibitem{Allen:1987tz}
B.~Allen and A.~Folacci, {\it {The massless minimally coupled scalar field in
  de Sitter space}},  {\em Phys. Rev.}
  \href{http://dx.doi.org/10.1103/PhysRevD.35.3771}{{\bf D35}}
  \href{http://dx.doi.org/10.1103/PhysRevD.35.3771}{\href{http://dx.doi.org/10%
.1103/PhysRevD.35.3771}{(1987)}}
  \href{http://dx.doi.org/10.1103/PhysRevD.35.3771}{\href{http://dx.doi.org/10%
.1103/PhysRevD.35.3771}{\href{http://dx.doi.org/10.1103/PhysRevD.35.3771}{3771%
}}}.

\bibitem{Allen:1985wd}
B.~Allen and T.~Jacobson, {\it {Vector two point functions in maximally
  symmetric spaces}},  {\em Comm. Math. Phys.}
  \href{http://dx.doi.org/10.1007/BF01211169}{{\bf 103}}
  \href{http://dx.doi.org/10.1007/BF01211169}{\href{http://dx.doi.org/10.1007/%
BF01211169}{(1986)}}
  \href{http://dx.doi.org/10.1007/BF01211169}{\href{http://dx.doi.org/10.1007/%
BF01211169}{\href{http://dx.doi.org/10.1007/BF01211169}{669}}}.

\bibitem{Allen:1986tt}
B.~Allen and M.~Turyn, {\it {An evaluation of the graviton propagator In De
  SITTER space}},  {\em Nucl. Phys.}
  \href{http://dx.doi.org/10.1016/0550-3213(87)90672-9}{{\bf B292}}
  \href{http://dx.doi.org/10.1016/0550-3213(87)90672-9}{\href{http://dx.doi.or%
g/10.1016/0550-3213(87)90672-9}{(1987)}}
  \href{http://dx.doi.org/10.1016/0550-3213(87)90672-9}{\href{http://dx.doi.or%
g/10.1016/0550-3213(87)90672-9}{\href{http://dx.doi.org/10.1016/0550-3213(87)9%
0672-9}{813}}}.

\bibitem{Higuchi:2001uv}
A.~Higuchi and S.~S. Kouris, {\it {The Covariant graviton propagator in de
  Sitter space-time}},  {\em Class. Quant. Grav.}
  \href{http://dx.doi.org/10.1088/0264-9381/18/20/311}{{\bf 18}}
  \href{http://dx.doi.org/10.1088/0264-9381/18/20/311}{\href{http://dx.doi.org%
/10.1088/0264-9381/18/20/311}{(2001)}}
  \href{http://dx.doi.org/10.1088/0264-9381/18/20/311}{\href{http://dx.doi.org%
/10.1088/0264-9381/18/20/311}{4317--4328}},
  \href{http://dx.doi.org/10.1088/0264-9381/18/20/311}{[\href{http://xxx.lanl.%
gov/abs/gr-qc/0107036}{{\tt gr-qc/0107036}}]}.

\bibitem{Garriga:2007zk}
J.~Garriga and T.~Tanaka, {\it {Can infrared gravitons screen Lambda?}},  {\em
  Phys. Rev.} \href{http://dx.doi.org/10.1103/PhysRevD.77.024021}{{\bf D77}}
  \href{http://dx.doi.org/10.1103/PhysRevD.77.024021}{\href{http://dx.doi.org/%
10.1103/PhysRevD.77.024021}{(2008)}}
  \href{http://dx.doi.org/10.1103/PhysRevD.77.024021}{\href{http://dx.doi.org/%
10.1103/PhysRevD.77.024021}{024021}},
  \href{http://dx.doi.org/10.1103/PhysRevD.77.024021}{[\href{http://xxx.lanl.g%
ov/abs/0706.0295}{{\tt arXiv:0706.0295}}]}.

\bibitem{Tsamis:2007is}
N.~Tsamis and R.~Woodard, {\it {Comment on 'Can infrared gravitons screen
  Lambda?'}},  {\em Phys. Rev.}
  \href{http://dx.doi.org/10.1103/PhysRevD.78.028501}{{\bf D78}}
  \href{http://dx.doi.org/10.1103/PhysRevD.78.028501}{\href{http://dx.doi.org/%
10.1103/PhysRevD.78.028501}{(2008)}}
  \href{http://dx.doi.org/10.1103/PhysRevD.78.028501}{\href{http://dx.doi.org/%
10.1103/PhysRevD.78.028501}{028501}},
  \href{http://dx.doi.org/10.1103/PhysRevD.78.028501}{[\href{http://xxx.lanl.g%
ov/abs/0708.2004}{{\tt arXiv:0708.2004}}]}.

\bibitem{Tsamis:2011ep}
N.~Tsamis and R.~Woodard, {\it {A Gravitational Mechanism for Cosmological
  Screening}},  {\em Int. Jour. Mod. Phys.}
  \href{http://dx.doi.org/10.1142/S0218271811020652}{{\bf D20}}
  \href{http://dx.doi.org/10.1142/S0218271811020652}{\href{http://dx.doi.org/1%
0.1142/S0218271811020652}{(2011)}}
  \href{http://dx.doi.org/10.1142/S0218271811020652}{\href{http://dx.doi.org/1%
0.1142/S0218271811020652}{2847--2851}},
  \href{http://dx.doi.org/10.1142/S0218271811020652}{[\href{http://xxx.lanl.go%
v/abs/1103.5134}{{\tt arXiv:1103.5134}}]}.

\bibitem{Higuchi:2012vy}
A.~Higuchi, {\it {Equivalence between the Weyl-tensor and gauge-invariant
  graviton two-point functions in Minkowski and de Sitter spaces}},
  \href{http://xxx.lanl.gov/abs/1204.1684}{{\tt arXiv:1204.1684}}.

\bibitem{gauge}
D.~Marolf and I.~A. Morrison, ``{Gauge symmetries protect asymptotic behvior in
  de Sitter space}.'' In preparation.

\bibitem{Maldacena:2002vr}
J.~M. Maldacena, {\it {Non-Gaussian features of primordial fluctuations in
  single field inflationary models}},  {\em JHEP}
  \href{http://dx.doi.org/10.1088/1126-6708/2003/05/013}{{\bf 0305}}
  \href{http://dx.doi.org/10.1088/1126-6708/2003/05/013}{\href{http://dx.doi.o%
rg/10.1088/1126-6708/2003/05/013}{(2003)}}
  \href{http://dx.doi.org/10.1088/1126-6708/2003/05/013}{\href{http://dx.doi.o%
rg/10.1088/1126-6708/2003/05/013}{013}},
  \href{http://dx.doi.org/10.1088/1126-6708/2003/05/013}{[\href{http://xxx.lan%
l.gov/abs/astro-ph/0210603}{{\tt astro-ph/0210603}}]}.

\bibitem{Harlow:2011ke}
D.~Harlow and D.~Stanford, {\it {Operator Dictionaries and Wave Functions in
  AdS/CFT and dS/CFT}},  \href{http://xxx.lanl.gov/abs/1104.2621}{{\tt
  arXiv:1104.2621}}.

\bibitem{Anninos:2011ui}
D.~Anninos, T.~Hartman, and A.~Strominger, {\it {Higher Spin Realization of the
  dS/CFT Correspondence}},  \href{http://xxx.lanl.gov/abs/1108.5735}{{\tt
  arXiv:1108.5735}}.

\bibitem{Ng:2012xp}
G.~S. Ng and A.~Strominger, {\it {State/Operator Correspondence in Higher-Spin
  dS/CFT}},  \href{http://xxx.lanl.gov/abs/1204.1057}{{\tt arXiv:1204.1057}}.

\bibitem{Bateman:1955}
A.~Erdelyi, ed., {\em Higher transcendental functions}, vol.~1 of {\em Bateman
  Manuscript Project}.
\newblock McGraw-Hill, New York, 1953.

\bibitem{Smirnov:2004ym}
V.~A. Smirnov, {\it {Evaluating Feynman integrals}},  {\em Springer Tracts Mod.
  Phys.} {\bf 211} (2005), no.~0 1--244.

\end{thebibliography}\endgroup

\end{document}